\documentclass[12pt]{article}
\usepackage{epsfig,amssymb}
\usepackage{latexsym}

\newcommand{\bq}{\begin{equation}}
\newcommand{\eq}{\end{equation}}
\newcommand{\bqa}{\begin{eqnarray}}
\newcommand{\eqa}{\end{eqnarray}}
\newcommand{\ben}{\begin{enumerate}}
\newcommand{\een}{\end{enumerate}}
\newcommand{\bc}{\begin{center}}
\newcommand{\ec}{\end{center}}
\newcommand{\bqb}{\begin{eqnarray*}}
\newcommand{\eqb}{\end{eqnarray*}}

\def\lsim{\lesssim}

\def\pr#1#2#3{ Phys. Rev. ${\bf{#1}}$,#2 (#3)}

\def\pl#1#2#3{ Phys. Lett. ${\bf{#1}}$,#2 (#3)}
\def\prep#1#2#3{ Phys. Rep. ${\bf{#1}}$,#2 (#3)}

\def\np#1#2#3{ Nucl. Phys. ${\bf{#1}}$,#2 (#3)}
\def\zp#1#2#3{ Z. f. Phys. ${\bf{#1}}$,#2 (#3)}
\def\epj#1#2#3{ Eur. Phys. J. ${\bf{#1}}$,#2 (#3)}
\def\ijmp#1#2#3{ Int. J. Mod. Phys. ${\bf{#1}}$,#2 (#3)}

\def\cpc#1#2#3{Comput. Phys. Commun. ${\bf{#1}}$,#2 (#3)}

\def\ie{{\it i.e. ~}}
\def\eg{{\it e.g. ~}}

\def\etal{{\it et.al.~}}

\global\nulldelimiterspace = 0pt

\def\what#1{\widehat {#1}}

\def\L{ {\cal L }}
\def\Bcal{\tilde {\cal B}}

\def\sw{s_W}
\def\cw{c_W}
\def\swd{s^2_W}
\def\cwd{c^2_W}

\def\mwd{m_W^2}
\def\mw{m_W}
\def\mz{m_Z}
\def\mzd{m_Z^2}

\def\Sn#1{\mathrm{Sign} #1 }
\def\tchi{\tilde \chi}

\def\cbeta{\cos\beta}
\def\sbeta{\sin \beta}
\def\ssf{s_{\tilde \theta_f}}
\def\csf{c_{\tilde \theta_f}}
\def\ssu{s_{\tilde \theta_u}}
\def\csu{c_{\tilde \theta_u}}
\def\ssd{s_{\tilde \theta_d}}
\def\csd{c_{\tilde \theta_d}}

\begin{document}
\pagenumbering{arabic}
\thispagestyle{empty}
\def\thefootnote{\fnsymbol{footnote}}
\setcounter{footnote}{1}

\begin{flushright}
PM/02-16 \\
THES-TP 2002/02 \\
hep-ph/0207273 \\
July 2002\\

 \end{flushright}
\vspace{2cm}
\begin{center}
{\Large\bf Description of
$e^-e^+\to\gamma\gamma, Z \gamma, ZZ$ in SM and
MSSM\footnote{Programme d'Actions Int\'egr\'ees Franco-Hellenique,
 Platon 04100 UM}}
 \vspace{1.5cm}  \\
{\large G.J. Gounaris$^a$, J. Layssac$^b$  and F.M. Renard$^b$}\\
\vspace{0.2cm}
$^a$Department of Theoretical Physics, Aristotle
University of Thessaloniki,\\
Gr-54124, Thessaloniki, Greece.\\
\vspace{0.2cm}
$^b$Physique
Math\'{e}matique et Th\'{e}orique,
UMR 5825\\
Universit\'{e} Montpellier II,
 F-34095 Montpellier Cedex 5.\\

\vspace*{1.cm}

{\bf Abstract}
\end{center}
We present  a complete analysis of the one loop electroweak corrections
to  $e^-e^+\to\gamma\gamma, ~Z\gamma, ~ZZ$ in the
Standard (SM)  and the Minimal Supersymmetric Standard Model (MSSM).
Analytic expressions are written for the contributions to the
helicity amplitudes. Several observables accessible for polarized
or unpolarized beams and transverse, longitudinal or unpolarized
final states are computed. We show that in the few hundred GeV region,
these observables provide a test of the various SM or MSSM components.
For the dominant TT amplitude at high energy, the sensitivity to
the details of the various sectors disappears, but the energy
dependence fixed by leading logarithmic contributions, provides a
model independent signature discriminating SM from MSSM. Subdominant
TL or LL amplitudes though, remain sensitive to the details of the
SM or MSSM sectors. Numerical illustrations are given for energies
up to several TeV. The analysis may also be used  to search for new
physics characterized by anomalously strong interactions among
the neutral gauge bosons.\\


\def\thefootnote{\arabic{footnote}}
\setcounter{footnote}{0}
\clearpage

\section{Introduction}

The search for new
physics (NP) beyond the Standard Model (SM) has strongly motivated
projects of high energy $e^-e^+$ colliders (LC, CLIC)
\cite{LC, CLIC}. This NP search should proceed either in a direct way
(production of new
particles), or in an indirect way (observation of departures from
SM predictions in processes where the external particles are
standard, and  NP effects only arise from virtual exchanges).

In this paper we are addressing  the indirect way. The experimental
accuracy that should be available at the high luminosity $e^-e^+$
machines is expected to be very high; better than the percent level.
This means that the SM predictions, from which departures will
be searched, should be made with a comparable high accuracy, requiring
computations of high order effects of electroweak interactions.

One already knows that the electroweak radiative corrections
to several standard processes strongly increase with the energy.
This arises due to the presence already at the   1-loop level,
of large double (DL) and single (SL)
logarithm terms behaving like\footnote{In process like
$\gamma \gamma \to \gamma \gamma, ~ Z\gamma,~ ZZ$, which
do not contain any Born contribution, only
single logarithm terms caused by the imaginary part of  DL
contributions remain; the rest is canceled \cite{ggZZ-first}.}
$(\alpha/\pi) \ln^2s$,  $(\alpha/\pi) \ln s$,
\cite{Sud1, Sud2, log, Denner2}.
In the TeV range such terms reach the
several percent level, which renders them  observable
at the future colliders. Alternatively, these large logarithmic effects
may also be viewed as large  background
contributions to possible
NP signals. It will therefore
 be essential to have a full control on them, and
to analyse precisely the various virtual contributions they get from
each dynamical  sector.

The relevance of these large logarithmic effects
at high energy colliders, has been stressed recently
for the process
$e^+e^-\to f\bar f$ in the SM  \cite{log} and
MSSM \cite{MSSMlog} cases, and for the process
$e^+e^-\to \tilde f \bar{\tilde f}$ \cite{BMRV}.
As these 1-loop  effects are known to reach the
10\% level at the multi-TeV range,
the need for a two loop  computation and
even a resummation of the higher order leading effects arises;
 attempts in this direction have already started  \cite{BMRV, resum}.

A very important property of these large logarithms is that they
offer  a  striking signature for studying   the  underlying
dynamics \cite{class}. Depending on the interaction
sector (gauge, Yukawa) from which they originate, these large
logarithmic terms may be isotropic and universal with well defined
relative coefficients, or they may present  very specific
angular dependencies \cite{class}.
This has allowed a classification
of all such  log-terms and their possible physical origins \cite{class}.
In particular,  the logarithmic behaviour of the
$e^-e^+\to f \bar f,~\tilde f \bar{\tilde f}$ cross sections
 at high energy  reflects in an observable way
the gauge and Higgs structures of the
interactions, and even  differentiates between  SM and MSSM,
in a way which is largely independent of
the specific values of the MSSM parameters \cite{reality}.

Similar properties for the leading logarithmic SM and MSSM
contributions at high energies are  also expected in
 $\gamma\gamma\to f\bar{f}$  \cite{LR}, which should be
measurable at  photon-photon colliders \cite{ggcoll}.

The inverse process
$e^+e^-\to \gamma\gamma$ and
well as the  neutral gauge boson  production ones
$e^+e^-\to \gamma Z,~ZZ$ have been calculated in SM a long time ago
\cite{Denner1}, and received recently considerable
theoretical \cite{NAGCt, NAGCt1} and experimental \cite{NAGCe}
interest motivated by the search for anomalous neutral gauge boson
self couplings (NAGC).
At tree level, there is no NAGC coupling among three
neutral gauge bosons ($\gamma$ or $Z$) in SM or MSSM; \ie
no contribution of the
type $e^+e^-\to (\gamma,Z) \to \gamma Z, ~ ZZ$; (real
$\gamma\gamma$ final states are forbidden).

Non vanishing  NAGC  couplings
first arise at one loop, through fermionic triangles
involving leptons and quarks   in SM, and additional
chargino and neutralino  triangles in MSSM \cite{NAGCt1}.
Additional contributions may also come from NP forms  containing
\eg heavier fermions, non perturbative structures, or  even
direct neutral boson couplings. Since such NP effects may be
rather small, a complete and accurate
computation of the high order SM and MSSM contributions is needed,
in order to  identify them. \\

The aim of this paper is to discuss these various points. Thus,
we  analyze the content of the complete 1-loop contributions
to the $e^+e^-\to \gamma\gamma, ~\gamma Z, ~ZZ$ amplitudes,
firstly  within SM, and secondly within the MSSM.
 Since the  exact 1-loop formulae are rather complicated,
the study of the high energy behaviour of the amplitudes,
helps supplying a
 clear intuitive picture. We therefore study  in detail
the relative importance of each
type of asymptotic and non asymptotic contributions
(double log (DL), single log (SL),
angular independent and angular dependent terms)
in the gauge, Higgs and particle and sparticle sectors,
indicating  how these  sectors
conspire, to produce the correct high energy behaviour.
This should also be instructive for the
discussion of possible modifications due to NP.

As in the
fermion and sfermion production cases mentioned above
\cite{log,MSSMlog,BMRV}, the
numerical value of the SL coefficient  may serve as signature
discriminating between   SM and MSSM, in way which is
largely independent of the   specific values of the
MSSM parameters. In other words,  the dependence
on the specific  values of  the MSSM
parameters largely disappears,  once the
MSSM thresholds are overpassed\footnote{Only a
dependence in the overall  MSSM scale may remain in
some cases.}. These  discussions are done
in parallel for the three neutral processes
$e^+e^-\to \gamma\gamma,\gamma Z, ZZ$.
Many numerical  illustrations are also given.
An asymptotic energy  treatment for the SM case
of such amplitudes has also recently appeared
in \cite{Denner2}; we have checked that our results agree with
those of this reference.

We then concentrate on the role of the NAGC couplings
in  $e^+e^-\to \gamma Z,~ZZ$ and
compare their effect to the one of the other sectors  of
electroweak corrections, as well as to possible new
additional NP contributions.

Finally, we  discuss the role of  longitudinal $Z_L$ production.
The production of this state
is strongly depressed at the high energy. Moreover,  for  $Z_LZ_L$
production above 1TeV, the Born contribution is found  to be
negligible compared to the 1-loop one.
Such effects render the above processes very sensitive to virtual
contributions and provide  interesting checks of possible
anomalous NP contribution arising, for example, from
a strongly interacting Higgs sector.
We make this study at various energies,
showing the road to asymptopia, from the LEP2 energy range
to the LC and to the CLIC one.

The paper is organized as follows.
Section 2 contains the kinematics
for the three considered processes.
The one loop electroweak contributions
to the amplitudes are written in Section 3;
renormalized Born,  triangle  and box contributions.
Section 4 is devoted to the asymptotic properties.
Numerical applications are given in Section 5, while
the physics issues and conclusions are presented in Section 6.
Useful technical details are given in several Appendices;
details on kinematics in Appendix A;
the chargino and neutralino  mixing matrices in Appendix B;
the gauge and electron self-energies and renormalization constants
in Appendix C; details of triangle contributions in Appendix D;
asymptotic self-energy and triangle
contributions  in Appendix E and
 Box ones in Appendix F.

\section{Kinematics and Observable quantities}

We consider the process
\bq
e^-(\lambda, l)~+~e^+(\lambda', l')
\to V(e,p)+ V'(e',p') ~~ , \label{process}
\eq
where  $(l,~l')$ are the  incoming electron and positron momenta,
and $(\lambda,~\lambda')$ their corresponding helicities.
Since the electron mass is throughout  neglected, we have
$\lambda'= - \lambda= \pm 1/2 $.

Correspondingly,
$V$ and $V'$ denote  the outgoing neutral gauge bosons $Z$ or
$\gamma$,  whose momenta are described as   $(p,~p')$ respectively,
while $(e, ~ e')$ denote the complex conjugate of
their polarization vectors and
$(\mu, \mu')$ the corresponding  helicities. We also define
\bqa
&& q=l-p=p'-l' ~~~~ ,~~~~  q'=l-p'=p-l' ~~
\nonumber \\
&& s=(l+l')^2=(p+p')^2~~,~~t=q^2~~,~ ~
u=q'^2~ .
\nonumber
\eqa
The c.m. scattering angle between $\vec l$ and $\vec p$ is denoted
by $\theta$. The  helicity amplitude of the above process
(\ref{process}) is written as
\bqa
F_{\lambda,\mu,\mu'}&\equiv&
F[ e^-( \lambda , l )~ e^+(\lambda'=-\lambda,
 ~ l')~ \to ~V (e(\mu),p) ~V'(e'(\mu'),p') ]\nonumber\\
&&=\sum_{j=1,9} \bar{v}(\lambda', l')~I_j~N_j(s,t,u, \lambda)
~u(\lambda, l)~~, \label{hel}
\eqa
in terms of  nine  Lorentz invariant forms
 $I_j$, ~($j=1,9$)   defined in Appendix A. Their coefficients
 may then be split, according to the electron-helicity, as
\bq
N_j(s,t,u,\lambda )
 \equiv N^L_j(s,t,u)P_L+N^R_j(s,t,u)P_R ~~, \label{NjLR}
\eq
\noindent
where
\bq
P_L=\frac{1}{2} -\lambda ~~~,~~~P_R=\frac{1}{2} +\lambda ~~,
\label{proj}
\eq
while $N^{L,R}_j(s,t,u)$ are scalar functions.\\

\noindent
{\bf Observables}\\
The polarized angular distribution is obtained in terms of the helicity
amplitudes as:
\bq
{d\sigma(\lambda,\mu,\mu')\over d\cos\theta}
={\beta\over 32 \pi s}~C_{stat}
~|F_{\lambda,\mu,\mu'}|^2~~,
\eq
\noindent
with $C_{stat}=1/2,~1/2,~1$ for $\gamma\gamma,~ZZ,~\gamma Z$,
respectively. The corresponding integrated cross sections are
\bq
\sigma(\lambda,\mu,\mu')=\int^{c}_{-c} d\cos\theta~~
{d\sigma(\lambda,\mu,\mu')\over d\cos\theta} ~~,
\eq
\noindent
where $c\equiv \cos\theta_{min}$ is an angular cut (fixed
at $\theta_{min}=30^0$ in the numerical applications).\\

The cross section for unpolarized $e^{\pm}$ beams is
\bq
\sigma(\mu,\mu')_{\rm unp} ={1\over4}~
\sum_{\lambda =\pm 1/2}~\sigma(\lambda,\mu,\mu')~~,
\eq
\noindent
while we refer to
final transverse (T) or longitudinal (L) gauge bosons by
taking $\mu=\pm 1 $,  $ (\mu'=\pm1) $ or $\mu=0$, $(\mu'=0)$,
respectively.

For longitudinally polarized $e^{\pm}$ beams, the Left-Right
polarization asymmetry is defined as:
\bq
A_{LR}(\mu,\mu')=
\frac{\sigma(-{1\over2},\mu,\mu')-\sigma(+{1\over2},\mu,\mu')}
{\sigma(-{1\over2},\mu,\mu')+\sigma(+{1\over2},\mu,\mu')} ~~~.
 \label{ALR-definition}
\eq
In the numerical examples below, we only consider the Left-Right
asymmetry $A_{LR}$, where all possible final gauge boson
polarizations are summed over.

\section{The Born and 1-loop Amplitudes}

\subsection{The Born terms}

These  are due to electron exchange in the $t$ and $u$ channels.
In terms of the invariant functions defined in (\ref{NjLR})
and in Appendix A, they are written as
\bq
N_j^{\rm Born} = N_j^{\rm Born,~t} +N_j^{\rm Born,~u} ~~,
\label{Nj-Born}
\eq
which  give:\\
$\bullet$~ \underline{$e^-e^+\to\gamma\gamma$.}
\bqa
&& N^{\rm Born,~t}_1=N^{\rm Born,~t}_2=N^{\rm  Born,~t}_4
=  -~\frac{e_L^2}{t}P_L
-~\frac{e_R^2}{t}P_R
 ~~ , \nonumber \\
&& N^{\rm Born,~u}_1=N^{\rm Born,~ u}_2=
- N^{\rm Born,~ u}_4
=  -~\frac{e_L^2}{u}P_L
-~\frac{ e_R^2}{u}P_R
 ~~, \label{Nj-Born-gg}
\eqa
\noindent
with
\bq
e_L= e_R=~-e ~~; \label{eLR}
\eq
\noindent
$\bullet$~ \underline{$e^-e^+\to Z Z$.}
\bqa
&& N^{\rm Born,~t}_1
=  N^{\rm Born,~t}_2=N^{\rm Born,~t}_4
=-\, \frac{s}{t-\mzd} \, N^{\rm Born,~t}_5
  = -\, \frac{s}{2}\, N^{\rm Born,~t}_6\nonumber \\
&&
=-\, \frac{s}{s-t+\mzd }\, N^{\rm Born,~t}_7=
 \, \frac{s}{2}\, N^{\rm Born,~t}_8=
 -\, \frac{g^2_{ZL}}{t}\,-
 \, \frac{g^2_{ZR}}{t}\, ,
 \nonumber \\
&& N^{\rm Born,~u}_1
= N^{\rm Born,~u}_2=-N^{\rm Born,~u}_4
= \frac{-s}{s-u +\mzd} \, N^{\rm Born,~u}_5
\nonumber \\
&& =  \frac{-s}{2}\, N^{\rm Born,~u}_6=
\, \frac{-s}{u -\mzd }\,  N^{\rm Born,~u}_7=
   \frac{s}{2}~ N^{\rm Born,~u}_8=
 -\, \frac{g^2_{ZL}}{u}\, -\,
 \frac{g^2_{ZR}}{u}~  ,
\label{Nj-Born-ZZ}
\eqa
\noindent
with
\bq
g_{ZL}=e ~{(2s^2_{W}-1)\over2s_{W}c_{W}}~~~~~~~~~~~
g_{ZR}=e ~{s_{W}\over c_{W}} ~~;
\label{gZLR}
\eq
\noindent
$\bullet$~ \underline{$e^-e^+\to Z \gamma $.}
\bqa
&&\frac{s}{s-\mzd} ~ N^{\rm Born,~ t}_1
= \frac{s}{s +\mzd}~ N^{\rm Born,~ t}_2= N^{\rm Born,~t}_4
\nonumber \\
&&
=-~ \frac{s}{t-\mzd}~ N^{\rm Born,~t}_5
= - ~ \frac{s}{2}~ N^{\rm Born,~t}_6=
 -~\frac{e_Lg_{ZL}}{t}P_L
-~\frac{e_Rg_{ZR}}{t}P_R ~~ ,
 \nonumber \\
&&
\left (\frac{s}{s-\mzd}\right ) N^{\rm Born,~u}_1
= \left (\frac{s}{s +\mzd}\right ) N^{\rm Born,~u}_2
= -  N^{Born,~u}_4\nonumber \\
&&
=  \left (\frac{-s}{s -u}\right ) N^{\rm Born,~u}_5
 = \frac{-s}{2}~ N^{\rm Born,~u}_6=
 -~\frac{e_Lg_{ZL}}{u}P_L ~ -~\frac{ e_Rg_{ZR}}{u}P_R ~~ .
 \label{Nj-Born-Zg}
\eqa

Note that for all  processes
\bq
N^{\rm Born,~t}_{3,9}=N^{\rm Born,~u}_{3,9} =0 ~~ .
\label{N3-Born-ZZ}
\eq

In the following Sections,
the complete $e^-e^+\to VV'$ amplitudes at one loop are
obtained applying  the usual renormalization program
in the on-shell scheme \cite{Hollik}. The renormalized Lagrangian is
obtained from the unrenormalized one by
the substitution
\bqa
&& \psi_{eL} \to \sqrt{Z_{eL}}~~~,~~~\psi_{eR} \to \sqrt{Z_{eR}}~~,
\label{e-wave-function-renormalization} \\
&& B_\mu \to \sqrt{Z_B}\, B_\mu ~~~,~~~
\vec W_\mu \to \sqrt{Z_W}\, \vec W_\mu
\label{gauge-wave-function-renormalization} \\
&& g' \to  \frac{1}{\sqrt{Z_B}}\, g' ~~~,~~~
g \to \frac{\tilde Z_2}{\sqrt{Z_W}}~ g ~~~ ,
\label{gauge-renormalization-constants}
\eqa
where (\ref{e-wave-function-renormalization},
\ref{gauge-wave-function-renormalization}) describe the
wave function renormalization for the electron
and the gauge-bosons. Eqs.(\ref{gauge-renormalization-constants})
supply the renormalization of the gauge couplings,
taking  into account that
the $U(1)_Y$ Ward identity  guarantees  that $g'$ does not
need any additional renormalization, at least, at the 1-loop
level  \cite{Hollik}.
Contrary to these, the $SU(2)$
coupling $g$ does need the  additional renormalization
described by $\tilde Z_2$, which in the 'tHoof-Feynman gauge
 is determined by the $W$ and Goldstone loop contributions to the
$\gamma Z$ mixed self-energy.

The SM and MSSM contributions to the
various renormalization constants  are given in Sect.3.2
and Appendix C.
The explicit expressions of the 1-loop amplitudes are  given
in terms of $N_j$-functions containing the contributions
from the renormalized Born terms, the triangle and the
box diagrams according to
\bq
N_j(s,t,u) =N^{\rm ren+Born}_j+N^{\rm  Tri}_j+N_j^{\rm Box} ~ ~,
\label{Nj}
\eq
\noindent
which are  computed  in the subsequent
  subsections.

\vspace{0.5cm}
\subsection{The renormalized Born contribution}

The on-shell renormalization procedure \cite{Hollik}
uses as input
the electric charge $e(0)\equiv\sqrt{ 4\pi\alpha(0)}$,
 the physical masses $\mw,~\mz$,
and the  Weinberg angle defined\footnote{We follow
the usual convention
$W_\mu^3=\cw Z_\mu+\sw A_\mu$ and $B_\mu=-\sw Z_\mu+\cw A_\mu$,
which has  a sign difference compared to the one in \cite{Hollik}.}
by $c^2_W=1-s^2_W=m^2_W/\mzd$.
The renormalization introduces modifications to the Born
amplitude induced by the substitutions
(\ref{e-wave-function-renormalization},
\ref{gauge-wave-function-renormalization},
\ref{gauge-renormalization-constants}),
and  the self-energies given in Appendix C.

We separate the finite renormalized self-energy contributions
denoted with a "hat",  from the
divergent ones. The former are absorbed  in $N^{\rm ren+Born}_j$; while
the later are  put in $N^{\rm Tri }_j$,
 together  with the divergent triangle contributions
 presented in Sect.3.3. Thus, both $N^{\rm ren+Born}_j$ and
$N^{ Tri }_j$ are finite.

The finite hat-quantities entering $N^{\rm ren+Born}_j$ stem
from the renormalized electron self energy,
and the renormalized  $Z$ self energy and
$\gamma Z$ mixing on the Z-mass shell \cite{Hollik}.

In analogy to (\ref{Nj-Born}), we  write
\bq
N_j^{\rm ren+Born} = N_j^{\rm ren+Born,~t} +N_j^{\rm ren+Born,~u} ~~,
\label{Nj-Born-ren}
\eq
where the two terms in the r.h.s. arise from electron exchanges in
the $t$- and $u$-channel respectively.\\

\noindent
$\bullet$~ \underline{$e^-e^+\to\gamma\gamma$.}\\
According to the aforementioned conventions,
$N_j^{\rm ren+Born}$ are  obtained from
 (\ref{Nj-Born-gg},\ref{eLR}) by the  replacement
\bqa
&&
e_L^2 \Longrightarrow  \hat e_L^2(x) \equiv  4\pi\alpha(0)
[1-\hat\Sigma_{Le}(x) ] ~~, \nonumber \\
&&e_R^2  \Longrightarrow  \hat e_R^2(x) \equiv  4\pi\alpha(0)
[1-\hat\Sigma_{Re}(x) ] ~~, \label{rebar}
\eqa
where $x$ stands for $t$ or $u$.
The counter terms for the gauge boson-electron vertices
which could contribute to  (\ref{rebar})
through the  divergent factors
\bq
[1+2 \delta Z_{Le} +\delta \tilde Z_2 ]~~~~~~,~~~~~~~
[1+2\delta Z_{Re}]~~~~,
\label{ctgg}\eq
\noindent
for $e_L^2$, $e_R^2$ respectively, are (as said above) absorbed in the
triangle contributions $N_j^{\rm Tri}$; see (\ref{TLR-gg}).\\

\noindent
$\bullet$~ \underline{$e^-e^+\to Z Z$.}\\
The replacement to be made now  in (\ref{Nj-Born-ZZ},\ref{gZLR}) is:
\bqa
&& g_{ZL}^2 \Longrightarrow \what{g^2}_{ZL}(x)
\equiv
\frac{4\pi\alpha(0)(1-2s^2_W)^2}{4s^2_Wc^2_W}
\Big [1-\hat\Sigma_{Le}(x) -\,\hat \Sigma_{ZZ}'(\mzd)
+\, \frac{4s_Wc_W \hat \Sigma_{\gamma Z}(\mzd)}{(2s^2_W-1)\mzd} \Big ]
~, \nonumber \\
&& g_{ZR}^2 \Longrightarrow \what{g^2}_{ZR}(x)
\equiv\frac{4\pi\alpha(0)s^2_W}{c^2_W}
\Big [1-\hat\Sigma_{Re}(x)
-~\hat \Sigma_{ZZ}'(\mzd)
+~\frac{2c_W\hat\Sigma_{\gamma Z}(\mzd) }{s_W\mzd}\Big ]~,
\label{rcouplings-Born-ZZ}
\eqa
while the $Zee$ counter terms which
would  contribute through the additional factors
\bq
 \Big [1+ 2\delta Z_{Le}
 +\, \frac{2 \cwd }{(1-2s^2_W)}~ \delta \tilde Z_2
 \Big ]~~~~~,~~~~~~~~  [1+ 2\delta Z_{Re}]~~~, \label{ctZZ}
\eq
\noindent
 will be put together with  the
triangle contributions, in order to make finite
quantities; see   (\ref{TLR-ZZ}).\\

\noindent
$\bullet$~ \underline{$e^-e^+\to Z \gamma $.}\\
The replacement in (\ref{Nj-Born-Zg}) is now
\bqa
&& e_Lg_{ZL}\Longrightarrow \what {e_Lg_{ZL}}(x)
\equiv\frac{4\pi\alpha(0)(1-2s^2_W)}{2s_Wc_W}
\Big [1-\hat\Sigma_{Le}(x)
 -\, \frac{ \hat \Sigma_{ZZ}'(\mzd)}{2}
+\, \frac{2s_Wc_W\hat\Sigma_{\gamma Z}(\mzd) }{(2s^2_W-1)\mzd}\Big ]
~, \nonumber\\
&&e_Rg_{ZR} \Longrightarrow \what{e_Lg_{ZR}}(x)
\equiv  -\, \frac{4\pi\alpha(0)s_W}{c_W}
\Big [1-\hat\Sigma_{Re}(x)
-\,\frac{\hat \Sigma_{ZZ}'(\mzd)}{2}
+\,\frac{c_W\hat \Sigma_{\gamma Z}(\mzd) }{s_W\mzd}\Big ]~,
\label{rcouplings-Born-Zg}
\eqa
while the additional divergent factors
\bqa
\Big [1+ 2\delta Z_{Le}
 +\, \frac{(3-4\swd )}{2(1-2s^2_W)}~ \delta \tilde Z_2
\Big ], ~~~~~~~~~~~  [1+ 2\delta Z_{Re}  ]~
\label{ctZg}
\eqa
generated by the counter terms, will again  be put together with the
triangle contributions  in (\ref{Tt-Zg}).\\

The complete expressions of the various self-energy functions and
counter terms are given in Appendix C.

\subsection{Triangle contributions}

These arise from   triangle diagrams of the type  depicted in
Fig.1d, 1e and from the diagrams in Fig.1f
which induce  "anomalous neutral gauge couplings" NAGC \cite{NAGCt}.
Below and in Appendix D, we give the complete expressions
for these contributions, while in Appendix E
we quote their dominant leading logarithmic contribution
when  $s,~t,~u $ are all much larger
than all internal and external  masses.
As already said,  these "Triangle" contributions to the various
$N_j^{\rm Tri}$ amplitudes, also include the counter term factors
in (\ref{ctgg}, \ref{ctZZ}, \ref{ctZg}), which guarantee their
finiteness. This has been checked using the expressions
in  Appendix D.

\vspace{0.5cm}
\noindent
$\bullet$ ~ \underline{$e^-e^+\to \gamma\gamma$}\\
In this case, the  diagram of the type Fig.1d
gives the SM contributions generated by loops involving
the particle-strings
\[
 (abc) \equiv (\gamma e e) , ~~(Zee), ~~(\nu_eWW) ~~,
 \]
 while the MSSM contributions involve
\[
 (abc) \equiv (\tilde \nu_e \tchi^+_i \tchi^+_j),
 ~~(\tchi^0_i\tilde e \tilde e)~.
\]

The generic diagram  Fig.1e  only induces an  SM contribution
involving the particle-strings
\[
(abc) \equiv (\nu_eWW) ~~,
\]
containing  the 4-leg $WW\gamma\gamma$ coupling.
There is no NAGC contribution from Fig.1f, for two on-shell
final photons.

The resulting contributions to the $N^{Tri}_j$ functions
are
\bqa
&&N^{Tri}_1=N^{Tri}_2 = \alpha^2~
\Big [{2T^{\gamma}_t+T'^{\gamma}_t\over t}
+{2T^{\gamma}_u+T'^{\gamma}_u\over u}+N''^{ \gamma}_1P_L\Big ]
\nonumber\\
&&N^{Tri}_3=0~~,
\nonumber \\
&& N^{Tri}_4 =  \alpha^2~
\Big [{2T^{\gamma}_t\over t}-{2T^{\gamma}_u\over u}\Big ] ~.
\label{N-Tri-gg}
\eqa
Separating the $L,R$ parts of $T$ terms in (\ref{N-Tri-gg}),
and adding the divergent counter term
corrections  generated from (\ref{ctgg}),  we write
\bq
T^{\gamma}\equiv [T^{\gamma L}+\delta^{\gamma L}]P_L+
[T^{\gamma R}+\delta^{\gamma R}]P_R ~,
\label{TLR-gg}
\eq
\noindent
with
\bqa
\delta^{\gamma L}& =& -~{4\pi\over\alpha}[
\delta Z_{Le} +{1\over2}\delta \tilde Z_2 ]
~~, \label{dgL} \\
\delta^{\gamma R} &=& -~{4\pi\over\alpha}
\, \delta Z_{Re} ~~~~\label{dgR}
\eqa
obtained from (\ref{ctgg}), and
\bq
T'^{\gamma}\equiv T'^{\gamma L}P_L+
T'^{\gamma R}P_R ~.
\label{TpLR-gg}
\eq

The SM contributions arising from the triangles involving
$\gamma,~Z,~W$ exchanges are then written as
\bqa
T^{\gamma L~SM}_t&=& b^L_{\gamma}(t)
+~{(2s^2_W-1)^2\over4s^2_Wc^2_W}b^L_{Z}(t)
+~{1\over2s^2_W}b^L_{W}(t)~, \label{Tt-SM-L-gg} \\
T^{\gamma R~SM}_t&=& b^R_{\gamma}(t)
+~{s^2_W\over c^2_W}b^R_{Z}(t)~, \label{Tt-SM-R-gg} \\
T'^{\gamma L~SM}_t&=& a^L_{\gamma}(t)
+~{(2s^2_W-1)^2\over4s^2_Wc^2_W}a^L_{Z}(t)
+~{1\over2s^2_W}a^L_{W}(t) ~, \label{Ttp-SM-L-gg}\\
T'^{\gamma R~SM}_t&=& a^R_{\gamma}(t)
+~{s^2_W\over c^2_W}a^R_{Z}(t) ~, \label{Ttp-SM-R-gg}
\eqa
while the MSSM contributions due to triangles involving chargino or
neutralino exchanges are
\bqa
T^{\gamma L~MSSM}_t&=&- ~{1\over s^2_W}b^L_{2\tchi}(t) -~
{1\over s^2_Wc^2_W}b^L_{1\tchi}(t)~, \label{Tt-MSSM-L-gg} \\
T^{\gamma R~MSSM}_t&=&-~
{4\over c^2_W}b^R_{1\tchi}(t) ~,\label{Tt-MSSM-R-gg} \\
T'^{\gamma L~MSSM}_t&=&-~ {1\over s^2_W}a^L_{2\tchi}(t)
-~{1\over s^2_Wc^2_W}a^L_{1\tchi}(t )~,  \label{Ttp-MSSM-L-gg}\\
T'^{\gamma R~MSSM}_t&=& -~{4\over c^2_W}a^R_{1\tchi}(t)
~  \label{Ttp-MSSM-R-gg},
\eqa
where  $a^{L,R}_i,~b^{L,R}_i$ and $N''^{\gamma}_1$
corresponding to each triangle
diagram are given in (\ref{fig1d-gg}, \ref{fig1e-gg})
in terms of  Passarino-Veltman
functions \cite{Passarino}, in which the internal
propagator masses are determined
by the particle-strings mentioned above. The analogous
$u$-channel expressions are obtained correspondingly.

The results for the asymptotic regime where
 $s,~t,~u $ are much larger
than all internal propagator  masses, are given in Appendix E.

\vspace{0.5cm}
\noindent
$\bullet $ ~ \underline{$e^-e^+\to ZZ$}\\
The diagrams of the type of  Fig.1d supply the SM
contributions  due to the particle-strings
\[
 (abc)  \equiv (\gamma e e), ~~(Zee), ~~(W\nu_e\nu_e),
 ~~ (\nu_eWW)~,
\]
and the  MSSM ones through
\[
 (abc)\equiv (\tilde{\nu}_e \tchi^+_i \tchi^+_j), ~~
(\tilde{e} \tchi^0_i \tchi^0_j),
~~(\tchi^+_i\tilde \nu_e \tilde \nu_e),
~~(\tchi^0_i\tilde{e}\tilde{e}) ~~.
\]
The diagram of Fig.1e induces just an  SM contribution for
\[
(abc)\equiv (\nu_eWW) ~,
\]
 involving the $WWZZ$ coupling.
Finally,  Fig.1f, which can only involve  a fermionic triangle
(leptons and quarks in SM, and charginos and neutralinos in MSSM),
supplies the NAGC contribution to the couplings $f^{\gamma,Z}_5$
\cite{NAGCt}.

The  set of these triangular contributions is described as
\bqa
&&N^{Tri}_1=N^{Tri}_2 = \alpha^2~
\Big [{2T^Z_t+T'^Z_t\over t}
+{2T^Z_u+T'^Z_u\over u}+N''^{Z}_1P_L \Big ] ~,
\nonumber\\
&&N^{Tri}_3=N^{Tri}_9=0~~,
\nonumber \\
&& N^{Tri}_4 =  \alpha^2~
\Big [{2T^Z_t\over t}-{2T^Z_u\over u}\Big ]~,
\nonumber\\
&&
N^{Tri}_5 = \alpha^2~
\Big [\Big ({\mzd-t\over s}\Big )~{2T^Z_t\over t}+~{1\over u}~
\Big ({(u-s-\mzd) \over s} 2T^Z_u-T'^Z_u\Big )+N''^{Z}_5P_L\Big ]
+N^{AGC}_5 ~,
\nonumber\\
&&N^{Tri}_6 = -N^{Tri}_8=\alpha^2~
\Big [-~{2\over s}\Big ({2T^Z_t\over t}+{2T^Z_u\over u}\Big )\Big ]
+N^{AGC}_6 ~,
\nonumber\\
&&N^{Tri}_7 = \alpha^2~\Big [{1\over t}
\Big ({(t-\mzd-s) \over s}\, 2T^Z_t-T'^Z_t\Big )
+\Big ({\mzd-u\over s}\Big )~{2T^Z_u\over u}
+N''^{Z}_5P_L \Big ] +N^{AGC}_7  . \label{N-Tri-ZZ}
\eqa

The $(L,R)$ decompositions for $T$ and $T'$ above,
after including also the divergent  $Zee$ counter terms
from  (\ref{ctZZ}), become
\bq
T^Z\equiv {2s^2_W-1\over2s_Wc_W}~[T^{ZL}+\delta^{Z L}] P_L
+{s_W\over c_W}
~[T^{ZR}+\delta^{Z R}] P_R ~, \label{TLR-ZZ}
\eq
\noindent
with
\bqa
\delta^{Z L}& = & -~
\left ({ 2\pi(2s^2_W-1)\over \alpha s_Wc_W}\right )
[\delta Z_{Le} +{c^2_W\over1-2s^2_W}\delta \tilde Z_2 ]
~~ ~, \label{dZL} \\
\delta^{Z R} &= & -~\left ({4\pi \sw\over\alpha \cw}\right )
\, \delta Z_{Re} ~~~, \label{dZR}
\eqa
and
\bq
T'^Z\equiv {2s^2_W-1\over2s_Wc_W}~T'^{ZL} P_L+{s_W\over c_W}
~T'^{ZR} P_R ~. \label{TpLR-ZZ}
\eq

The SM Figs.1d,e triangle contributions to them are  given by
\bqa
T^{ZL~SM}_t&=& {2s^2_W-1\over2s_Wc_W}~
b^L_{\gamma}(t)
+{(2s^2_W-1)^3\over8s^3_Wc^3_W}b^L_{Z}(t)+{1\over4s^3_Wc_W}
b''^L_{W}(t)
-~{c_W\over2s^3_W}b'^L_{W}(t) ~, \label{Tt-SM-L-ZZ} \\
T^{ZR~SM}_t&=& {s_W\over c_W}b^R_{\gamma}(t)
+{s^3_W\over c^3_W}b^R_{Z}(t) ~, \label{Tt-SM-R-ZZ} \\
T'^{ZL~SM}_t&=&
{2s^2_W-1\over2s_Wc_W}a^L_{\gamma}(t)+~
{(2s^2_W-1)^3\over8s^3_Wc^3_W}a^L_{Z}(t)
+{1\over4s^3_Wc_W}a''^L_{W}(t)-{c_W\over2s^3_W}a'^L_{W}(t)
~, \label{Ttp-SM-L-ZZ} \\
T'^{ZR ~SM}_t&=& {s_W\over c_W}a^R_{\gamma}(t)+
{s^3_W\over c^3_W}a^R_{Z}(t) ~, \label{Ttp-SM-R-ZZ}
\eqa
\noindent
while the MSSM contributions are
\bqa
&& T^{ZL~MSSM }_t =
-~{1\over 4s^3_Wc^3_W}b'^L_{2\tchi}(t)+
{1\over 2s^3_Wc_W}b''^L_{2\tchi}(t)-~{1\over s^3_Wc_W}b'^L_{1\tchi}(t)
+{1\over 2s^3_Wc^3_W}b''^L_{1\tchi}(t) ~,  \label{Tt-MSSM-L-ZZ} \\
&& T^{ZR~MSSM }_t=
-~{1\over 4s^3_Wc^3_W}b'^R_{2\tchi}(t)
+{1\over 2s^3_Wc^3_W}b''^R_{1\tchi}(t) ~, \label{Tt-MSSM-R-ZZ}\\
&&T'^{ZL ~MSSM }_t= -~{1\over 4s^3_Wc^3_W}a'^L_{2\tchi}(t)
+{1\over 2s^3_Wc_W}a''^L_{2\tchi}(t)
-{1\over s^3_Wc_W}a'^L_{1\tchi}(t)
+{1\over 2s^3_Wc^3_W}a''^L_{1\tchi}(t) ~, \label{Ttp-MSSM-L-ZZ} \\
&&T'^{ZR~MSSM}_t =-~{1\over 4s^3_Wc^3_W}a'^R_{2\tchi}(t)~
+{1\over 2s^3_Wc^3_W}a''^R_{1\tchi}(t)~, \label{Ttp-MSSM-R-ZZ}
\eqa
where $a^{L,R}_i,~b^{L,R}_i$, $N''^Z_{1,5}$ are calculated
from the diagrams in Figs.1d,1e and given
in (\ref{fig1d-ZZ}, \ref{fig1e-ZZ}).

Finally,  the NAGC contribution induced from Fig.1f in
(\ref{N-Tri-ZZ}) is
\bqa
&&N^{AGC}_5=-N^{AGC}_7=
\Big [{2(\mzd-u)\over s}-1 \Big ]N^{AGC} ~, \nonumber\\
&&N^{AGC}_6= -N^{AGC}_8={4\over s} ~N^{AGC} ~,
\label{N-Tri-NAGC-ZZ}
\eqa
with
\bqa
N^{AGC}&=&
\Big ({e^2\over \mzd}\Big )\Big [f^{\gamma}_5(P_R-P_L)-
f^{Z}_5\Big ({1-2s^2_W\over2s_Wc_W}P_L+{s_W\over c_W}P_R\Big )\Big ]
~, \label{N-Tri-NAGC-ZZ-res}
\eqa
where  $f^{\gamma,Z}_5$ are taken from \cite{NAGCt1},
apart from the neutralino loop case with general  mixings
which was not considered in \cite{NAGCt1} and
 is given in (\ref{f5Z-neutralino}).

As for the $\gamma \gamma$ production case,
the dominant logarithmic terms in the asymptotic regime where
 $s,~t,~u $ are much larger
than all internal propagator  masses, are given in Appendix E.

\vspace{0.5cm}
\noindent
$\bullet$ ~ \underline{$e^-e^+\to  Z\gamma$}\\
Contributions in this case arise
from  diagrams of type Fig.1d already considered for the
$e^-e^+\to \gamma\gamma$ and
$e^-e^+\to Z Z$ process. In addition we also have
the $(\nu_eWW)$ contribution from the diagram of Fig.1e
with the 4-leg $WW\gamma Z$ vertex, and a Fig.1f
 NAGC contribution to the couplings
 $h^{\gamma,Z}_3$ \cite{NAGCt, NAGCt1}
containing a fermionic triangle consisting of leptons, quarks
in SM, and charginos, neutralinos in MSSM.
The whole set of these triangular contributions may be
written as
\bqa
&&N^{Tri}_1=
\alpha^2~
\Big [\Big (1-{\mzd\over s}\Big )\Big ({T^{\gamma Z}_t\over t}
+{T^{\gamma Z}_u\over u}\Big )
+{T'^{\gamma Z}_t\over t}+{T'^{\gamma Z}_u\over u}
+N''^{\gamma Z}_1\Big ]
+N^{AGC~\gamma Z}_1
~, \nonumber\\
&&N^{Tri}_2 = \alpha^2~
\Big[ \Big (1+{\mzd\over s}\Big )\Big({T^{\gamma Z}_t\over t}
+{T^{\gamma Z}_u\over u}\Big )
+{T'^{Z\gamma}_t\over t}+{T'^{Z\gamma}_u\over u}+N''^{\gamma Z}_1\Big ]
+N^{AGC~\gamma Z}_2
~, \nonumber \\
&&N^{Tri}_3=N^{Tri}_9=0~~,
~, \nonumber \\
&& N^{Tri}_4 = \alpha^2~
\Big [{T^{\gamma Z}_t\over t}-{T^{\gamma Z}_u\over u}\Big]
 ~, \nonumber\\
&& N^{Tri}_5 = \alpha^2~
\Big [\Big ({u-s\over s}\Big ){T^{\gamma Z}_u\over u}
+\Big ({\mzd-t\over s}\Big ){T^{\gamma Z}_t\over t}-
{T'^{\gamma Z}_u\over u}
+N''^{\gamma Z}_5\Big ] +N^{AGC~\gamma Z}_5
~, \nonumber\\
&& N^{Tri}_6 =\alpha^2~
\Big [-~{2\over s}\Big ({T^{\gamma Z}_t\over t}
+{T^{\gamma Z}_u\over u}\Big ) \Big]  +N^{AGC~\gamma Z}_6  ~.
\label{N-Tri-Zg}
\eqa
Decomposing $T^{\gamma Z}$, $T'^{\gamma Z}$,
$T'^{Z\gamma}$ in their L,R components as
\bq
T\equiv T^L P_L+ T^RP_R ~,
\label{TLR-Zg}
\eq
in analogy with (\ref{TLR-gg}) and including also the
divergent contributions  from Sect.3.2,  already defined in
(\ref{dgL}, \ref{dgR}, \ref{dZL}, \ref{dZR}), we get
\bqa
T^{\gamma ZL}_t& = &- [T^{ZL}_t+\delta^{Z L}]-
\Big({2s^2_W-1\over 2s_Wc_W}\Big )
[T^{\gamma L}_t+\delta^{\gamma L}] ~, \nonumber \\
T^{\gamma ZR}_t & =& -[T^{ZR}_t+\delta^{ZR }]
-~{s_W\over c_W}\, [T^{\gamma R}_t+\delta^{\gamma R}]
~, \label{Tt-Zg}
\eqa
\noindent
and
\bqa
&&T'^{\gamma ZL}_t =- T'^{ZL}_t~~~~~~~~~~~~~~
T'^{\gamma ZR}_t = - T'^{ZR}_t\nonumber \\
&&T'^{Z\gamma L}_t =-\Big({2s^2_W-1\over 2s_Wc_W}\Big )T'^{\gamma L}_t
~~~~~~~~~~~~~~~
T'^{Z\gamma L}_t =-~{s_W\over c_W}\, T'^{\gamma R}_t
~, \label{Tpt-Zg}
\eqa
\noindent
using the triangle functions
already defined for  the $\gamma\gamma$ and  $ZZ$ cases
in (\ref{Tt-SM-L-gg}-\ref{Ttp-MSSM-R-gg},
\ref{Tt-SM-L-ZZ}-\ref{Ttp-MSSM-R-ZZ}).
The quantities $N''^{\gamma Z}_{1,5}$ are derived from
Fig.1e and given in (\ref{fig1e-Zg})

Finally, the NAGC parts induced by the diagram in Fig.1f are
\bqa
N^{AGC~\gamma Z}_1 &= &-~ N^{AGC~\gamma Z}_2~=~
\Big (1-{\mzd\over s}\Big )
N^{AGC~\gamma Z}~, \nonumber \\
N^{AGC~\gamma Z}_5 &= & {u\over s}\, N^{AGC~\gamma Z}
~, \nonumber \\
N^{AGC~\gamma Z}_6 &= & -~{2\over s}\, N^{AGC~\gamma Z}
~, \nonumber \\
N^{AGC~\gamma Z}&=&{e^2\over \mzd}\Big [h^{\gamma}_3(P_R-P_L)-
h^{Z}_3\Big ({1-2s^2_W\over2s_Wc_W}P_L+{s_W\over c_W}P_R\Big )\Big ]
~, \label{NAGC-Zg}
\eqa
with the form factors $h^{\gamma,~Z}_3$ obtained
from \cite{NAGCt1}.

The leading logarithmic terms in the asymptotic regime are
again given in Appendix E.

\subsection{Box contributions}

The generic box diagrams contributing to  the processes
$e^+e^-\to V'V$ are shown in Fig.1g,h labeled as $(abcd)$,
according to the particles in the four propagators.
There are seven  kinds  ($k=1,...7$) of such box
contributions, which combined with the nature  of particles
(fermion $f$, vector $V$, scalar $S$) running inside the
loop, create altogether 11 types of contributions labeled as

\begin{tabular}{cc}
type 1: Fig.1g$(Vfff)$; & type 2: Fig.1g$(fVVV)$; \\
type 3: Fig.1h$(VffV)$; &~~~~~~~~~~~~~~~
type 4A, 4B, 4C, 4D: Fig.1g$(Sfff)$; \\
type 5: Fig.1g$(fSSS)$; &~~~~ type 6A, 6B: Fig.1g$(SffS)$; \\
type 7: Fig.1g$(fVSV)$.  &
\end{tabular}

\noindent
Concerning the above list, we should note that the
separation of  the $k=4$ contributions into four parts
labeled 4A, 4B, 4C and 4D,
and the analogous  separation
of the $k=6$ ones  into 6A and 6B,
is induced by the appearance of  different combinations of
the kinetic and mass parts in the fermion propagators.

The generic contributions to each of these eleven types are
denoted as $\bar{N}^{k,~Box}_j(s,t,u)$ and expressed
 in terms of Passarino-Veltman functions \cite{Passarino}.
Because of their complexity, we only write in Appendix F
their leading logarithmic
contributions, which serve also to  define them.
MATHEMATICA and FORTRAN codes determining $\bar{N}^{k,~Box}_j(s,t,u)$
 in terms of the Passarino-Veltman functions, are available upon request.

Multiplying the $\bar{N}^{k,~Box}_j(s,t,u)$ functions
by the appropriate coupling combinations,
we obtain the contributions $N_j^{Box}$ to be inserted
in (\ref{Nj}). These are given below in
the  SM and  MSSM cases, for each of the three neutral gauge
boson production  processes  considered.

\vspace{0.5cm}
\noindent
$\bullet$ ~ \underline{$e^-e^+\to \gamma \gamma$}\\
The SM contributions arise from  the type 1 boxes:
$(\gamma eee)$ and $(Z eee)$, the
type 2: $(\nu_e W^+W^+W^+)$, and the
type 7: $(\nu_eW^+G^+W^+)$.
The additional MSSM contributions come from the type 4 boxes:
$(\tilde \nu_e \tchi^+_i\tchi^+_j\tchi^+_k)$, and the type 5:
$(\tchi^0_i\tilde{e}\tilde{e}\tilde{e})$. These are
\bqa
N^{\gamma\gamma~SM~Box}_j&= & \alpha^2
\Big \{\bar{N}^{1,Box}_j(\gamma)[P_L+P_R]+\bar{N}^{1,Box}_j(Z)
\Big [{(2s^2_W-1)^2\over4s^2_Wc^2_W}P_L+{s^2_W\over c^2_W}P_R\Big ]
\nonumber \\
&& +\bar{N}^{2,Box}_j(W)\Big [{1\over2s^2_W}P_L\Big ]
+\bar{N}^{7,Box}_j(G^+)\Big [{\mwd\over2s^2_W}P_L \Big ]
+{\rm "sym" } \Big \} ~, \label{N-Box-SM-gg} \\
N^{\gamma\gamma~MSSM~Box}_j&=& \alpha^2
\Big \{-~{1\over s^2_W}\sum_i|Z^+_{1i}|^2[\bar{N}^{4A}_j+M^2_{\tchi_i}
(\bar{N}^{4B}_j+\bar{N}^{4C}_j+\bar{N}^{4D}_j)]P_L\nonumber\\
&&+{1\over 2s^2_Wc^2_W}\sum_i|Z^N_{1i}s_W+Z^N_{2i}c_W|^2\bar{N}^{5}_j
(\tilde{e}_L)P_L \nonumber\\
&& +{2\over c^2_W}\sum_i|Z^N_{1i}|^2\bar{N}^{5}_j
(\tilde{e}_R)P_R~ +{\rm "sym"} \Big \} ~, \label{N-Box-MSSM-gg}
\eqa
where $+ \rm "sym"$ implies symmetrizations  of the form
\bqa
&&\bar{N}^{k,Box}_1+\bar{\tilde{N}}^{k,Box}_2,~~~
\bar{N}^{k,Box}_2+\bar{\tilde{N}}^{k,Box}_1,~~~
\bar{N}^{k,Box}_3-\bar{\tilde{N}}^{k,Box}_3,~~~
\bar{N}^{k,Box}_4-\bar{\tilde{N}}^{k,Box}_4,\nonumber\\
&& \bar{N}^{k,Box}_5+\bar{\tilde{N}}^{k,Box}_7,~~~
\bar{N}^{k,Box}_6-\bar{\tilde{N}}^{k,Box}_8,~~~
\bar{N}^{k,Box}_7+\bar{\tilde{N}}^{k,Box}_5,
\nonumber\\
&& \bar{N}^{k,Box}_8-\bar{\tilde{N}}^{k,Box}_6,~~~
\bar{N}^{k,Box}_9-\bar{\tilde{N}}^{k,Box}_9,
\label{+sym}\eqa
in which  $\bar{\tilde {N}}_j$ is constructed from
$\bar{N}_j$ by interchanging $t \Leftrightarrow u$
and $V \Leftrightarrow  V' $. The notation for the chargino and
neutralino mixing matrices appearing in (\ref{N-Box-MSSM-gg})
(and (\ref{N-Box-MSSM-ZZ},\ref{N-Box-MSSM-Zg}) below), is
summarized in Appendix B.

\vspace{0.5cm}
\noindent
$\bullet$ ~ \underline{$e^-e^+\to ZZ$}\\
The SM contributions arise from
type 1: $(\gamma eee)$, $(Z eee)$, $(W \nu_e\nu_e\nu_e)$;
type 2: $(\nu_eW W W)$;  type 7:
$(eZH_{SM}Z)$ and $(\nu_eW^+G^+W^+)$; and  type 3:
$(W\nu_e\nu_eW)$. These are
\bqa
&& N^{ZZ~SM~Box}_j=\alpha^2 \Big \{\bar{N}^{1,Box}_j(\gamma)
\Big [{(2s^2_W-1)^2\over 4s^2_Wc^2_W}P_L+{s^2_W\over c^2_W}P_R \Big ]
+\bar{N}^{1,Box}_j(Z)
\Big [{(2s^2_W-1)^4\over 16s^4_Wc^4_W}P_L \nonumber\\
&& +{s^4_W\over c^4_W}P_R \Big ]
+\bar{N}^{1,Box}_j(W)\Big [{1\over 8s^4_Wc^2_W}P_L \Big ]
+\bar{N}^{2,Box}_j(W)\Big [{c^2_W\over 2s^4_W}P_L\Big ]
+\bar{N}^{3,Box}_j(W) \Big [{1\over 4s^4_W}P_L \Big ]
\nonumber\\
&&+\bar{N}^{7,Box}_j(G^+)\Big [{\mwd\over 2c^2_W}P_L \Big ]+
\bar{N}^{7,Box}_j(H_{SM})
\Big [ \frac{\mwd [(2s^2_W-1)^2 P_L+4s^4_W P_R]}
{4s^4_W c^6_W} \Big ]  + {\rm "sym"} \Big \} ~.
\label{N-Box-SM-ZZ}
\eqa

The additional MSSM contributions arise from  type 4:
$(\tilde \nu_e \tchi^+_i\tchi^+_j\tchi^+_k)$,
$(\tilde{e}\tchi^0_i\tchi^0_j\tchi^0_k)$;
type 5: $(\tchi^+_i\tilde \nu_e\tilde \nu_e \tilde \nu_e)$,
$(\tchi^0_i\tilde{e}\tilde{e}\tilde{e})$;
and from\footnote{In order to
get the "additional MSSM contribution due to $H^0,~h^0$" which
should  added to the SM one without making
a double counting of the Higgs sector, one has to subtract the
$H_{SM}$ contribution.}  type 3:
[$(eZH^0Z)$+$(eZh^0Z)$~-~$(eZH_{SM}Z)$],
and type 6: $(\tilde \nu_e\tchi^+_i\tchi^+_j\tilde \nu_e)$,
 $(\tilde{e}\tchi^0_i\tchi^0_j\tilde{e})$. They are given by
\bqa
&& N^{ZZ~MSSM~Box}_j=\alpha^2 \Big
\{-\,{1\over4s^4_Wc^2_W}\sum_{ilk}Z^{+*}_{1i}Z^{+}_{1k}
[\bar{N}^{4A}_j(Z^{+}_{1i}Z^{+*}_{1l}+\delta_{il}(1-2s^2_W))
(Z^{+}_{1l}Z^{+*}_{1k}\nonumber\\
&& +\delta_{lk}(1-2s^2_W))+M_{\tchi_i}M_{\tchi_k}\bar{N}^{4B}_j
(Z^{-}_{1i}Z^{-*}_{1l}+\delta_{il}(1-2s^2_W))
(Z^{-}_{1l}Z^{-*}_{1k}+\delta_{lk}(1-2s^2_W))\nonumber\\
&&+M_{\tchi_i}M_{\tchi_l}\bar{N}^{4C}_j
(Z^{-}_{1i}Z^{-*}_{1l}+\delta_{il}(1-2s^2_W))
(Z^{+}_{1l}Z^{+*}_{1k}+\delta_{lk}(1-2s^2_W))\nonumber\\
&&+M_{\tchi_l}M_{\tchi_k}\bar{N}^{4D}_j
(Z^{+}_{1i}Z^{+*}_{1l}+\delta_{il}(1-2s^2_W))
(Z^{-}_{1l}Z^{-*}_{1k}+\delta_{lk}(1-2s^2_W))~]P_L\nonumber\\
&&-\,{1\over8s^4_Wc^4_W}\sum_{ilk}(Z^{N*}_{1i}s_W+Z^{N*}_{2i}c_W)
(Z^{N}_{1k}s_W+Z^{N}_{2k}c_W) \nonumber\\
&& \cdot [\bar{N}^{4A}_j(\tilde{e}_L)
(Z^{N}_{4i}Z^{N*}_{4l}-Z^{N}_{3i}Z^{N*}_{3l})
(Z^{N}_{4l}Z^{N*}_{4k}-Z^{N}_{3l}Z^{N*}_{3k})\nonumber\\
&&+M_{\tchi^0_i}M_{\tchi^0_k}\bar{N}^{4B}_j(\tilde{e}_L)
(Z^{N*}_{4i}Z^{N}_{4l}-Z^{N*}_{3i}Z^{N}_{3l})
(Z^{N*}_{4l}Z^{N}_{4k}-Z^{N*}_{3l}Z^{N}_{3k})\nonumber\\
&&-M_{\tchi^0_i}M_{\tchi^0_l}\bar{N}^{4C}_j(\tilde{e}_L)
(Z^{N*}_{4i}Z^{N}_{4l}-Z^{N*}_{3i}Z^{N}_{3l})
(Z^{N}_{4l}Z^{N*}_{4k}-Z^{N}_{3l}Z^{N*}_{3k})\nonumber\\
&&-M_{\tchi^0_l}M_{\tchi^0_k}\bar{N}^{4D}_j(\tilde{e}_L)
(Z^{N}_{4i}Z^{N*}_{4l}-Z^{N}_{3i}Z^{N*}_{3l})
(Z^{N*}_{4l}Z^{N}_{4k}-Z^{N*}_{3l}Z^{N}_{3k})~]P_L\nonumber\\
&&-\,{1\over2s^2_Wc^4_W}\sum_{ilk}Z^{N}_{1i}Z^{N*}_{1k}
[\bar{N}^{4A}_j(\tilde{e}_R)
(Z^{N*}_{4i}Z^{N}_{4l}-Z^{N*}_{3i}Z^{N}_{3l})
(Z^{N*}_{4l}Z^{N}_{4k}-Z^{N*}_{3l}Z^{N}_{3k})\nonumber\\
&&+M_{\tchi^0_i}M_{\tchi^0_k}\bar{N}^{4B}_j(\tilde{e}_R)
(Z^{N}_{4i}Z^{N*}_{4l}-Z^{N}_{3i}Z^{N*}_{3l})
(Z^{N}_{4l}Z^{N*}_{4k}-Z^{N}_{3l}Z^{N*}_{3k})\nonumber\\
&&-M_{\tchi^0_i}M_{\tchi^0_l}\bar{N}^{4C}_j(\tilde{e}_R)
(Z^{N}_{4i}Z^{N*}_{4l}-Z^{N}_{3i}Z^{N*}_{3l})
(Z^{N*}_{4l}Z^{N}_{4k}-Z^{N*}_{3l}Z^{N}_{3k})\nonumber\\
&&-M_{\tchi^0_l}M_{\tchi^0_k}\bar{N}^{4D}_j(\tilde{e}_R)
(Z^{N*}_{4i}Z^{N}_{4l}-Z^{N*}_{3i}Z^{N}_{3l})
(Z^{N}_{4l}Z^{N*}_{4k}-Z^{N}_{3l}Z^{N*}_{3k})~]P_R\nonumber\\
&&+{1\over4s^4_Wc^2_W}\sum_{i}|Z^{+}_{1i}|^2
\bar{N}^{5}_j(\tilde{\nu}_L)P_L
+{(1-2s^2_W)^2\over8s^4_Wc^4_W}\sum_{i}|Z^{N}_{1i}s_W
+Z^{N}_{2i}c_W|^2\bar{N}^{5}_j(\tilde{e}_L)P_L
\nonumber\\
&&+{(2s^2_W)\over c^4_W}\sum_{i}|Z^{N}_{1i}|^2
\bar{N}^{5}_j(\tilde{e}_R)P_R
+\,{1\over 4s^4_Wc^2_W}\sum_{il}Z^{+*}_{1i}Z^{+}_{1l}
~(\bar{N}^{6A}_j(Z^{+}_{1i}Z^{+*}_{1l}
+\delta_{il}(1-2s^2_W))\nonumber\\
&&
-M_{\tchi_i}M_{\tchi_l}\bar{N}^{6B}_j
(Z^{-}_{1i}Z^{-*}_{1l}+\delta_{il}(1-2s^2_W))~)P_L
\nonumber\\
&&+\,{1-2s^2_W\over8s^4_Wc^4_W}\sum_{il}(Z^{N*}_{1i}s_W
+Z^{N*}_{2i}c_W)(Z^{N}_{1l}s_W+Z^{N}_{2l}c_W)
[\bar{N}^{6A}_j(\tilde{e}_L)
(Z^{N}_{4i}Z^{N*}_{4l}-Z^{N}_{3i}Z^{N*}_{3l})\nonumber\\
&&
+M_{\tchi^0_i}M_{\tchi^0_l}\bar{N}^{6B}_j(\tilde{e}_L)
(Z^{N*}_{4i}Z^{N}_{4l}-Z^{N*}_{3i}Z^{N}_{3l})]P_L
\nonumber\\
&&+~{1\over c^4_W}\sum_{il}Z^{N}_{1i}Z^{N*}_{1l}
[\bar{N}^{6A}_j(\tilde{e}_R)
(Z^{N*}_{4i}Z^{N}_{4l}-Z^{N*}_{3i}Z^{N}_{3l})\nonumber\\
&&
+M_{\tchi^0_i}M_{\tchi^0_l}\bar{N}^{6B}_j(\tilde{e}_R)
(Z^{N}_{4i}Z^{N*}_{4l}-Z^{N}_{3i}Z^{N*}_{3l})]P_R
\nonumber\\
&&+[\bar{N}^{7,Box}_j(H^0)cos^2(\beta-\alpha)+
\bar{N}^{7,Box}_j(h^0)sin^2(\beta-\alpha)-\bar{N}^{7,Box}_j(H_{SM})]
\nonumber\\
&&
\cdot [{\mwd((2s^2_W-1)^2P_L+4s^4_WP_R) \over4s^4_Wc^6_W}]
  +{\rm "sym"} \Big \} ~. \label{N-Box-MSSM-ZZ}
\eqa

\vspace{0.5cm}
\noindent
$\bullet$ ~ \underline{$e^-e^+\to Z\gamma$}\\
The SM contributions come
from  type 1:  $(\gamma eee)$,  $(Zeee)$;
type 2: $(\nu_eW^+W^+W^+)$;
type 7: $(\nu_eW^+G^+W^+)$;
and from  type 3:
$(\nu_eWW\nu_e)$. They  give
\bqa
&& N^{\gamma Z~SM~Box}_j= \alpha^2
\Bigg \{\Big [ \bar{N}^{1,Box}_j(\gamma)
\Big ({(1-2s^2_W)\over2s_Wc_W}P_L-{s_W\over c_W}P_R \Big )
+\bar{N}^{1,Box}_j(Z)
\Big ({(1-2s^2_W)^3\over8s^3_Wc^3_W}P_L
\nonumber\\
&& -{s^3_W\over c^3_W}P_R \Big )
 +\bar{N}^{2,Box}_j(W) {c_W\over2s^3_W}P_L
-\bar{N}^{7,Box}_j(G^+){\mwd\over2s_Wc_W}P_L  +{\rm ~"sym"}
\Big]\nonumber\\
&&
+\bar{N}^{3,Box}_j(W) {1\over4s^3_Wc_W}P_L ~~
{\rm (no~"sym")}~\Bigg \} ~. \label{N-Box-SM-Zg}
\eqa

The additional MSSM contributions are from  type 4:
$(\tilde \nu_e\tchi^+_i\tchi^+_j\tchi^+_k)$;
type 5: $(\tchi^0_i\tilde{e}\tilde{e}\tilde{e})$;
and    type 6:
$(\tilde \nu_e \tchi^+_i\tchi^+_j\tilde \nu_e)$,
$(\tchi^0_i\tilde{e}\tilde{e}\tchi^0_j)$.
\bqa
 &&N^{\gamma Z~MSSM~Box}_j=\alpha^2
\Bigg \{ -~{1\over2s^3_Wc_W} \sum_{ik}Z^{+*}_{1i}Z^{+}_{1k}
~\{~\bar{N}^{4A}_j(Z^{+}_{1i}Z^{+*}_{1k} +\delta_{ik}(1-2s^2_W))
\nonumber\\
&& +M_{\tchi_i}M_{\tchi_k}(\bar{N}^{4B}_j+\bar{N}^{4D}_j)
(Z^{-}_{1i}Z^{-*}_{1k} +\delta_{ik}(1-2s^2_W))
+M_{\tchi_i}^2\bar{N}^{4C}_j
(Z^{+}_{1i}Z^{+*}_{1k} +\delta_{ik}(1-2s^2_W))  \nonumber\\
&&
+
[\bar{N}^{4A}_{j}(Z^{+}_{1i}Z^{+*}_{1k}+\delta_{ik}(1-2s^2_W))
+M_{\tchi_i}M_{\tchi_k}(\bar{N}^{4B}_{j}
+\bar{N}^{4C}_{j})
(Z^{-}_{1i}Z^{-*}_{1k}+\delta_{ik}(1-2s^2_W))\nonumber\\
&&
+M_{\tchi_k}^2\bar{N}^{4D}_{j}
(Z^{+}_{1i}Z^{+*}_{1k}+\delta_{ik}(1-2s^2_W))]^{sym}~\}~P_L
\nonumber\\
&&+ \Big [ {(1-2s^2_W)\over4s^3_Wc^3_W}
\sum_{i}|Z^{N}_{1i}s_W+Z^{N}_{2i}c_W|^2
\bar{N}^{5}_j(\tilde{e}_L)P_L
-~{2s_W\over c^3_W}\sum_{i}|Z^{N}_{1i}|^2
\bar{N}^{5}_j(\tilde{e}_R)P_R+~{\rm "sym"}\Big ]
\nonumber\\
&&+~{1\over 2s^3_Wc_W}\sum_{i}|Z^{+}_{1i}|^2
[\bar{N}^{6A}_j-|M_{\tchi_i}|^2\bar{N}^{6B}_j]P_L
+~{1\over4s^3_Wc^3_W}\sum_{il}(Z^{N*}_{1i}s_W+Z^{N*}_{2i}c_W)
(Z^{N}_{1l}s_W
\nonumber\\
&& +Z^{N}_{2l}c_W) [(Z^{N}_{4i}Z^{N*}_{4l}-Z^{N}_{3i}Z^{N*}_{3l})
\bar{N}^{6A}_j(\tilde{e}_L)
+M_{\tchi^0_i}M_{\tchi^0_l}\bar{N}^{6B}_j(\tilde{e}_L)
(Z^{N*}_{4i}Z^{N}_{4l}-Z^{N*}_{3i}Z^{N}_{3l})]^{sym}P_L\nonumber\\
&&-~{1\over s_Wc^3_W}\sum_{il}Z^{N}_{1i}
Z^{N*}_{1l}
[(Z^{N*}_{4i}Z^{N}_{4l}-Z^{N*}_{3i}Z^{N}_{3l})
\bar{N}^{6A}_j(\tilde{e}_R) \nonumber \\
&&
+M_{\tchi^0_i}M_{\tchi^0_l}\bar{N}^{6B}_j(\tilde{e}_R)
(Z^{N}_{4i}Z^{N*}_{4l}-Z^{N}_{3i}Z^{N*}_{3l})]^{sym}~P_R~~~
 \Bigg \}~~. \label{N-Box-MSSM-Zg}
\eqa

In all cases, the symmetrization
"sym" is done according to the rules given in (\ref{+sym}).

\section{Asymptotic amplitudes at one loop}

It is interesting to construct asymptotic expressions for the
$N_j$ invariant amplitudes for the processes
$e^-e^+\to \gamma \gamma, ~Z \gamma, ZZ$, which
should in principle be valid
when $(s,~ t, ~u)$ are much larger than $\mzd$ and any of the
masses of the particles in the loop. Such asymptotic expressions
for the $N^{\rm ren+Born}_j$, $N^{\rm Tri}_j$ and  $N^{\rm Box}_j$
parts of these invariant amplitudes, are given
in Appendices  E and F.

These asymptotic expressions  are very  interesting
since they provide a simple  picture for
the physical amplitudes, which turns out to approximate the exact
1-loop results  at the percent level, as soon as we pass
the one TeV energy range.  We give them below for each
of the $N_j$ amplitudes, always omitting  $N_3$ and $N_9$, which never
receive any leading-log contribution.

\vspace{0.5cm}
\noindent
$\bullet$ ~ \underline{$N_j(e^-e^+\to\gamma\gamma)$; ~$(j=1,2,4)$}
\bqa
N_j& \simeq &N^{ Born,L}_j~[1+c^{(\gamma)}_L+c^{(Z)}_L
+c^{(W)}_L+c^{(MSSM)}_L]
+d^{(W)}_{j,L}\nonumber\\
&&+ N^{ Born,R}_j~[c^{(\gamma)}_R+c^{(Z)}_R+c^{(MSSM)}_R] ~.
\label{N-asym-gg}
\eqa
The structure of this expression is very intuitive. It consists of
the Born term (\ref{Nj-Born-gg}), to which the universal leading-log
correction factors generated by
\bqa
&& c^{(\gamma)}_L=c^{(\gamma)}_R={\alpha\over4\pi}
\Big [3 \ln{s\over M^2_{\gamma}}-\ln^2{s\over M^2_{\gamma}}\Big ]
~~, \label{c-asym-g-gg} \\
&& c^{(Z)}_L= \frac{\alpha (2s^2_W-1)^2}{16\pi s^2_Wc^2_W}
\Big [3 \ln{s\over \mzd}-\ln^2{s\over \mzd}\Big ]~~,~~
c^{(Z)}_R=\frac{\alpha s^2_W}{4\pi  c^2_W}
\Big [3 \ln{s\over \mzd}-\ln^2{s\over \mzd} \Big ] ~,
\label{c-asym-Z-gg} \\
&& c^{(W)}_L={\alpha\over 8\pi s^2_W}
\Big [(3\ln{s\over \mwd}-\ln^2{s\over \mwd})~ -2\ln^2{s\over \mwd}\Big ] ~,
\label{c-asym-W-gg} \\
&& c^{(MSSM)}_L=-~{\alpha (1+2c^2_W)\over 16\pi s^2_Wc^2_W}
 \ln{s\over \mwd}~~, ~~
c^{(MSSM)}_R=-~{\alpha\over4\pi  c^2_W}
\ln{s\over \mwd} ~, \label{c-asym-MSSM-gg}
\eqa
are applied. These corrections are generated by
diagrams involving photon, $Z$, $W$ and
MSSM partner exchanges. In the photon correction
(\ref{c-asym-g-gg}), the quantity $M_{\gamma}$ has been introduced,
 which separates the ultraviolet and infrared
contributions,  the latter being  generally
absorbed to the so-called electromagnetic radiative
corrections.  In addition to these corrections,
the $W$ exchange boxes also induce an angular dependent
contribution
\bqa
&& d^{(W)}_{j,L}= {\alpha^2 \over  s^2_W}P_L~\Big \{\eta^j_t
\Big [2 \ln{s\over m^2_W} \ln {1-\cos\theta\over 2} +
\ln^2 {1-\cos\theta\over 2}\Big ]\nonumber\\
&& +\eta^j_u~
\Big [2 \ln{s\over m^2_W}\ln{1+\cos\theta\over 2}
+\ln^2{1+\cos\theta\over 2}\Big ]\Big \} ~~, \label{d-asym-gg}
\eqa
\noindent
where
\bq
\eta^{1,2}_t={1\over t}~~,~~\eta^{4}_t={1\over t}+{2\over s}
~~, ~~\eta^{1,2}_u={1\over u}~~,~~\eta^{4}_u=-~{1\over u}-~{2\over s}
~~.
\eq

Taking  $M_{\gamma}\simeq m_Z\simeq m_W$, one gets from
(\ref{c-asym-g-gg}-\ref{c-asym-MSSM-gg})
the \underline{universal} SM combinations \cite{class,reality}
\bqa
&& c^{(\gamma)}_L+c^{(Z)}_L+c^{(W)}_L=
{\alpha (1+2c^2_W)\over 16\pi s^2_Wc^2_W}
\Big [3 \ln{s\over \mwd}-\ln^2{s\over \mwd} \Big]
-{\alpha\over4\pi s^2_W} \ln^2{s\over \mwd} ~~,
\nonumber \\
&& c^{(\gamma)}_R+c^{(Z)}_R=
{\alpha\over 4\pi c^2_W}
\Big [3 \ln{s\over \mwd}-\ln^2{s\over \mwd}\Big ] ~~,
\label{gg-SMasym}
\eqa
and the MSSM ones \cite{BMRV}
\bqa
&& c^{(\gamma)}_L+c^{(Z)}_L+c^{(W)}_L+c^{(MSSM)}_L=
{\alpha (1+2c^2_W) \over 16 \pi s^2_Wc^2_W}
\Big [2 \ln{s\over \mwd}-\ln^2{s\over \mwd}\Big ]
-{\alpha\over 4\pi s^2_W} \ln^2{s\over \mwd} ~,
\nonumber \\
&&c^{(\gamma)}_R+c^{(Z)}_R+c^{(MSSM)}_R=
[{\alpha\over4\pi}][{1\over c^2_W}]
[2 \ln{s\over \mwd}-\ln^2{s\over \mwd}]~~, \label{gg-MSSMasym}
\eqa
which satisfy the rules established in \cite{reality,LR}.
Indeed, we find again  that the radiative corrections
 associated to the electron line create the logarithmic
factors $[3\ln (s/\mwd)-\ln^2(s/ \mwd) ]$ in SM,
and $[2\ln(s/ \mwd)-\ln^2(s/ \mwd) ]$ in MSSM; while their
coefficients are determined by\footnote{Here $I_e$ refers to the
total isospin of the electron $e$,  while  $Y_e=2(Q_e-I_e^{(3)})$.
The same formulae should apply also to any quark or lepton, and to
 their supersymmetric partners.}
$\alpha /(4\pi) [I_e(I_e+1)/s^2_W +Y_e^2/(4c^2_W)]$,  which equals to
$\alpha (1+2c^2_W)/(16\pi s^2_Wc^2_W)$ for the Left case,
and $\alpha /(4\pi c^2_W)$ for the Right case \cite{reality,LR}.
The photon lines supply the
additional term  $-\alpha/(4\pi s^2_W) \ln^2(s/ \mwd)$
in (\ref{gg-SMasym},\ref{gg-MSSMasym}) \cite{reality,LR}.

The non universal angular dependent term
$d^{(W)}_{j,L}$ in (\ref{d-asym-gg}),
 is a specific  SM gauge W box contribution whose
  coefficient is fixed by the $\gamma WW$ coupling
 \cite{reality,LR}.

\vspace{0.5cm}
\noindent
$\bullet$~ \underline{$N_j(e^-e^+\to Z \gamma )$;
~$(j=1,2,4,5,6)$}\\
The asymptotic expressions now are
\bqa
N_j& \simeq
&N^{ Born,L}_j~[c^{(\gamma)}_L+c^{(Z)}_L+c^{(W)}_L+c^{(MSSM)}_L]
+d^{(W)}_{j,L}\nonumber\\
&&+
N^{ Born,R}_j~[c^{(\gamma)}_R+c^{(Z)}_R+c^{(MSSM)}_R]
~~, \label{N-asym-Zg}
\eqa
with
\bqa
&& c^{(\gamma)}_L=c^{(\gamma)}_R={\alpha\over4\pi}
\Big [3 \ln{s\over M^2_{\gamma}}-\ln^2{s\over M^2_{\gamma}}\Big ]
~~, \nonumber \\
&& c^{(Z)}_L={\alpha (2s^2_W-1)^2 \over 16 \pi s^2_Wc^2_W}
\Big [3 \ln{s\over \mzd}-\ln^2{s\over \mzd}\Big ]~~, ~~
c^{(Z)}_R= {\alpha s^2_W \over4\pi c^2_W}
[3\ln{s\over \mzd}-\ln^2{s\over \mzd}]
~~, \nonumber \\
&& c^{(W)}_L={\alpha\over 8\pi s^2_W}
\Big [\Big (3\ln{s\over \mwd}-\ln^2 {s\over \mwd}\Big )
+{(3-4s^2_W)\over(2s^2_W-1)}\Big (\ln^2{s\over \mwd}\Big )\Big ]
~~, \nonumber \\
&& c^{(MSSM)}_L=-~{\alpha(1+2c^2_W) \over 16 \pi s^2_Wc^2_W}
\ln{s\over \mwd}~~, ~~
c^{(MSSM)}_R=-~{\alpha \over 4\pi c^2_W}
\ln^2{s\over \mwd} ~~, \label{c-asym-Zg} \\
&& d^{(W)}_{j,L}={\alpha^2 \over 4 s^3_Wc_W}P_L~\Big \{\eta^j_t
\Big [2\ln{s\over m^2_W}\ln{1-\cos\theta\over 2}
+\ln^2{1-\cos\theta\over 2}\Big ]
\nonumber\\
&&+{\eta^j_u}~ \Big [2\ln{s\over m^2_W}\ln{1+\cos\theta\over 2}
+\ln^2{1+\cos\theta\over 2}\Big ] \Big \} ~~,
\eqa
and
\bqa
\eta^{1,2}_t=-{s\over2}\eta^{6}_t={3-4s^2_W\over t}-{1\over u} &,&
\eta^{1,2}_u=-{s\over2}\eta^{6}_u={3-4s^2_W\over u}-{1\over t}
~, \nonumber \\
\eta^{4}_t={3-4s^2_W\over t}+{1\over u}+{8c^2_W\over s} &,&
\eta^{4}_u=-~{3-4s^2_W\over u}-{1\over t}-~{8c^2_W\over s}
~, \nonumber \\
\eta^{5}_t={1\over u}+{4c^2_W\over s} &,&
\eta^{5}_u=-~{3-4s^2_W\over u}-~{4c^2_W\over s} ~.
\label{d-asym-Zg}
\eqa
The universal SM
and MSSM contribution of the electron line
are again found to be in agreement with
(\ref{gg-SMasym},\ref{gg-MSSMasym}) \cite{reality}.
There exist an  additional $c^{(W)}_{L}$ contribution
though in (\ref{c-asym-Zg}), caused by the $Z\gamma$ final state.
The  angular dependent box  term $d^{(W)}_{j,L}$
  fixed by the $\gamma WW$ and the $ZWW$
couplings \cite{reality}. \\

\vspace{0.5cm}
\noindent
$\bullet$~ \underline{$N_j(e^-e^+\to Z Z)$;
~$(j=1,2,4,5,6,7,8)$}\\
In a similar way we have
\bqa
N_j& \simeq &
N^{Born,L}_j~[c^{(\gamma)}_L+c^{(Z)}_L+c^{(W)}_L+c^{(MSSM)}_L]
+d^{(W)}_{j,L}\nonumber\\
&&+
N^{Born,R}_j~[c^{(\gamma)}_R+c^{(Z)}_R+c^{(MSSM)}_R]
~, \label{N-asym-ZZ}
\eqa
with
\bqa
&& c^{(\gamma)}_L=c^{(\gamma)}_R={\alpha\over4\pi}
\Big [3\ln{s\over M^2_{\gamma}}-\ln^2{s\over M^2_{\gamma}}\Big ]
~~, \nonumber \\
&& c^{(Z)}_L={\alpha (2s^2_W-1)^2 \over 16 \pi s^2_Wc^2_W}
\Big [3\ln{s\over \mzd}-\ln^2{s\over \mzd}\Big ]~~, ~~
c^{(Z)}_R={\alpha s^2_W \over4\pi c^2_W}
\Big [3\ln{s\over \mzd}-\ln^2{s\over \mzd}\Big ]
~, \nonumber \\
&& c^{(W)}_L={\alpha\over 8\pi s^2_W}
\Big [\Big (3\ln{s\over \mwd}-\ln^2{s\over \mwd}\Big )
+{4c^2_W\over2s^2_W-1}\Big (\ln^2{s\over \mwd}\Big )\Big ]
~, \nonumber \\
&& c^{(MSSM)}_L=-~{\alpha (1+2c^2_W) \over 16 \pi s^2_Wc^2_W}
\ln{s\over \mwd} ~~,~~
c^{(MSSM)}_R=-~{\alpha\over 4\pi c^2_W}
\ln^2{s\over \mwd} ~~ , \label{c-asym-ZZ} \\
&& d^{(W)}_{j,L}= {\alpha^2 \over 2 s^4_W}P_L~\Big \{\eta^j_t
\Big [2 \ln{s\over m^2_W}\ln{1-\cos\theta\over 2}
+\ln^2{1-\cos\theta\over 2}\Big ]
\nonumber\\
&&+{\eta^j_u} \Big [2\ln{s\over m^2_W}\ln{1+\cos\theta\over 2}
+\ln^2{1+\cos\theta\over 2}\Big ]\Big \} ~~, \label{d-asym-ZZ}
\eqa
with
\bqa
\eta^{1,2}_t={1-2s^2_W\over t}-{1\over u} &,&
\eta^{1,2}_u={1-2s^2_W\over u}-{1\over t} ~, \nonumber \\
\eta^{4}_t={1-2s^2_W\over t}+{1\over u}+{4c^2_W\over s} &,&
\eta^{4}_u=-~{1-2s^2_W\over u}-{1\over t}-~{4c^2_W\over s}
 ~, \nonumber \\
\eta^{5}_t={1\over u}+{2c^2_W\over s} &,&
\eta^{5}_u=-~{1-2s^2_W\over u}-~{2c^2_W\over s}
 ~, \nonumber \\
\eta^{6}_t=-\eta^{8}_t=-~{2\over s}\Big [{1-2s^2_W\over t}
-~{1\over u}\Big ] &,&
\eta^{6}_u=-\eta^{8}_u=-~{2\over s}
\Big [{1-2s^2_W\over u}-~{1\over t}\Big ]  ~, \nonumber \\
\eta^{7}_t={2s^2_W-1\over t}-~{2c^2_W\over s} &,&
\eta^{7}_u={1\over t}+~{2c^2_W\over s} ~.
\eqa

As expected, the universal electron line SM and MSSM contributions
 to (\ref{c-asym-ZZ}), are  the same, as in the $\gamma\gamma$ case
   \cite{reality}.
The only modifications are due to    the $ZZ$ final state
 universal contribution  $c^{(W)}_{L}$,
and  the angular dependent term $d^{(W)}_{j,L}$
determined by the $ZWW$ coupling.

 Finally we should comment about the asymptotic behaviour of
 the \underline{longitudinal $ZZ$} production amplitudes.
 Contrary to the Born level  TT amplitudes which behave like a constant
at asymptotic energies, and the  TL ones which   vanish
 only like $m_Z/\sqrt{s}$,   the LL amplitudes  diminish like   $m^2_Z/s$.
This latter property  can be explicitly seen in
\bqa
F^{Born}_{\lambda 00}& \simeq &-~(2\lambda){16m^2_Z\over
s}~ {\cos\theta\over\sin\theta} \Big \{{(2s^2_W-1)^2
\over4s^2_Wc^2_W}P_L+{s^2_W\over c^2_W}P_R\Big \}~~.
\label{born-asym-ZZ00}
\eqa

When one loop effects are included, the asymptotic behaviours of the
TT  and TL remain largely the same, modified only by logarithmic
enhancements  determined  by (\ref{N-asym-ZZ}, \ref{c-asym-ZZ}),
and (to a lesser extent)  (\ref{d-asym-ZZ}).
But for the LL amplitudes a strikingly  different structure
arises, since  the rapidly vanishing $\sim m^2_Z/s$ Born behaviour
is   replaced by a logarithmically increasing one
involving $\ln^2{|t|/ M^2}$ and $\ln^2{|u|/ M^2}$ terms.
This structure is induced by  Higgs sector Box
diagrams, whose contribution asymptotically dominates the tree-level one.

The simplest way to obtain it,
is to use the equivalence theorem and consider the processes
$e^+e^-\to G^0 G^0$.
Since in the $m_e=0$ limit this later process has no Born term,
its only contribution comes from the boxes $(eZHZ)$
and $(\nu WGW)$, where $H$ stands for the
standard Higgs boson in SM, while  in
 MSSM  it represents a mixture of the CP-even states
 $H^0$ and the $h^0$.
 The resulting asymptotic helicity amplitudes then is
\bqa
&& F_{\lambda 00}
 \simeq
(2\lambda){\alpha^2 \sin\theta\over4}
\Big \{[\ln^2{|t|\over m^2_W}-\ln^2{|u|\over m^2_W}]\Big \}
\Big \{\Big ({1\over s^4_W}+{(2s^2_W-1)^2\over2s^4_Wc^4_W}\Big )P_L
+\Big ({2\over c^4_W}\Big )P_R\Big \} \nonumber \\
&& \simeq  (2\lambda){\alpha^2 \sin\theta\over 2}
 \ln \Big ({s\over \mwd}\Big )
  \ln\Big ({1-cos\theta\over 1+cos\theta}\Big )
\Big \{\Big ({1\over s^4_W}+{(2s^2_W-1)^2\over2s^4_Wc^4_W}\Big )P_L
+\Big ({2\over c^4_W}\Big )P_R\Big \}, \label{goldsm}
\eqa
in both, SM and MSSM.  Thus, at sufficiently  high energy,
the  order $\alpha^2$ contribution of (\ref{goldsm}),
becomes larger than the (suppressed) Born LL contribution
of (\ref{born-asym-ZZ00}).
We note that  the cross-over of these two terms
is around 1TeV.

Note also that, asymptotically, there is no
difference between the SM and the MSSM predictions for
longitudinal $ZZ$ production. This  is due to the fact
that the $H^0$ contribution is proportional to $\cos^2(\beta-\alpha)$
and the $h^0$ one proportional
to $\sin^2(\beta-\alpha)$, producing a result identical to
the SM one.

\section{Numerical Illustrations}

\noindent
\underline{{\bf Results for $e^+e^-\to \gamma\gamma$}}\\
Due to the electron exchange diagrams in the $t$ and $u$ channels,
the angular distribution
is strongly peaked in the forward and
backward directions. Because of detection difficulties
along the beam directions, we  only consider
scattering angles larger than $30^o$ and smaller than $150^o$.
The Born contribution is then shown for unpolarized beams and
energies at $0.5$ TeV and  $5$ TeV in Figs.\ref{gg-differential-fig}a,b.

The  1-loop\footnote{The numerical computation of the Passarino-Veltman
functions is done using the FF-package \cite{FF}.}
 radiative correction effects in the angular distribution,
are described in Figs.\ref{gg-differential-fig}c,d,  presenting
 the ratios of the unpolarized
differential cross sections to the Born contribution
$d \sigma/d\sigma_{\rm Born}$,
for SM and  a representative set of  MSSM SUGRA models
suggested in   \cite{Ellis-bench, Snowmass}.
As seen in these figures, the effect of the
radiative corrections is always negative
and increases with the energy and scattering angle.
In the SM case it is of about $-7\%$ at 0.5TeV, and
$-27\%$ at 5TeV. The differences among the various
 MSSM cases and the  SM one, are within $\pm 1\%$.

The total cross section is calculated
by integrating for scattering angles in
the region $30^o< \theta < 150^o$. The Born contributions to
it for all $e^-e^+ \to \gamma \gamma, ~\gamma Z,~ ZZ $ processes,
and various polarization of the final $Z$ states,
are  shown in Figs.\ref{Born-integrated-fig}. Since the dominant
Born amplitudes are
energy-independent at high energies, the integrated cross
sections  decrease like $1/s$, as shown in
Figs.\ref{Born-integrated-fig}.

The radiative corrections to  the integrated
cross sections are  described in Figs.\ref{gg-sig-ratio-fig}a,
 by  the  ratios $\sigma/\sigma_{Born}$ of the total cross
 section to the Born one,
 for SM and the aforementioned  set of MSSM models.
The energy behaviour of these  ratios
agrees with the asymptotic leading  log expressions
 (\ref{gg-SMasym}) and (\ref{gg-MSSMasym}), for SM and MSSM respectively.
According to them, the main difference between SM and MSSM at high energy,
stems from the respective factors $(3\ln s-\ln^2 s)$ and
$(2\ln s-\ln^2 s)$, and is independent of any other MSSM parameter.
A measurement of the coefficient of
the linear log term, could thus provide a  signature discriminating
between SM and MSSM. The magnitude of the effect is determined
by the remark that if one puts
an additional constant to the asymptotic  cross section expression,
 and fits its value so that it  agrees with the exact 1-loop result at 5TeV,
then the departure at 0.2TeV appears to be  at the permille
level.

We also note that the above agreement between the exact 1-loop and the
asymptotic predictions for  the $e^-e^+\to\gamma\gamma$ amplitudes,
turns out to be rather insensitive to the masses
of the virtual particles running along the loops.
This applies also to all cases involving production
of  transverse gauge bosons; see below.
On the contrary, as we will also see below, a large sensitivity
to  mass effects appears in the LT $e^-e^+\to Z\gamma$ amplitudes,
as well in the LT and LL $e^-e^+\to  ZZ$ ones.

The Left-Right polarization asymmetry $A_{LR}$ defined in
(\ref{ALR-definition}), with all final gauge polarizations summed over,
is shown in Fig.\ref{gg-sig-ratio-fig}b.
Since there  is no Born contribution in $e^+e^-\to\gamma\gamma$,
$A_{LR}$ is totally due to the electroweak loop-corrections.
Comparing Figs.\ref{gg-sig-ratio-fig}a,b, one can see that about
the same type of effects appear in both, the
cross sections and the $A_{LR}$ asymmetries;
(magnitude and sign of the SM and MSSM effects).
Since $A_{LR}$ should be not  affected by normalization
uncertainties though, its measurement may be experimentally
more interesting.

As a final comment we note that
in $e^-e^+\to\gamma\gamma$, there is no Higgs
or NAGC contributions. So this process
is  suited for  studying
the pure gauge and the gauge-lepton coupling
structures of the electroweak corrections
as well as their supersymmetric
(gaugino and  gaugino-slepton) counterparts. \\

\noindent
\underline{{\bf Results for $e^+e^-\to Z\gamma$}}\\
The unpolarized angular distribution in the Born approximation,
and its  radiative corrections described by
the ratios of the  differential cross sections
 $d \sigma/d \sigma_{\rm Born}$
at 0.5TeV and  5TeV, are shown in Figs.\ref{Zg-differential-fig}a-d.
Thus, the radiative correction effects   are now
of about $-8\%$ at 0.5TeV, and $-40\%$ at 5TeV in SM;
while  the sensitivity to the various MSSM models is larger
than in the $e^-e^+\to \gamma\gamma$ process, (several percent
at  0.5TeV, and few percent at 5TeV).
This increase of sensitivity is mainly due to the
LT amplitudes, and will disappear at asymptotic energies; see below.
In any case, it is therefore  interesting to study separately the
behaviour of the cross sections for  TT and LT
final $Z\gamma$ states.

The Born cross sections
$\sigma_{Born}(e^+e^-\to (Z \gamma)_{ TT,~LT})$,
(for producing TT or LT $Z\gamma$ final states),
are shown in Figs.\ref{Born-integrated-fig}a.
As seen there, the TT cross section behaves like $1/s$
and quickly dominates the LT one behaving as  $m^2_Z/s^2$.
As a result, the not-shown unpolarized Born cross section
$\sigma_{Born}(e^+e^-\to (Z \gamma)_{unp})$, is almost identical
to $\sigma_{Born}(e^+e^-\to (Z \gamma)_{TT})$
for most of the energy region of the figure.

The radiative correction effects on the unpolarized,
TT and LT integrated cross sections are described by the ratios
  in Figs.\ref{Zg-sig-ratio-fig}a,b,c. Correspondingly,
 the radiative corrections to the Left-Right
asymmetry are constructed in Fig.\ref{Zg-sig-ratio-fig}d, by
subtracting  the $A_{LR}(Born)=0.2181$  contribution, from the
1-loop result.

As one can see from Figs.\ref{Zg-sig-ratio-fig}a,b,d,
the radiative corrections to the unpolarized
and TT cross sections and  to the
 $A_{LR}$ asymmetry, have a similar structure,
which is also rather  close to that of  the
$\gamma\gamma$ case, for both the SM and MSSM aspects.
We have also checked that the high energy behaviour of the
TT cross section\footnote{This is also true for the unpolarized
cross section to which the TT part is by far the dominant one.}
agrees,  with the asymptotic logarithmic
expressions of (\ref{N-asym-Zg}-\ref{d-asym-Zg}),
and that its characteristics are similar to those
of  the $\gamma\gamma$ case.

The radiative correction to the $LT$ cross section presented in
Figs.\ref{Zg-sig-ratio-fig}c, requires a special discussion though;
since the relevant helicity amplitudes are suppressed, behaving
 like ${\cal M} /\sqrt{s}$. Probably because of this,
 they are also very sensitive to the
(real and virtual) masses involved in the various diagrams.
In SM, the scale ${\cal M} $ is determined essentially by $m_{W,Z}$
or  $m_t$ (for the NAGC contribution);
while  in MSSM, the various new  masses
generate a strong model dependence.
Illustrations are given for the same sample of models
of \cite{Ellis-bench} used in the previous figures.
On can then see in Fig.\ref{Zg-sig-ratio-fig}c
 various energy dependence structures, induced by
 chargino, neutralino or slepton  thresholds. These appear
 in an energy range where $\sigma_{LT}$ should still be
measurable. At higher energies, $\sigma_{LT}$ becomes very small
and marginally
observable except with very high luminosity colliders.

There exist  NAGC contributions to  $e^+e^-\to\gamma Z$ arising
from the diagram in Fig.1f involving a fermionic loop;
but no  Higgs contributions.
The magnitude of NAGC for SM and a
 representative set of MSSM models \cite{Ellis-bench}
  is shown  in Fig.\ref{Zg-NAGC-fig}a-c,
where one plots the difference between
the ratios to the Born cross sections, with  and without NAGC,
for unpolarized, as well as TT and LT final gauge boson states.
In  Fig.\ref{Zg-NAGC-fig}d, the difference between the
Left-Right asymmetry with  and without NAGC, is also shown.

As mentioned in Section 3, the magnitude of the NAGC
effects, (created   by fermionic triangular loops),
 decreases with energy faster than $1/s$ , \cite{NAGCt1}.
 Moreover,  at high energies,  there is no interference between the
 NAGC and Born amplitudes because the Born TT amplitudes
involve  opposite gauge helicities, while the NAGC TT amplitudes
concern equal gauge helicities only; in addition
the Born $LT$ amplitudes  drop down so quickly, that
 their NAGC interference is also forced  to  vanish quickly.
Thus,  the SM and MSSM NAGC effects for the models
in  Figs.\ref{Zg-NAGC-fig} are at the permille
level, and should be only marginally
observable, except with very high luminosity colliders.

It is conceivable, that forms of fermionic NP exist
(beyond SM or MSSM), that only contribute through the
NAGC diagram of Fig.1f.
We have  looked at the sensitivity of such
contributions to the $h^{\gamma}_3$ and $h^Z_3$ couplings
at first order, in a model independent way \cite{NAGCt}.   Assuming
a given experimental accuracy on the unpolarized integrated
cross section and the Left-Right asymmetry,
we obtain observability limits for the total NAGC contribution.
Such effects are  illustrated   in  Fig.\ref{NAGClim-fig}a,
assuming  $1\%$ accuracy, and taking the energies
 0.5TeV and 1TeV.

Note that the $\sigma_{unp}$ and $A_{LR}$ constraints are almost orthogonal,
allowing a good limitation on both NAGC couplings. This arises because
$\sigma_{np}$ mainly depends on  $h^{\gamma}_3$, whereas
$A_{LR}$ is more  sensitive to $h^{Z}_3$; which is just because the
photon couples vectorialy, while the  $Zee$ coupling is essentially purely
axial. As seen from Fig.\ref{NAGClim-fig}a,
the implied sensitivity is likely to increase with energy.
Thus, on the basis of Fig.\ref{NAGClim-fig}a, we conclude
that only  NP forms inducing
 \eg percent level NAGC effects, could be observable
 through $e^-e^+\to Z\gamma $ measurements.

Finally we come back to a point mentioned in the Introduction
concerning the search for NP through NAGC measurements.
Since the NAGC effects are intrinsically small,
it is essential to have a good
evaluation of the complete  SM or MSSM radiative corrections, before looking
for possible NP contributions. For example, at 1TeV these radiative
correction effects are of the order of $10\%$  on $\sigma$ or
$A_{LR}$, so that neglecting (or approximating) them,
would invalidate the bounds
one would put on the basis of NAGC.\\

\noindent
\underline{{\bf Results for $e^+e^-\to ZZ$}}\\
The unpolarized angular distributions at 0.5TeV and 5TeV are shown
in Fig.\ref{ZZ-differential-fig}a,b for the Born contribution,
and in Fig.\ref{ZZ-differential-fig}c,d
for the ratio $(d \sigma/d\cos\theta)/(d \sigma/d\cos\theta)_{\rm Born}$.
The radiative correction effects are now larger than in the
$\gamma\gamma$ and $Z\gamma$ production cases. In SM they reach
$-15\%$ at 0.5TeV, and $-58\%$ at 5TeV. The sensitivity
to MSSM models is also increased, up to an additional $-15\%$,
especially at large angles.

The integrated Born cross sections
(using  the same angular cut at $30^o$),
are given in Figs.\ref{Born-integrated-fig}b for TT, TL and LL final states.
As in the $\gamma \gamma$ and $Z\gamma $ cases,
 $\sigma_{Born}^{TT}\simeq 1/s$ at high energies, which    is much
larger than the TL or LL cross sections, and therefore almost identical
to the unpolarized one. In analogy to the $Z\gamma$ case, the TL cross
section is suppressed like   $ \mzd /s^2$; while
 $\sigma_{Born}^{LL} \simeq 1/s^3$, compare (\ref{born-asym-ZZ00}).

The energy dependence of the
radiative corrections to the unpolarized and TT cross sections
is  presented in Figs.\ref{ZZ-sig-ratio-fig}a,b;  it is
 similar to that observed for the other processes, and agrees with the
logarithmic analysis at high energy contained  in  (\ref{N-asym-ZZ}).

The same effects are also  found in the Left-Right polarization
asymmetry $A_{LR}$, (for unpolarized final $ZZ$ states).
The   $A_{LR}$  Born value  is $0.4164$,
 and   the radiative correction to this value
is shown in Fig.\ref{ZZ-higgs-fig}a.

The radiative correction to  $\sigma_{LT}$
is  presented in Fig.\ref{ZZ-sig-ratio-fig}c where,
one observes a strong model dependence, similar to the one seen in
Fig.\ref{Zg-sig-ratio-fig}c for the $Z_L\gamma$ case.

The case of $\sigma_{LL}$ is even more interesting,
because of the change of behaviour around 1TeV appearing
in Fig.\ref{ZZ-sig-ratio-fig}d. For $\sqrt{s} \lsim 1$TeV,
one observes a suppression
like ${\cal M}^4/s^3$, already mentioned in
connection with the Born LL contribution. Above 1TeV though,
a logarithmic increase  arises, caused by the
 Higgs sector and discussed at the end of Section 4;
 compare (\ref{goldsm}).

Finally, in Figs.\ref{ZZ-higgs-fig}b,c,d, we  show the Higgs mass
dependence of the radiative  corrections in the SM case,
for  TT, TL and LL productions.
We plot the ratios to the Born contribution,  of the differences
between the   $m_H=0.3$ TeV  or  $m_H=1$ TeV cross sections,
from  the $m_H=0.113$ TeV  case.  \\

\noindent
\underline{The role of the NAGC}\\
In the $e^+e^-\to ZZ$ process, the SM or MSSM NAGC
contributions $f^{\gamma,Z}_5$ only exist for  the TL
amplitudes \cite{NAGCt}.
In the above illustrations, the unpolarized or TL
cross sections, as well as the $A_{LR}$ asymmetry, are containing
these  contributions.
 The situation is  different
from the $e^+e^-\to Z\gamma$ case because the
roles of $\gamma-NAGC$ and of $Z-NAGC$
are interchanged due to the different chirality structure of the
Born terms which interfere with the NAGC amplitudes. Consequently
$\sigma_{unp}$ is mainly sensitive to $f^{Z}_5$
whereas $A_{LR}$ is mainly
sensitive to $f^{\gamma}_5$. The net NAGC effects are shown in
Fig.\ref{ZZ-NAGC-fig}a-c, and as in the $Z\gamma$ case, they are
at most at the permille level.

As for the $Z\gamma$ case, we have also made a model independent
analysis of the sensitivity
to unknown $f^{\gamma,Z}_5$ couplings, at first order in
$\sigma_{unp}$ and $A_{LR}$. The result is shown in
Fig.\ref{NAGClim-fig}b
for 0.5 and 1 TeV, assuming again $1\%$  accuracy
on these observables.
The orders of magnitude and the prospects for observability
are comparable to the $Z\gamma$ case. Thus, SM and MSSM contributions
will be marginally observable and only stronger NP
contributions may be constrained.

\section{Physics issues and Conclusions}

In this paper we have made a complete analysis of the processes
$e^+e^-\to\gamma\gamma, ~\gamma Z, ~ZZ$, including
electroweak corrections at the one loop level in the context  of SM and
MSSM.\par

These processes are particularly interesting in various aspects. From
an experimental point of view, the final states are easy to detect.
>>>From the  theoretical point of view, these
processes have a simple structure providing clean tests
of the properties of the electroweak interactions.
At tree level there is no s-channel term (contrary to the $WW$ case);
the Born terms are only due to electron exchanges in the
$t$ and $u$ channels.  There are no QCD or  Yukawa contributions;
the identification of the electroweak corrections
should then be very clean. \par

We have computed them completely, both within  SM and within
MSSM. We have given these results, in analytical
form, apart  from the exact 1-loop Box contributions to the
11 types of independent  contributions, for  which
we provide MATHEMATICA and FORTRAN codes upon request.
In all cases, we have shown how the corrections are constructed
in each sector; gauge neutral, gauge charged, Higgs, and the
supersymmetric counterparts. Special emphasis has been put to
the study of how these contributions behave with the energy, and how
 they match with the  high energy
logarithmic expressions expected from general rules.\par

We have then computed numerically the effects  on many
possible observables at variable energies;
\ie integrated cross sections,
Left-Right asymmetries and angular distributions,
for unpolarized and polarized initial and final states.
We next summarize the results and the physics issues.

The electroweak radiative corrections  are large and
grow with the energy. They are of a few percent in an
energy range of at few hundreds of GeV, reaching already $10\%$ at
1 TeV. They then continue to grow according to the logarithmic
rules. Such effects should be observable at the high precision
 future colliders \cite{LC,CLIC}, whose accuracy
should reach the percent level or even better.

In $e^+e^-\to\gamma\gamma$, the natural observables
(angular distribution at large angles, integrated cross section,
$A_{LR}$ asymmetry for unpolarized $\gamma\gamma$ final states)
reflect the gauge (gaugino) structure
of the electroweak interactions in a clean way. Below
1 TeV, the various considered  MSSM benchmark models of \cite{Ellis-bench},
 differ from SM within the $\pm1\%$ level.
 Above 1 TeV, the model dependence (for models involving relatively
 light supersymmetric particles) vanishes,
and the effects match the asymptotic
rules giving in MSSM a growing contribution
like $2\ln s-\ln^2s$ (times the Born amplitude),
instead of the $3\ln s-\ln^2s$ factor expected in SM.
So at asymptotic energies, we could in principle discriminate
between SM and MSSM; although we would have no means to choose
among     MSSM models involving relatively light
supersymmetric particles.
The $A_{LR}$ asymmetry shows the same effects as the
unpolarized cross section, a feature which may be experimentally
interesting.

The same properties can be observed
in the unpolarized or Transverse-Transverse
$e^+e^-\to Z\gamma,~ZZ$ cross sections and $A_{LR}$ asymmetries.
The model dependence is somewhat larger at low energies, but it
also vanishes in agreement with the logartthmic rules
at high energies. In these processes the
"mass suppressed" TL  cross sections are strongly
decreasing with the energy and
model dependent. Up to the 1 TeV range though,
these TL   cross sections should still be measurable, giving
 interesting tests of the MSSM models.

In $e^+e^-\to ZZ$, the LL cross section has
peculiar features associated to the Higgs sector. It is strongly
decreasing with the energy up to $1$ TeV, but above
$1$ TeV a flattening of the energy dependence appears which
depends on the value of the Higgs mass. However this happens
at a level which is only marginally observable with the
expected LC luminosities.\\

The $e^+e^-\to Z\gamma,~ZZ$ processes are sensitive to the
so-called NAGC, $h^{\gamma,Z}_3$ and $f^{\gamma,Z}_5$.
Below 1 TeV,  the SM and MSSM contributions to these
couplings should be marginally observable. But
the above processes could give interesting limits on
possible additional NP NAGC contributions, which (to the extent
they are described by local effective Lagrangians)
would lead to contributions growing with the energy.

In conclusions these three processes present a large
panel of interesting properties. They are extremely
simple at Born level, but extremely rich in
information at the one loop level.
The $\gamma\gamma$, $Z\gamma$, $ZZ$ final states
are complementary for the study of the gauge (gaugino) sector,
the MSSM models, the Higgs sector and the search for
Neutral Anomalous Gauge Couplings. They should
be considered as a part of the research program
at the future high energy colliders,  demanding for
the highest luminosities. In the very high energy range
(several TeV),  higher order effects (two loop effects
and/or resummation) should also be computed,
in order to make really accurate theoretical predictions,.
The several TeV domain indeed appears to be the region where the
electroweak interactions start becoming strong.

\newpage

\renewcommand{\theequation}{A.\arabic{equation}}
\renewcommand{\thesection}{A.\arabic{section}}
\setcounter{equation}{0}
\setcounter{section}{0}

{\large \bf Appendix A: Kinematical details.}\\

According to (\ref{process}, \ref{hel}, \ref{NjLR}),
the invariant amplitude of the process
\bq
e^-(\lambda, l)~+~e^+(\lambda', l')
\to V(e,p)+ V'(e',p') ~~ , \label{processa}
\eq
\noindent
 may be written as
\bq
F_{\lambda,\mu,\mu'}
=\sum_{j=1,9} \bar{v}(\lambda', l')~I_j~N_j(s,t,u, \lambda)
~u(\lambda, l) ~~ . \label{amp}
\eq
When  the electron mass is neglected, so that
$\lambda'=-\lambda $ in both the SM and MSSM models,
and the  9 invariant forms is (\ref{amp}) are
\bqa
I_1=(e \cdot l)(\gamma \cdot e') &,~~
I_2=(e' \cdot l)(\gamma \cdot e) & ,
~~ I_3=(e\cdot l)(e'\cdot l)(\gamma\cdot p),~~
\nonumber \\
I_4=(e\cdot e')(\gamma \cdot p) & , ~~
I_5=(e \cdot p')(\gamma \cdot e') & ,~~
I_6=(e \cdot p')(e'\cdot l)(\gamma \cdot p)~~,
\nonumber \\
I_7=(e' \cdot p)(\gamma \cdot  e) &,~~ I_8=(e'\cdot p)(e\cdot l)
(\gamma \cdot p)&,
~~ I_9=(e \cdot p')(e'\cdot p)(\gamma \cdot p) ~. \label{Ij}
\eqa
The related scalar  amplitudes $N_j(s,t,u, \lambda)$
may be split according to the electron helicity as  in
(\ref{NjLR},\ref{proj}).

The transverse and longitudinal amplitudes
implied by (\ref{amp}), are:\\
$\bullet$~
\underline{TT amplitudes, $\mu=\pm 1$ and $\mu' =\pm 1$}
\bqa
&& F_{\lambda,\mu,\mu'}=
{ s \sin\theta\over4}~\Big \{\mu(1-2\lambda\mu' \cos\theta)N_1
-\mu'(1+2\lambda\mu \cos\theta)N_2\nonumber\\
&&-~{\beta s\over4}\sin^2\theta
~(2\lambda\mu\mu') N_3
+\beta (2\lambda) (1+\mu\mu')N_4 \Big \}  ~~,  \label{TT-amp}
\eqa
\noindent
$\bullet$~
\underline{TL amplitudes, $\mu=\pm 1$, ~$\mu'=0$}
\bqa
&&F_{\lambda,\mu,0}={s\sqrt{2s}~\over 8M'}~\Big \{
-(2\lambda\mu)\beta'_0 \sin^2\theta N_1
+(\beta+\beta'_0 \cos\theta)(1+2\lambda\mu \cos\theta)N_2
\nonumber\\
&&
+{\beta s\over4}(\beta+\beta'_0 \cos\theta)
(2\lambda\mu) sin^2\theta N_3
+2\beta(1+2\lambda\mu \cos\theta)N_7
+{s~\beta^2 \sin^2\theta\over2} 2\lambda\mu N_8 \Big \}
, \label{TL-amp}
\eqa
\noindent
$\bullet$~
\underline{LT amplitudes, $\mu =0$, $\mu' =\pm1$}
\bqa
F_{\lambda,0,\mu'} &=&
{s\sqrt{2s} \over8M}~\Big \{~\beta_0 (2\lambda\mu') sin^2\theta N_2
-(\beta-\beta_0 \cos\theta)(1-2\lambda\mu' \cos\theta)N_1
\nonumber\\
&&
+{\beta s\over4}(\beta-\beta_0 \cos\theta)(2\lambda\mu')\sin^2\theta
N_3
-2\beta(1-2\lambda\mu' cos\theta)N_5\nonumber\\
&&
+{s~\beta^2 \sin^2\theta\over2} (2\lambda\mu') N_6\Big \}
~~, \label{LT-amp}
\eqa
\noindent
$\bullet$~
\underline{LL amplitudes, $\mu =0$, $\mu'=0$}
\bqa
F_{\lambda,0,0}&=&{s^2~\sin\theta(2\lambda)\over32MM'}~
\Big \{~4\beta'_0(\beta-\beta_0\cos\theta)N_1
-4\beta_0(\beta+\beta'_0\cos\theta)N_2\nonumber\\
&&
-\beta s(\beta-\beta_0\cos\theta)
(\beta+\beta'_0 \cos\theta)N_3-4\beta(\beta_0\beta'_0+\beta^2)N_4
\nonumber\\
&&+8\beta(\beta'_0N_5-\beta_0N_7)
-2s\beta^2(\beta+\beta'_0\cos\theta)N_6\nonumber\\
&&-2s\beta^2(\beta-\beta_0\cos\theta)N_8
-4s\beta^3N_9\Big \} ~~ , \label{LL-amp}
\eqa
where $\theta$ is the c.m. scattering angle and
\bqa
&& \beta \equiv \frac{2 |\vec p|}{\sqrt{s}} ~~~,~~~
\beta_0 \equiv \frac{2 p_0}{\sqrt{s}} ~~~,~~~
\beta'_0 \equiv \frac{2 p'_0}{\sqrt{s}} ~~, \nonumber
\eqa
with $|\vec p|$ being  the magnitude of the c.m. momenta of the final
gauge bosons and $p_0,~ p'_0$ their energies.

In the specific case of   the process
\underline{$e^-e^+\to\gamma\gamma$},
only 4 TT amplitudes appear involving the ($N_1$, $N_2$, $ N_3$, $N_4$)
functions. In this case
\[
s+t+u=0 ~~~,~~~ \beta=\beta_0=\beta'_0=1~~.
\]

In the case of   \underline{$e^-e^+\to Z \gamma $},
where the gauge boson polarization and momenta are
defined by  $Z(e(\mu),p)$ and $\gamma(e'(\mu'),p')$,
there exist only  6 TT and LT amplitudes receiving
contributions from  $(N_1, ...~ N_6)$. In this case
\[
s+t+u=\mzd ~~, ~~
\beta=\beta'_0=1-{\mzd\over s} ~~,~~
\beta_0=1+{\mzd\over s}~~~.
\]

Finally,  for \underline{$e^-e^+\to ZZ$} the complete set of
 ${N_1, ... ~N_9}$ contributes to the 9  TT, TL, LT and LL
 amplitudes with
\[
s+t+u=2\mzd ~~, ~~
\beta^2=1-{4\mzd\over s} ~~, ~~ \beta_0=\beta'_0=1 ~~.
\]


\vspace{2.cm}
\renewcommand{\theequation}{B.\arabic{equation}}
\renewcommand{\thesection}{B.\arabic{section}}
\setcounter{equation}{0}
\setcounter{section}{0}

{\large \bf Appendix B: The chargino and
neutralino  mixing matrices.}\\

\noindent
{\bf The chargino mixing.}\\
The Left flavor space chargino fields of  positive and negative
electric charge\footnote{See \eg Eqs.(A.26-A.35) in
\cite{ggZZ-second}.}
\bq
\tilde \psi_L^+=
\left (\matrix {\tilde W^+ \cr \tilde H_2^+}\right )_L
~~, ~~
\tilde \psi_L^-=
\left (\matrix {\tilde W^- \cr \tilde H_1^-}\right )_L
\label{chargino-matrix1}
\eq
are related to the mass-eigenstate chargino fields by
\bq
\tilde \psi_{\alpha L}^+=\sum_{j=1}^2 V_{j \alpha }\eta_{cj}
\tchi^+_{jL} ~~,~~
\tilde \psi_{\alpha L}^-=\sum_{j=1}^2 U_{j \alpha }
\tchi^-_{jL} ~~,~~ \label{chargino-matrix2}
\eq
where $\alpha =1,~2$ counts the charginos in the flavor space,
while $j=1,~2$ in space of the mass eigenstates.
Assuming that the  MSSM breaking parameters $M_1, ~ M_2, ~\mu$ are
real and choosing the arbitrary phases so that $M_2>0$, the
chargino physical masses may be written as
\bq
M_{ \tchi_1,  \tchi_2}
=\frac{1}{\sqrt{2}} [M_2^2 +\mu^2 +2\mwd \mp \tilde D ]^{1/2}
~ , \label{chi-mass}
\eq
where
\bq
\tilde D \equiv  \left [
(M_2^2+\mu^2+ 2\mwd)^2- 4 (M_2\mu-\mwd \sin(2\beta))^2
\right ]^{1/2} ~ , \label{Dtilde}
\eq
while the mixing matrices defined in (\ref{chargino-matrix2})
for the negative and positive Left-charginos are
\bq
U =
 \left (\matrix{ \cos\phi_L & \Bcal_L \sin\phi_L \cr
-\Bcal_L \sin\phi_L  &  \cos\phi_L  \cr } \right )
 ~~  , ~~
V =
 \left (\matrix{ \cos\phi_R & \Bcal_R \sin\phi_R \cr
-\Bcal_R \sin\phi_R  &  \cos\phi_R  \cr } \right )
  \label{UV-chargino}
\eq
where $\phi_L, \phi_R$  are defined as
\bqa
\cos\phi_L &=& -~ \frac{1}{\sqrt{2\tilde D}}
[\tilde D-M_2^2+\mu^2 +2\mwd \cos 2\beta ]^{1/2} ~~ ,
\nonumber \\
\cos\phi_R &=& -~ \frac{1}{\sqrt{2\tilde D}}
[\tilde D-M_2^2+\mu^2 -2\mwd \cos 2\beta ]^{1/2} ~~ ,
\label{chi-angles}
\eqa
 so that they always lie in the second quarter
\bq
\frac{\pi}{2} \leq \phi_L < \pi  ~~~~ , ~~~~
\frac{\pi}{2} \leq \phi_R  < \pi  ~~~ . \label{chi-angle-range}
\eq
Because of this definition $\sin\phi_{L,R}$ are always positive
demanding that the appearance of the  sign coefficients
\bqa
\Bcal_L & = &\Sn (\mu \sin\beta +M_2 \cos\beta)  ~,
\nonumber \\
\Bcal_R & = & \Sn (\mu \cos\beta +M_2 \sin \beta) ~,
\label{B-signs}
\eqa
appear in (\ref{UV-chargino}). In addition the sign-coefficients
 \bqa
\eta_{c1}  &= &
\Sn (M_2 [\tilde D-M_2^2+\mu^2-2\mwd] -2 \mwd \mu \sin 2\beta )  ~, ~
\nonumber \\
 \eta_{c2} &=&
\Sn (\mu [\tilde D-M_2^2+\mu^2 +2\mwd] +2 \mwd M_2 \sin 2\beta ) ~ ,
\eqa
also enter (\ref{chargino-matrix2}), determining the
way the left and right charginos combine in  the
 Dirac field.  \par

For comparison with the notation of  \cite{Rosiek} we  note
that the $Z^\pm$ matrices defined there are given by
\bq
Z^+_{\alpha j}= V_{j \alpha }\eta_{cj} ~~,~~
Z^-_{\alpha j}=U_{j \alpha } ~~, \label{Rosiek-chargino}
\eq
for real $M_1$, $M_2$ and $\mu$ parameters.

Using (\ref{Rosiek-chargino}, \ref{UV-chargino}), the
chargino contribution to (\ref{Sigma-e-gaugino}) is then
determined from
\bq
|Z^+_{1j}|^2 = |V_{j1}|^2 ~~.
\eq\\

\noindent
{\bf The neutralino  mixing.}\\
We follow the notation of \cite{LeMouel} and continue restricting
to real $M_1$, $M_2$ and $\mu$ parameters. In the space
of the Left neutralino
fields
\bq
\psi^0_L \equiv \left ( \matrix{\tilde B_L \cr \tilde
W^{(3)}_L \cr \tilde H_{1L}^0 \cr  \tilde H_{2L}^0 \cr } \right )
~~ , \label{Psi0L}
\eq
the mass-matrix  is of course symmetric and given by
\bq
Y=  \left ( \matrix{  M_1  & 0  & -\mz \sw \cbeta &   \mz \sw
\sbeta \cr 0  & M_2  & \mz \cw \cbeta  & -\mz \cw \sbeta \cr -\mz
\sw \cbeta  & \mz \cw \cbeta  & 0  & -\mu \cr
 \mz \sw \sbeta  &  -\mz \cw \sbeta  & -\mu  & 0 \cr}
\right ) ~~ . \label{Y-matrix}
\eq
This is
diagonalized through the  real orthogonal transformation $U^0$
giving
\bq
U^{0\top} Y U^0 =\left (  \matrix{\tilde M_{\tchi_1^0} & 0 & 0&
0 \cr 0 & \tilde M_{\tchi_2^0} & 0 & 0 \cr 0 & 0 & \tilde
M_{\tchi_3^0} & \cr 0 & 0 & 0& \tilde M_{\tchi_4^0}  \cr }\right ) ~~
, ~~ \label{Yd-matrix}
\eq
where the real eigenvalues  $\tilde M_{\tchi_j}$ can be of either
sign and have been ordered so that
\bq
|\tilde M_{\tchi_1^0}| \leq |\tilde M_{\tchi_2^0}| \leq |\tilde
M_{\tchi_3^0}| \leq |\tilde M_{\tchi_4^0}| ~~ . \label{mj-ordering}
\eq
The quantities $\tilde M_{\tchi_j^0}$ are
the  "signed" neutralino masses which are
directly determined by solving the characteristic equation
implied by (\ref{Yd-matrix}) using \eg  the formalism in \cite{Cairo}
 or Eqs.(10-18) in \cite{LeMouel}.
Their absolute
values determine the physical neutralino masses $M_{\tchi_j^0}$,
while the related signs $\eta_j$ are determined by
\bq
\tilde M_{\tchi_j^0} =\eta_j M_{\tchi_j^0}~~~~ {\rm with} ~~~~~
~~~~  \eta_j=\pm 1 ~~~ ~. ~  \label{etaj}
\eq

Following  (\cite{LeMouel},  the definition
$(\tilde \eta_j =1~ ~ {\rm or} ~~i)$ is also introduced,
so that  $ \eta_j = \tilde \eta_j^2$.
The relation between the flavor and mass-eigenstate
neutralino fields is then given by
\bq
\psi^0_{\alpha L}= \sum_{j=1}^4
U^0_{\alpha j}\tilde \eta_j\tchi_{jL}^0
 ~~~ \label{U0-neutralino}
\eq
where the index $\alpha$ (as well as $\beta$ in the next paragraph)
 counts the neutralino flavor components, while
the index $j$ refers to  the mass-eigenstate ones.
Of course both indices run from 1 to 4. The above $U^0$
neutralino matrix is related to the $Z^N$ one defined
in \cite{Rosiek} by absorbing it in the $\tilde \eta_j$ phases as
\bq
Z^N_{\alpha j} = U^0_{\alpha j} \tilde \eta_j ~~~.
\label{Rosiek-neutralino}
\eq

As shown in \cite{LeMouel}, all neutralino related
physical observables in the case
of real $(M_1$, $M_2$ and $\mu)$, can then be  expressed in terms of
the signs $\eta_j$, and the four density matrices $P_j,~ (j=1,...4)$
describing the flavor composition of each of the four neutralinos.
These density matrices act in the flavor space and  are given by
\bq
P^0_{j\alpha\beta}=U^0_{\alpha j}U^0_{ \beta j} ~~ .
\label{Projector-a}
\eq
As expected for any density matrix describing  pure states,
they have the mathematical properties of projection
operators and may be immediately calculated
 from \cite{LeMouel, Jarlskog}
\bqa
P_1^0 &=& \frac{(\tilde M_{\tchi_4^0} -Y) (\tilde M_{\tchi_3^0}
-Y)(\tilde M_{\tchi_2^0} -Y)} {(\tilde M_{\tchi_4^0} -\tilde
M_{\tchi_1^0}) (\tilde M_{\tchi_3^0} -\tilde M_{\tchi_1^0}) (\tilde
M_{\tchi_2^0} -\tilde M_{\tchi_1^0})} ~~ ~~, \nonumber \\[0.3cm]
P_2^0 &=& \frac{(\tilde M_{\tchi_4^0} -Y) (\tilde M_{\tchi_3^0} -Y)
(Y- \tilde M_{\tchi_1^0} )} {(\tilde M_{\tchi_4^0} -
\tilde M_{\tchi_2^0})
(\tilde M_{\tchi_3^0} -\tilde M_{\tchi_2^0}) (\tilde M_{\tchi_2^0}
-\tilde M_{\tchi_1^0})} ~~ ~~, \nonumber \\[0.3cm]
P_3^0 &=&
\frac{(\tilde M_{\tchi_4^0} -Y) (Y- \tilde M_{\tchi_2^0} )(Y- \tilde
M_{\tchi_1^0} )} {(\tilde M_{\tchi_4^0} -\tilde M_{\tchi_3^0}) (\tilde
M_{\tchi_3^0} -\tilde M_{\tchi_2^0}) (\tilde M_{\tchi_3^0} -\tilde
M_{\tchi_1^0})} ~~ ~~, \nonumber \\[0.3cm]
P_4^0 &=& \frac{(Y- \tilde
M_{\tchi_3^0} ) (Y- \tilde M_{\tchi_2^0} )(Y- \tilde M_{\tchi_1^0} )}
{(\tilde M_{\tchi_4^0} -\tilde M_{\tchi_3^0}) (\tilde M_{\tchi_4^0}
-\tilde M_{\tchi_2^0}) (\tilde M_{\tchi_4^0} -\tilde M_{\tchi_1^0})} ~~
~~, \label{Projector-b}
\eqa
where $Y$ is given in (\ref{Y-matrix}).

In terms  of by-linear forms of the $Z^N$ mixing matrix defined in
(\ref{Rosiek-neutralino})  \cite{Rosiek},
the above neutralino density matrices and signs satisfy
\bqa
&& Z^N_{\alpha j}Z^{N *}_{\beta j}=
Z^{N *}_{\alpha j}Z^{N}_{\beta j}=
P^0_{j\alpha \beta}=P^0_{j \beta \alpha} ~~, \\
&& Z^N_{\alpha j}Z^N_{\beta j}=
Z^{N*}_{\alpha j}Z^{N *}_{\beta j}=
P^0_{j\alpha \beta}\eta_j =P^0_{j \beta \alpha}\eta_j  ~~ ,
\label{Rosiek-projector}
\eqa
which fully describe all neutralino loop contributions.

\vspace{1cm}
\noindent
{\bf The chargino and neutralino couplings.}\\
In terms of the chargino and neutralino mixings
defined above, we  list
explicitly below the couplings needed for describing
the charginos and neutralino contributions to the
$Z$ and $W$ self-energies. They are given by the
interaction Lagrangian
\bqa
\L&=& -~\frac{e}{2 \sw \cw} Z_\mu \Bigg \{
\sum_{i,j=1}^2\Big [ O_{ij}^{ZL}
\bar{\tchi}^+_{iL}\gamma^\mu \tchi^+_{jL}
+O_{ij}^{ZR} ~\bar{\tchi}^+_{iR}\gamma^\mu \tchi^+_{jR} \Big ]
\nonumber \\
&& -\sum_{i,j=1}^4  O_{ij}^{0ZL}
~\bar{\tchi}^0_{iL}\gamma^\mu \tchi^0_{jL}
\Bigg \}
\nonumber \\
&& + g W^+_\mu \sum_{i=1}^2\sum_{j=1}^4 \Big \{
  \bar{\tchi}^+_{i}\gamma^\mu
  \Big [O_{ij}^{L}~\frac{(1-\gamma^5)}{2}
  +O_{ij}^{R}~\frac{(1+\gamma^5)}{2} \Big ]\tchi^0_{j}
  \nonumber \\
&& + g W^-_\mu \sum_{i=1}^2\sum_{j=1}^4 \Big \{
  \bar{\tchi}^0_{j}\gamma^\mu
  \Big [O_{ij}^{L*}~\frac{(1-\gamma^5)}{2}
  +O_{ij}^{R*}~\frac{(1+\gamma^5)}{2} \Big ]\tchi^+_{i}
  \Big \} ~~. \label{Z-W-gaugino-Lagrangian}
\eqa

The $Z$-chargino couplings in (\ref{Z-W-gaugino-Lagrangian})
are given by
(compare (\ref{Rosiek-chargino})
\bqa
O_{ij}^{ZL}&=& Z^{+*}_{1i}Z^{+}_{1j}+\delta_{ij}(1- 2\swd)=
\eta_{ci}\eta_{cj}V_{i1}V_{j1} +\delta_{ij}(1- 2\swd) ~~ ,
\nonumber \\
O_{ij}^{ZR}&=& Z^-_{1i}Z^{-*}_{1j}+\delta_{ij}(1- 2\swd)=
U_{i1}U_{j1} +\delta_{ij}(1- 2\swd) ~~ ,
\label{Z-chargino-coupling}
\eqa
where (\ref{UV-chargino}), \ref{chi-angles}) are needed.
For the  neutralino  couplings
\bqa
O_{ij}^{0ZL}=O_{ji}^{0ZL*}=-O_{ji}^{0ZR}=-O_{ij}^{0ZR*}
=Z_{4i}^{N*}Z_{4j}^N -Z_{3i}^{N*}Z_{3j}^N
\nonumber \\
= \tilde \eta_i^* \tilde \eta_j
(U_{4i}^0 U_{4j}^0-U_{3i}^0 U_{3j}^0)~~,
\label{Z-neutralino-coupling}
\eqa
the needed bilinear terms are
\bqa
&&O_{ij}^{0ZL}O_{ji}^{0ZL}=O_{ij}^{0ZR}O_{ji}^{0ZR}
=P^0_{i33}P^0_{j33}+P^0_{i44}P^0_{j44}-
2P^0_{i34}P^0_{j34}~~, \nonumber \\
&& O_{ij}^{0ZL}O_{ji}^{0ZR}=O_{ij}^{0ZR}O_{ji}^{0ZL}
=-\eta_i \eta_j [P^0_{i33}P^0_{j33}+P^0_{i44}P^0_{j44}-
2P^0_{i34}P^0_{j34}]~~ ,
\label{Z-neutralino-coupling-bilinears}
\eqa
fully defined in terms of the neutralino density matrices in
(\ref{Projector-b}).\par

Finally, the $W$-couplings in (\ref{Z-W-gaugino-Lagrangian}) are
given using (\ref{UV-chargino}, \ref{Rosiek-chargino},
\ref{U0-neutralino}, \ref{Rosiek-neutralino})
\bqa
&&O_{ij}^L=Z^+_{1i} Z^N_{2j}-\frac{1}{\sqrt{2}}~Z^+_{2i} Z^N_{4j}
=\eta_{ci}\tilde \eta_j
(V_{i1}U^0_{2j} -\frac{1}{\sqrt{2}} V_{i2}U^0_{4j}) ~~ ,
\nonumber \\
&&O_{ij}^R=Z^-_{1i} Z^{N*}_{2j}+\frac{1}{\sqrt{2}}
~Z^-_{2i} Z^{N*}_{3j}
=\tilde \eta_j^*
(U_{i1}U^0_{2j} +\frac{1}{\sqrt{2}} U_{i2}U^0_{3j}) ~~ ,
\label{W-gaugino-coupling}
\eqa
where the first index counts the chargino and the second the
neutralino. The corresponding
bilinears needed for the $W$-self-energies are
\bqa
 O_{ij}^L O_{ij}^{L*}+O_{ij}^R O_{ij}^{R*}&= &
[(V_{i1})^2+(U_{i1})^2]P^0_{j22}
+~\frac{1}{2}~[(V_{i2})^2P^0_{j44}+(U_{i2})^2P^0_{j33}]
\nonumber \\
&& -\sqrt{2} V_{i1}V_{i2}P^0_{j24} +
\sqrt{2} U_{i1}U_{i2}P^0_{j23} ~~,
\nonumber \\
 O_{ij}^L O_{ij}^{R*}+O_{ij}^{L*} O_{ij}^R&= &
2 \eta_{ci}\eta_j \Big [V_{i1}U_{i1}P^0_{j22}
+~\frac{1}{\sqrt{2}}~V_{i1}U_{i2}P^0_{j23}
\nonumber \\
&&- ~\frac{1}{\sqrt{2}}~V_{i2}U_{i1}P^0_{j24}
-~\frac{1}{2}V_{i2}U_{i2}P^0_{j34} \Big ] ~~.
\label{W-gaugino-coupling-bilinears}
\eqa

The use of
(\ref{Rosiek-neutralino}) allows to express the neutralino
contribution to (\ref{Sigma-e-gaugino}) in terms of the neutralino
density matrix elements defined in (\ref{Projector-b}) through
\bqa
|Z^N_{1j}s_W+Z^N_{2j}c_W|^2&=&P^0_{j11}\swd+P^0_{j22}\cwd+
2\sw\cw P^0_{j12} ~~ \nonumber \\
|Z^N_{1j}|^2 &=&P^0_{j11} ~~.  \label{e-gaugino-contribution}
\eqa\\

Finally we should emphasize that
for calculating the virtual  chargino and neutralino contributions,
it is essential that all masses and couplings are
calculated together at the \underline{same} accuracy,
from the "electroweak scale" values of
$M_2,~ M_1, ~\mu,~ \tan\beta$; otherwise the chargino-neutralino
contribution to \eg the renormalized gauge self energies will not be
finite, inducing  spurious scale dependencies to the numerical
results.


\vspace{2.cm}
\renewcommand{\theequation}{C.\arabic{equation}}
\renewcommand{\thesection}{C.\arabic{section}}
\setcounter{equation}{0}
\setcounter{section}{0}

{\large \bf Appendix C: Gauge self-energies, Electron self-energies
and renormalization constants.}\\

\noindent
\underline{{\bf C1) Gauge self-energies.}}\\
The needed gauge renormalization constants are expressed (using the
renormalization conditions, \cite{Hollik}) in terms of gauge
self-energies denoted by\footnote{To define their phase
we  give their relation to the S-matrix element as
$S_{fi}=-i g_{\mu \nu} \Sigma_{VV'}$.

 The various couplings
are defined as in \cite{Rosiek}. See also \eg
\cite{ggZZ-first, ggZZ-second}.} $\Sigma_{VV'}$
are given in Appendix B, for SM and the generic minimal MSSM
case. \par

Using the unrenormalized gauge self-energies  and
 (\ref{gauge-renormalization-constants}),
 the gauge wave-function renormalization constants satisfy
(compare (\ref{gauge-wave-function-renormalization}))
\bqa
\delta Z_W &=& -\, \Re \Big ( \Sigma_{\gamma \gamma}'(0)
-~\frac{2\cw}{\sw \mzd} \Sigma_{\gamma Z}(0)+
\frac{1}{\swd \mzd}\, \Big [ \Sigma_{WW}(\mwd)-
\cwd \Sigma_{ZZ}(\mzd) \Big ] \Big ) ~,
\nonumber \\
\delta Z_B &=& -\, \Re \Big ( \Sigma_{\gamma \gamma}'(0)
+\,\frac{2\sw}{\cw \mzd} \Sigma_{\gamma Z}(0)
-\, \frac{1}{\cwd \mzd}\, \Big [ \Sigma_{WW}(\mwd)-
\cwd \Sigma_{ZZ}(\mzd) \Big ] \Big ) ~,
\label{del-ZW-ZB}
\eqa
while the additional renormalization needed for the $SU(2)$
gauge coupling is
\bq
\delta \tilde Z_2=\frac{1}{\mzd \sw\cw} \Sigma_{\gamma Z}(0)
~~. \label{del-tilde-Z2}
\eq

Using (\ref{del-ZW-ZB}), we then
write the derivative of the renormalized Z-self-energy
and the $\gamma Z$-mixing at the $Z$-shell contributions entering
(\ref{eLR}, \ref{gZLR}) as
\bqa
\hat \Sigma_{ZZ}'(\mzd)&=& \Sigma_{ZZ}'(\mzd)+
\cwd \delta Z_W +\swd \delta Z_B ~~,
\nonumber \\
\hat \Sigma_{\gamma Z}(\mzd)&=& \Sigma_{\gamma Z}(\mzd)
+ \mzd \sw\cw  ( \delta Z_W -\delta Z_B -\delta \tilde  Z_2)
~. \label{Z-propagator}
\eqa \par

We now give the expressions for
the various contributions
to the transverse unrenormalized gauge self-energies in the MSSM
(without CP-violation) and at the end of this part
we give the recipe for
restricting to the SM case.
The relevant MSSM couplings for the chargino
and neutralino loops
are summarized in Appendix A.
In each case, we first give the contributions from
the gauge bosons, from the two Higgs doublets,
from the  standard quarks and leptons with isospin
$I_3^f$, charge $Q_f$  and Z-couplings
\bq
v_f=\frac{I_3^f-2 Q_f \swd}{2\sw\cw} ~~~,~~~
a_f=\frac{I_3^f}{2\sw\cw} ~~ , \label{vf-af}
\eq
subsequently  the contribution from a sfermion $\tilde f$
 whose mixing angles we denote as
 \[
 \csf=\cos(\tilde \theta_f) ~~~,~~~
 \ssf=\sin(\tilde \theta_f) ~~,
 \]
 following the same notation as in \cite{ggZZ-first},
 and finally the contribution from the chargino and/or neutralino
loop.

All soft MSSM breaking parameters and $\mu$
 are taken as real, and the  phases of the appropriate fields
 are selected so that $M_2>0$.

\vspace{1.cm}
\noindent
{\bf Contributions to  $\Sigma_{ZZ}(k^2)$.}

The respective $W$ plus Higgs, fermion and sfermion contributions
to the $Z$-self-energy are
\bqa
&&\Sigma_{ZZ}(k; {\rm gauge + 2H})=\frac{\alpha}{4\pi \swd\cwd}
\Bigg \{ \sin^2(\beta-\alpha)
[\mzd B_0^{Zh^0}-B_{22}^{Zh^0}-B_{22}^{A^0H^0}]
\nonumber \\
&& +\cos^2(\beta-\alpha)[\mzd B_0^{ZH^0}-B_{22}^{ZH^0}-B_{22}^{A^0h^0}]
-\cos^2(2\theta_W) B_{22}^{H^+H^+}+
\frac{1}{4}[ A^{h^0}+A^{H^0} \nonumber \\
&&  +A^{A^0}+A^{Z}] +\frac{\cos^2(2\theta_W)}{2}\, A^{H^+}
-[8 \cw^4+\cos^2(2\theta_W)]B_{22}^{WW}
\nonumber \\
&& -[4\cw^4 k^2+2\mwd \cos(2\theta_W)]B_0^{WW}
+\frac{1}{2}[12 \cw^4 -4\cwd+1]A^W -\, \frac{2}{3}\,
\cw^4 k^2 \Bigg \}~, \label{ZZ-gauge-H}
\\
&& \Sigma_{ZZ}(k; f)= -\, \frac{\alpha}{\pi} \sum_f N_c^f  \Big \{
(v_f^2+a_f^2)[-2B_{22}^{ff}+A^f+(m_f^2-\,\frac{k^2}{2})B_0^{ff}]
\nonumber \\
&& -(v_f^2-a_f^2)m_f^2B_0^{ff}  \Big \}~, \label{ZZ-f}
\\
&& \Sigma_{ZZ}(k;\tilde f)=
 -\, \frac{\alpha}{4 \pi\swd \cwd} \sum_f N_c^f  \Bigg \{
 4 [I_3^f \csf^2-Q_f \swd]^2 B_{22}^{\tilde f_1 \tilde f_1}
 +\ssf^2\csf^2[B_{22}^{\tilde f_1 \tilde f_2}
 +B_{22}^{\tilde f_2 \tilde f_1}]
 \nonumber \\
 && +4[I_3^f \ssf^2-Q_f \swd]^2 B_{22}^{\tilde f_2 \tilde f_2}
 -2[(I_3^f-Q_f \swd)^2\csf^2 +Q_f^2 \sw^4 \ssf^2] A^{\tilde f_1}
 \nonumber \\
 &&-2[(I_3^f-Q_f \swd)^2\ssf^2 +Q_f^2 \sw^4 \csf^2] A^{\tilde f_2}
  \Bigg \} ~~. \label{ZZ-sfermion}
\eqa
In (\ref{ZZ-f}, \ref{ZZ-sfermion}) $N_c^f$ is 3 or
1, depending on whether $f$ is a quark or a lepton respectively.
For the neutralino and chargino contributions we get
\bqa
&& \Sigma_{ZZ}(k; \tchi_j^0)=-\, \frac{\alpha}{8\pi \swd\cwd}
\sum_{i,j=1}^4\Big (P^0_{i33}P^0_{j33}+P^0_{i44}P^0_{j44}
-2P^0_{i34}P^0_{j34} \Big ) \cdot \Big
[ -2B_{22}^{\tchi_i^0 \tchi_j^0}
\nonumber \\
&& + A^{\tchi_j^0}
 + M_{\tchi_i^0}(M_{\tchi_i^0}+\eta_i\eta_j M_{\tchi_j^0})
B_0^{\tchi_i^0 \tchi_j^0}+k^2 B_1^{\tchi_i^0 \tchi_j^0}
\Big ] ~~, \label{ZZ-chi0} \\
&& \Sigma_{ZZ}(k; \tchi_j^+)=-\, \frac{\alpha}{8\pi \swd\cwd}
\sum_{i,j=1}^2 \Bigg \{
\Big ( O_{ij}^{ZL}O_{ji}^{ZL}+O_{ij}^{ZR}O_{ji}^{ZR} \Big )\Big [
-2 B_{22}^{\tchi_i^+ \tchi_j^+} + A^{\tchi_j^+}
\nonumber \\
&& + M_{\tchi_j^+}^2 B_0^{\tchi_i^+ \tchi_j^+}+
k^2B_1^{\tchi_i^+ \tchi_j^+} \Big ]-
\Big ( O_{ij}^{ZL}O_{ji}^{ZR}+O_{ij}^{ZR}O_{ji}^{ZL} \Big )
M_{\tchi_i^+} M_{\tchi_j^+} B_0^{\tchi_i^+ \tchi_j^+}
\Bigg \} ~~. \label{ZZ-chip}
\eqa

\noindent
{\bf Contributions to  $\Sigma_{\gamma \gamma}(k^2)$.}

The respective $W$ plus Higgs, fermion, sfermion and chargino
 contributions to the photon self-energy are
\bqa
\Sigma_{\gamma \gamma }(k; {\rm gauge + 2H})
&=& - \frac{\alpha}{2\pi} \Big \{ 6B_{22}^{WW}+
2 B_{22}^{H^+H^+}- A^{H^+}-3A^W
\nonumber \\
&& + 2 k^2 B_0^{WW}+ \frac{ k^2}{3}\Big \} ~~, \label{gg-W-H}
\\
\Sigma_{\gamma \gamma }(k; f) &=&-\, \frac{\alpha}{\pi}
\sum_f N_c^f Q_f^2 \Big \{ -2 B_{22}^{ff}+A^f-\,
\frac{k^2}{2}\, B_0^{ff}\Big \} ~~, \label{gg-f}
\\
\Sigma_{\gamma \gamma }(k;  \tilde f) &=& -\, \frac{\alpha}{2\pi}
\sum_f N_c^f Q_f^2 \Big \{ -A^{\tilde f_1}-A^{\tilde f_2}
+2 B_{22}^{\tilde f_1 \tilde f_1}+2 B_{22}^{\tilde f_2 \tilde f_2}
\Big \} ~~, \label{gg-sfermion}
\\
\Sigma_{\gamma \gamma }(k; \tchi_j^+ ) &=&-\, \frac{\alpha}{\pi}
\sum_{j=1}^2  \Big \{ -2 B_{22}^{\tchi_j^+\tchi_j^+  }
+A^{\tchi_j^+} -\,
\frac{k^2}{2}\, B_0^{\tchi_j^+ \tchi_j^+ }\Big \}
~~.  \label{gg-chargino}
\eqa

\vspace{1.cm}
\noindent
{\bf Contributions to  $\Sigma_{\gamma Z}(k^2)$.}

The respective $W$ plus Higgs, fermion, sfermion and chargino
 contributions to the photon-$Z$ mixing are
\bqa
&& \Sigma_{\gamma Z }(k; {\rm gauge + 2H})
=- \,  \frac{\alpha}{4\pi} \Bigg \{
\frac{\cos(2\theta_W)}{\sw\cw}\, \Big [-A^W-A^{H^+}+2B_{22}^{H^+H^+}+
2B_{22}^{WW}\Big ]
\nonumber \\
&& +\frac{\cw}{\sw}\Big [ 8B_{22}^{WW}-4A^W+(2\mzd +4k^2)B_0^{WW}+
\frac{2k^2}{3} \Big ]+2\mzd\sw\cw B_0^{WW} \Bigg \} ~~, \label{gZ-W-H}
\\
&& \Sigma_{\gamma Z}(k; f) =-\, \frac{\alpha}{\pi}
\sum_f N_c^f Q_f v_f [A^f-2B_{22}^{ff}-\frac{k^2}{2}B_0^{ff} ]
~~, \label{gZ-f}
\\
&& \Sigma_{\gamma Z}(k;\tilde f) =-\, \frac{\alpha}{2\pi\sw\cw}
\sum_f N_c^f Q_f \Big \{ (I_3^f \csf^2-Q_f \swd)
(2 B_{22}^{\tilde f_1 \tilde f_1}-A^{\tilde f_1})
\nonumber \\
&& + (I_3^f \ssf^2-Q_f \swd)
(2 B_{22}^{\tilde f_2 \tilde f_2}-A^{\tilde f_2}) \Big \}
~~,\label{gZ-sfermion}
\\
&& \Sigma_{\gamma Z}(k; \tchi_j) =-\, \frac{\alpha}{4\pi\sw\cw}
\sum_{j=1}^2 (O_{jj}^{ZL}+O_{jj}^{ZR})
 [A^{\tchi_j^+} -2B_{22}^{\tchi_j^+\tchi_j^+}
 -\frac{k^2}{2}B_0^{\tchi_j^+\tchi_j^+} ]
~~. \label{gZ-chargino}
\eqa

\vspace{1.cm}
\noindent
{\bf Contributions to  $\Sigma_{W W}(k^2)$.}

The respective gauge plus Higgs, fermion and sfermion
 contributions to the $W$ self-energy are
\bqa
&& \Sigma_{WW }(k; {\rm gauge + 2H})=\frac{\alpha}{4 \pi \swd}
\Bigg \{ \frac{1}{4} [A^{H^0}+A^{h^0}+(1+8\cwd)A^Z
+ A^{A^0}+10 A^W+2A^{H^+}]
\nonumber \\
&& + \cos^2(\beta-\alpha)
[\mwd B_0^{H^0W}-B_{22}^{H^0W}-B_{22}^{h^0H^+} ]
+\sin^2(\beta-\alpha)[\mwd B_0^{h^0W}-B_{22}^{h^0W}-B_{22}^{H^0H^+}]
\nonumber \\
&&-B_{22}^{A^0H^+}-(1+8\cwd)B_{22}^{ZW}-8\swd B_{22}^{\gamma W}
+(\mzd -3\mwd -4 \cwd k^2)B_0^{ZW}
\nonumber \\
&& -4\swd k^2 B_0^{\gamma W}-\, \frac{2 k^2}{3}\Bigg \}
~~, \label{WW-gauge-H}
\\
&& \Sigma_{WW}(k; f)= \frac{\alpha}{4 \pi\swd}
\sum_{f_{\rm boublet}} N_c^f  \Big \{
2 B_{22}^{du}-\, \frac{A^u+A^d}{2}+
\frac{(k^2- m_d^2-m_u^2)}{2}\, B_0^{du}\Big \} ~~ , \label{WW-f}
\\
&& \Sigma_{WW}(k; \tilde f)= -\, \frac{\alpha}{2 \pi\swd}
\sum_{ f_{\rm boublet}} N_c^f  \Big \{
\csu^2 \csd^2 B_{22}^{\tilde d_1 \tilde u_1}
+\csu^2 \ssd^2 B_{22}^{\tilde d_2 \tilde u_1}
+ \ssu^2 \csd^2 B_{22}^{\tilde d_1 \tilde u_2}
\nonumber \\
&& +\ssu^2 \ssd^2 B_{22}^{\tilde d_2 \tilde u_2}
-\, \frac{1}{4}[\csu^2 A^{\tilde u_1}+\ssu^2 A^{\tilde u_2}
+\csd^2 A^{\tilde d_1}+\ssd^2 A^{\tilde d_2}]
\Big \} ~~ , \label{WW-sfermion}
\eqa
where the summation in (\ref{WW-f}, \ref{WW-sfermion}) is over the
fermion doublets $f_{\rm doublet}=(u,~d)$ with color factor
$N_c^f$. Finally the chargino-neutralino loop gives
\bqa
&& \Sigma_{WW}(k;~ \tchi_i^+,\, \tchi_j^0)=-\,
\frac{\alpha}{2\pi \swd} \sum_{i,j}\Big \{
\Big (O_{ij}^L O_{ij}^{L*}+O_{ij}^R O_{ij}^{R*}\Big )
\Big [-2 B_{22}^{\tchi_i^+\tchi_j^0} +\frac{1}{2}(A^{\tchi_i^+}
+A^{\tchi_j^0} ) \nonumber \\
&& -\frac{1}{2}(k^2-M_{\tchi_i^+}^2
-M_{\tchi_j^0}^2 )B_0^{\tchi_i^+\tchi_j^0} \Big ]
-\Big (O_{ij}^L O_{ij}^{R*}+O_{ij}^{L*} O_{ij}^R \Big )
M_{\tchi_i^+}M_{\tchi_j^0}B_0^{\tchi_i^+\tchi_j^0} \Big \}
~~. \label{WW-gaugino}
\eqa

\vspace{0.5cm}
The above expressions refer to the MSSM case. In
the SM case, one should suppress the sfermion, chargino,
neutralino, $H^+$, $A^0$ and $H^0$ contributions. The
$h^0$ contribution is then identified with the $H_{SM}$ one, provided we
put $\alpha=\beta-\pi/2$.

\vspace{0.5cm}
\noindent
\underline{{\bf C2)  Electron self-energies}}\\
The unrenormalized electron self-energy\footnote{Its
phase is related to the corresponding S-matrix
element by $S_{ee}= i \Sigma_e $.}
defined by

\bq
\Sigma_e(q) =\rlap /q ~\frac{(1-\gamma^5)}{2}
~ \Sigma_{Le}(q^2)
+\rlap /q ~\frac{(1+\gamma^5)}{2}
~\Sigma_{Re}(q^2)  ~~, \label{Sigma-e-unren}
\eq
\noindent
receives contributions from SM
(photon, $Z$ and $W$ loops)
\bqa
\Sigma_{Le}^{\rm SM }(q^2)&=& -~{\alpha\over 2\pi}~
\Big [ ~ B^{(e\gamma)}_1(q^2)+
{(2s^2_W-1)^2\over 4s^2_Wc^2_W} B^{(eZ)}_1(q^2)+
{1\over 2s^2_W} B^{(\nu W)}_1(q^2) +\frac{1+2\cwd}{8\swd\cwd}
\Big ] ~~ ,
\nonumber \\
\Sigma_{Re}^{\rm SM }(q^2)&=& -~{\alpha\over 2\pi}~
\Big [ ~ B^{(e\gamma)}_1(q^2)+
{\swd \over \cwd } B^{(eZ)}_1(q^2) +\frac{1}{2\cwd} \Big ]
~~ , \label{Sigma-e-SM}
\eqa
and  from the MSSM chargino and neutralino loops
\bqa
 \Sigma_{Le}^{\tchi^\pm,\tchi^0 }(q^2)&=&
 -~ \frac{\alpha}{4\pi \swd } \Big [
~ {1\over 2 \cwd }\sum_{j=1}^4 |Z^N_{1j}s_W+Z^N_{2j}c_W|^2
B^{(\tchi^0_j \tilde e_L }_1(q^2)
\nonumber \\
&& + \sum_{j=1}^2|Z^+_{1j}|^2 B^{(\tchi_j\tilde \nu_L)}_1(q^2)
\Big ] ~~ , \nonumber \\
\Sigma_{Re}^{\tchi^0 }(q^2)&=& -~
\frac{\alpha}{2\pi  c^2_W} \sum_{j=1}^4 |Z^N_{1j}|^2
B^{(\tchi^0_j \tilde e_R)}_1(q^2)~ ~.
\label{Sigma-e-gaugino}
\eqa
The sum of (\ref{Sigma-e-SM}) and (\ref{Sigma-e-gaugino})
gives of course the total contribution to the electron
self energy at the 1-loop level.\par

The electron renormalization constants are given by
\bq
\delta Z_{Le}\equiv Z_{Le}-1=-\Sigma_{Le}(0) ~~~,~~~
\delta Z_{Re}\equiv Z_{Re}- 1 =-\Sigma_{Re}(0) ~~~, \label{ZLRe}
\eq
\noindent
and the renormalized electron self-energies are
as:
\bqa
\hat\Sigma_{Le}(q^2)& = &\Sigma_{Le}(q^2)+\delta Z_{Le}~~,
\nonumber \\
\hat\Sigma_{Re}(q^2)& = & \Sigma_{Re}(q^2)+\delta Z_{Re}
~~ . \label{Sigma-e-LR-ren}
\eqa

All these  contributions from the electron and gauge self energies
have been included the renormalized Born contributions
of Sect.3.2.

We may also remark that the renormalized electron self-energy
contributions
are induced by  the  $t$- and $u$-channel electron
exchanges in Fig.1c and the related counter-terms.
Since the residue of
the renormalized electron
propagator implied by (\ref{ZLRe}) is unity, the
contribution from a diagrams like in Fig.1a is cancelled
by  that induced from the electron self-energy counter term.

The terms involving the renormalized
 gauge boson self energies in Section 3.2  are generated by
the non-unit  residue of the renormalized $Z$-propagator
and the  non-vanishing $\gamma Z$-mixing at the $Z$-mass shell.
These contributions arise from diagrams  like Fig.1b and the related
gauge self-energy counter terms. Finally, the
remaining renormalization
contributions involving $\delta Z_{Le}$, $\delta Z_{Re}$
and $\delta \tilde Z_2$ arise from the counter terms to the $Vee$
vertices.

\vspace{0.5cm}
\noindent
\underline{Asymptotic expression of the internal electron
self-energies}\\
For $x\equiv t,~u $ much larger than all masses $M$ in the loop,
electron renormalized self-energies behave like
\bqa
&&  \Sigma_{Le}(x)~\to~ -\, {\alpha\over 4\pi}~\Big [
\Big ({1+2c^2_W \over4s^2_Wc^2_W}\Big )_{\rm SM}
+ \Big ({1+2c^2_W \over4s^2_Wc^2_W}\Big )_{\rm MSSM}\Big ]
~\ln \frac{|x|}{M^2} ~~ , \\
&&   \Sigma_{Re}(x)~\to~ -\, {\alpha\over 4\pi}~\Big [
\Big ({1\over c^2_W}\Big )_{\rm SM}+
\Big ({1\over c^2_W}\Big )_{\rm MSSM}
\Big ] \ln \frac{|x|}{M^2} ~~ .
\eqa
These expressions will be useful for calculating the asymptotic
expressions for the amplitudes $e^-e^+ \to V V'$.


\vspace{2.cm}
\renewcommand{\theequation}{D.\arabic{equation}}
\renewcommand{\thesection}{D.\arabic{section}}
\setcounter{equation}{0}
\setcounter{section}{0}

{\large \bf Appendix D: Details of triangle contributions.}\\

We express here, in terms of
Passarino-Veltman functions,  the
triangle contributions to the quantities  defined in Sect.3.3.
The expressions are labeled by referring to the particles
$(abc)$ running inside the loops in Fig.1d,e,f.
Through an arrow ($\to$) we also indicate the
leading logarithmic terms arising in the asymptotic regime,
together with the
divergent part $\Delta=\frac{1}{\epsilon} -\gamma +\ln(4\pi)$.

\vspace{0.5cm}
\noindent
$\bullet$~ \underline{$e^-e^+\to \gamma\gamma$}\\
The contributions from Fig.1d,  where the three
particles running along
the loop are indicated as upper indices,  are:
\bqa
&&
b^L_{(\gamma~or~Z)}(t)=b^R_{(\gamma~or~Z)}(t)=
-2[t(C_{12}+C_{23}+C_{11}+C_0)+2C_{24}
-4]^{(\gamma ee~or~Zee)}\nonumber\\
&&\to -~(\Delta + \ln  {|t|\over \mzd} ) ~~, \nonumber \\
&&
b^L_{W}(t)=[-t(C_{12}+2C_{23}+2C_{11})-12C_{24}+8]^{(\nu_eWW)}
\to -~3\Delta + \ln^2{|t|\over \mwd} ~~, \nonumber \\
&&
b^L_{2\tchi}(t)=\sum_i|Z^+_{1i}|^2[t(C_{12}+C_{13})+2C_{24}
-{1\over2}-M^2_{\tchi_i}C_0]^{(\tilde \nu_e \tchi^+_i\tchi^+_i)}
 \to {1\over2}(\Delta - \ln{|t|\over M^2}) ~~ ,\nonumber \\
&&
b^L_{1\tchi}(t)=\sum_i|Z^N_{1i}s_W+Z^N_{2i}c_W|^2
[C_{24}]^{(\tchi^0_i \tilde e_L \tilde e_L)}
\to {1\over4}(\Delta - \ln{|t|\over M^2}) ~~, \nonumber \\
&&
b^R_{1\tchi}(t)=\sum_i|Z^N_{1i}|^2[C_{24}]
^{(\tchi^0_i \tilde e_R \tilde e_R)}
\to {1\over4}(\Delta - \ln{|t|\over M^2}) ~~, \nonumber \\
&&
a^L_{(\gamma~or~Z)}(t)=a^R_{(\gamma~or~Z)}(t)
=4t[C_0+C_{11}+C_{12}+C_{23}]^{(\gamma ee~or~Zee)}
 \to 4\ln{|t|\over \mzd} ~~, \nonumber \\
&&
a^L_{W}(t)=2t[C_{11}-C_{12}-2C_{23}]^{(\nu_eWW)}
\to -~\ln^2{|t|\over \mwd} ~~, \nonumber \\
&&
a^L_{2\tchi}(t) = -2t\sum_i|Z^+_{1i}||^2[C_{12}+C_{23}]
^{(\tilde \nu_e \tchi^+_i\tchi^+_i)} \to ~0 ~~, \nonumber \\
&&
a^L_{1\tchi}(t)=t\sum_i|Z^N_{1i}s_W+Z^N_{2i}c_W|^2[C_{12}+C_{23}]
^{(\tchi^0_i\tilde e_L\tilde e_L)} \to ~0 ~~, \nonumber \\
&& a^R_{1\tchi}(t)=t\sum_i|Z^N_{1i}||^2[C_{12}+C_{23}]
^{(\tchi^0_i\tilde e_R \tilde e_R)} \to ~0 ~. \label{fig1d-gg}
\eqa
The only other triangular  contribution, arising
from Fig.1e and involving  the  $(\nu_eWW)$ string and
 the 4-leg $WW\gamma\gamma$ coupling,  is
\bq
N''^{\gamma}_1 =-
~{1\over s^2_W} [C_0+C_{11}-C_{12}]^{(\nu_eWW)}
\to - ~{2\over  s^2_W s} \ln{s\over \mwd} ~~. \label{fig1e-gg}
\eq

\vspace{0.5cm}
\noindent
$\bullet$~ \underline{$e^-e^+\to ZZ$}\\
The diagram of  Fig.1d now gives
\bqa
&&
b^L_{(\gamma~or~Z)}(t)=b^R_{(\gamma~or~Z)}(t)=
-2[t(C_{12}+C_{23}+C_{11}+C_0)+2C_{24}-4
\nonumber \\
&& +\mzd(C_{22}-C_{23}-C_{11}-C_0)]^{(\gamma ee~or~Zee)}
\to -~(\Delta + \ln {|t|\over \mzd}) ~, \nonumber \\
&&
b''^L_{W}(t)=-2[t(C_{12}+C_{23}+C_{11}+C_0)
+\mzd(C_{22}-C_{23}-C_{11}-C_0)
\nonumber\\
&& +2C_{24}-4]^{(W\nu_e\nu_e)}
~ \to  ~~-(\Delta + \ln{|t|\over \mwd}) ~~, \nonumber \\
&&
b'^L_{W}(t)=[-t(C_{12}+2C_{23}+2C_{11})
+2\mzd(C_{11}+C_{23}-C_{22})
-12C_{24}+8]^{(\nu_eWW)}
\nonumber\\
&& \to -~3\Delta + ln^2{|t|\over \mwd} ~~,\nonumber \\
&&
b'^L_{2\tchi}(t)=\sum_{ij}(Z^{N*}_{1i}s_W+Z^{N*}_{2i}c_W)
(Z^{N}_{4i}Z^{N*}_{4j}-Z^{N}_{3i}Z^{N*}_{3j})(Z^N_{1j}s_W+Z^N_{2j}c_W)
[t(C_{12}+C_{23}) \nonumber \\
&&  +\mzd(C_{22}-C_{23})+2C_{24}-{1\over2}
+M_{\tchi^0_i}
M_{\tchi^0_j}(Z^{N*}_{4i}Z^{N}_{4j} -Z^{N*}_{3i}Z^{N}_{3j})
C_0]^{(\tilde e_L\tchi^0_i\tchi^0_j)}
 ~ \to~0 ~~, \nonumber \\
&&
 b''^L_{2\tchi}(t)= \sum_{ij}Z^{+*}_{1i}Z^{+}_{1j}~\{~
[Z^{+}_{1i}Z^{+*}_{1j}+\delta_{ij}(1-2s^2_W)][t(C_{12}+C_{23})
+\mzd(C_{22}-C_{23})+\nonumber\\
&& +2C_{24}-{1\over2} ] -
[Z^{-*}_{1i}Z^{-}_{1j}+\delta_{ij}(1-2s^2_W)]M_{\tchi^+_i}
M_{\tchi^+_j}C_0~\}^{(\tilde \nu_e \tchi^+_i\tchi^+_j)}
~ \to~c^2_W(\Delta- \ln{|t|\over M^2})~~, \nonumber  \\
&&
b'^L_{1\tchi}(t)=
\sum_{i}|Z^+_{1i}|^2[C_{24}]^{(\tchi^+_i\tilde \nu_e
\tilde \nu_e )} \to {1\over4}(\Delta - \ln{|t|\over M^2})
~~, \nonumber \\
&&
b''^L_{1\tchi}(t)=(1-2s^2_W)\sum_i|Z^N_{1i}s_W+Z^N_{2i}c_W|^2[C_{24}]
^{(\tchi^0_i\tilde e_L \tilde e_L)}
 \to {(1-2s^2_W)\over4}(\Delta - \ln{|t|\over M^2})
 ~~, \nonumber  \\
&&
b''^R_{1\tchi}(t)=-8s^4_W\sum_i|Z^N_{1i}|^2
[C_{24}]^{(\tchi^0_i\tilde e_R \tilde e_R)}
~ \to -~ 2s^4_W(\Delta - \ln{|t|\over M^2}) ~, \nonumber \\
&&
b'^R_{2\tchi}(t)=-4s^2_W~\sum_{ij}Z^{N}_{1i}Z^{N*}_{1j}~
[Z^{N*}_{4i}Z^{N}_{4j}-Z^{N*}_{3i}Z^{N}_{3j}]~[~t(C_{12}+C_{23})
+\mzd(C_{22}-C_{23})\nonumber\\
&& +2C_{24}-{1\over2}+M_{\tchi^0_i}
M_{\tchi^0_j}[Z^{N}_{4i}Z^{N*}_{4j}-Z^{N}_{3i}Z^{N*}_{3j}]
C_0~]^{(\tilde e_R \tchi^0_i\tchi^0_j)}
~ \to ~~ 0 ~~, \nonumber \\
&&
a'^L_{W}(t)=2t[C_{11}-C_{12}-2C_{23}]^{\nu_eWW)}
~ \to -~\ln^2{|t|\over \mwd} ~, \nonumber  \\
&&
a''^L_{W}(t)=4t[C_0+C_{11}+C_{12}+C_{23}]^{(W\nu_e\nu_e)}
~~ \to ~~ 4\ln{|t|\over m^2_W} ~~, \nonumber \\
&&
a'^L_{2\tchi}(t)=-2t  \sum_{ij}(Z^{N*}_{1i}s_W+Z^{N*}_{2i}c_W)
(Z^{N}_{4i}Z^{N*}_{4j}-Z^{N}_{3i}Z^{N*}_{3j})
(Z^{N}_{1j}s_W+Z^{N}_{2j}c_W) \nonumber \\
&& \cdot [C_{12}+C_{23}]^{(\tilde e_L\tchi^0_i\tchi^0_j)}
 \to 0 ~, \nonumber \\
&&
a''^L_{2\tchi}(t)= -2t\sum_{ij}Z^+_{1i}Z^{+*}_{1j}
[Z^{+}_{1i}Z^{+*}_{1j}+\delta_{ij}(1-2s^2_W)][C_{12}+C_{23}]
^{(\tilde \nu_e \tchi^+_i\tchi^+_j)}
 \to 0 ~~, \nonumber  \\
&& a'^L_{1\tchi}(t)=t~\sum_{i}|Z^+_{1i}||^2[C_{12}+C_{23}]
^{(\tchi^+_i\tilde \nu_e \tilde \nu_e )}
~~\to ~~0 ~~, \nonumber \\
&&
a''^L_{1\tchi}(t)=t~\sum_i(1-2s^2_W)|Z^N_{1i}s_W+Z^N_{2i}c_W|^2
[C_{12}+C_{23}]^{(\tchi^0_i\tilde e_L \tilde e_L)}
 \to 0 ~,\nonumber  \\
&&a'^R_{2\tchi}(t) = 8s^2_Wt~\sum_{ij}(Z^{N*}_{1i}Z^{N}_{1j})
(Z^{N*}_{4i}Z^{N}_{4j}-Z^{N*}_{3i}Z^{N}_{3j})
[C_{12}+C_{23}]^{(\tilde e_R \tchi^0_i\tchi^0_j)}
~~ \to ~0 ~~ , \nonumber \\
&& a''^R_{1\tchi}(t)= -8s^4_Wt~\sum_{i}|Z^{N}_{1i}|^2[C_{12}+C_{23}]
^{(\tchi^0_i\tilde e_R \tilde e_R)}
~~ \to ~~0 ~~, \label{fig1d-ZZ}
\eqa
while  Fig.1e and the 4-leg $WWZZ$ coupling give
\bqa
&& N''^Z_1 =-~{c^2_W\over s^4_W}~
[C_0+C_{11}-C_{12}]^{(\nu_eWW)}
 ~~ \to -~{2c^2_W\over s^4_W s} \ln{s\over \mwd} ~~,\nonumber  \\
&& N''^Z_5= -~{c^2_W\over s^4_W}~
[C_{12}]^{(\nu_eWW)}
~~ \to ~ ~{c^2_W\over s^4_W s} \ln{s\over \mwd} ~~.
\label{fig1e-ZZ}
\eqa

The NAGC contribution corresponding to Fig.1f and discussed
in Sect.3.3, has been calculated in \cite{NAGCt1}; except for the
neutralino contribution to  the $ZZZ$-NAGC coupling, for which
only restricted   Z-neutralino couplings were
considered  allowing only one or at most
two  different neutralinos
to be running along the triangular loop. Following the same
formalism, we give bellow the  neutralino contribution
to the $ZZZ$-NAGC couplings for the most general
CP-conserving $Z$-neutralino
couplings of (\ref{Z-neutralino-coupling}). This is
\bqa
&& f_5^Z = \frac{e^2}{16 \pi^2 \sw^3 \cw^3}
~\frac{\mzd}{(s-\mzd)(s -4 \mzd)}~ \cdot
\sum_{j_k} \Bigg [
{\rm Re} \Big (O^{0ZL}_{j_3j_1}O^{0ZL}_{j_1j_2}O^{0ZL}_{j_2j_3}
\Big ) \nonumber \\
&& \cdot \Bigg \{- \frac{(s-\mzd)(s+2 \mzd)}{3s}
~ [B_0(\mzd; j_1, j_2)(M_{\tchi^0_{j1}}^2
+M_{\tchi^0_{j2}}^2-2 M_{\tchi^0_{j3}}^2)
\nonumber \\
&& +B_0(s; j_1, j_3)( M_{\tchi^0_{j1}}^2
+ M_{\tchi^0_{j3}}^2-2 M_{\tchi^0_{j2}}^2)
+B_0(\mzd; j_2, j_3)( M_{\tchi^0_{j2}}^2
+ M_{\tchi^0_{j3}}^2-2 M_{\tchi^0_{j1}}^2)]
\nonumber \\
&& +\frac{C_{ZZ}(s;j_1j_2j_3)}{2 s (s-4\mzd)}
\Big [ 2 M_{\tchi^0_{j1}}^2 s^3 +s^2
[ 4 ( M_{\tchi^0_{j2}}^4+\mz^4-M_{\tchi^0_{j2}}^2
\mzd)+2 M_{\tchi^0_{j1}}^2 M_{\tchi^0_{j3}}^2
\nonumber \\
&& -3 ( M_{\tchi^0_{j1}}^2
+M_{\tchi^0_{j3}}^2)(M_{\tchi^0_{j2}}^2 +\mzd)]
+4 \mzd s [M_{\tchi^0_{j1}}^4
+ M_{\tchi^0_{j3}}^4 -M_{\tchi^0_{j1}}^2 M_{\tchi^0_{j3}}^2
\nonumber \\
&& -( M_{\tchi^0_{j2}}^2-\mzd)^2 ]
 - 4 \mz^4 ( M_{\tchi^0_{j1}}^2
- M_{\tchi^0_{j3}}^2)^2 \Big ]-\frac{(s-\mzd)}{3}
\Bigg \}\nonumber \\
&& -~\frac{1}{2}
\Big [  M_{\tchi^0_{j1}} M_{\tchi^0_{j2}} {\rm Re}
\Big (O^{0ZL}_{j_3j_1}O^{0ZL}_{j_2j_3}O^{0ZL}_{j_2j_1}
\Big ) + M_{\tchi^0_{j2}} M_{\tchi^0_{j3}} {\rm Re}
\Big (O^{0ZL}_{j_1j_2}O^{0ZL}_{j_3j_1}O^{0ZL}_{j_3j_2}
\Big )\Big ] \nonumber \\
&& \cdot \Big \{ 2 [B_0(\mzd;j_1j_2)-B_0(s;j_1j_2)]
+(M_{\tchi^0_{j1}}^2
+  M_{\tchi^0_{j3}}^2-2 M_{\tchi^0_{j2}}^2+2 \mzd-s)
C_{ZZ}(s;j_1j_2j_3) \Big \}
\nonumber \\
&& -~\frac{1}{2} M_{\tchi^0_{j3}} M_{\tchi^0_{j1}} {\rm Re}
\Big (O^{0ZL}_{j_2j_3}O^{0ZL}_{j_1j_2}O^{0ZL}_{j_1j_3}
\Big )\Big \{ 2 [B_0(\mzd;j_1j_2)-B_0(s;j_1j_2)]
\nonumber \\
&& +( M_{\tchi^0_{j1}}^2
+ M_{\tchi^0_{j3}}^2-2 M_{\tchi^0_{j2}}^2- 2 \mzd)
C_{ZZ}(s;j_1j_2j_3) \Big \} \Bigg ] ~~.
\label{f5Z-neutralino}
\eqa

\vspace{0.5cm}
\noindent
$\bullet$~ \underline{$e^-e^+\to Z \gamma $}\\
The quantities corresponding to Fig.1d are the same as those
defined for the $\gamma\gamma$ and $ZZ$ final states.
We only need to add the specific contributions from Fig.1e, with
$(abc)=(\nu_eWW)$ and the 4-leg $WW\gamma Z$ coupling
\bqa
&& N''^{\gamma Z}_1= -~{c_W\over s^3_W}
[C_0+C_{11}-C_{12}]^{(\nu_eWW)}
~~ \to -~ {2 c_W\over s^3_W s} \ln{s\over \mwd} ~~, \nonumber \\
&&N''^{\gamma Z}_5=-~{c_W\over s^3_W}
[C_{12}]^{(\nu_eWW)} ~~ \to ~ {c_W\over s^3_W s}
\ln{s\over \mwd} ~~ . \label{fig1e-Zg}
\eqa


\vspace{2.cm}
\renewcommand{\theequation}{E.\arabic{equation}}
\renewcommand{\thesection}{E.\arabic{section}}
\setcounter{equation}{0}
\setcounter{section}{0}

{\large \bf Appendix E: {Asymptotic renormalized Born and
triangle contributions.}}\\

We list here the single and double logarithmic
leading contribution for $N^{ren+Born}_j$ and $N^{Tri}_j$
entering (\ref{Nj}), when $(s,~ t, ~u) $ are  much
larger than the internal propagator  masses.
The Born terms $N^{Born, L}_j$ and $N^{Born, R}_j$ appearing below
have already been defined in Sec.3.1.

\vspace{0.5cm}
\noindent
{\bf Asymptotic contributions to $N^{ren+Born}_j$ for
$(e^-e^+ \to \gamma \gamma, ~ZZ, ~Z\gamma)$. }
\bqa
&& N^{ren+Born}_j~ \to ~ {\alpha\over4\pi} ~N^{Born,L}_j
\Big [ \ln{s\over M^2_{\gamma}}
+{(2s^2_W-1)^2\over4s^2_Wc^2_W} \ln{s\over \mzd}
+{1\over2s^2_W} \ln{s\over \mwd} \nonumber \\
&&+
\Big ({1\over4s^2_Wc^2_W}+{1\over2s^2_W} \Big )
\ln{s\over M^2_{MSSM}} \Big ] P_L \nonumber\\
&&+~{\alpha\over4\pi} N^{Born,R}_j \Big [
\ln{s\over M^2_{\gamma}}+{s^2_W\over c^2_W} \ln{s\over \mzd}
+~{1\over c^2_W} \ln{s\over M^2_{MSSM}} \Big ]P_R ~~.
\label{N-ren-Born-leading}
\eqa

\vspace{0.5cm}
\noindent
{\bf Asymptotic contributions to  $N^{Tri}_j$.}

\vspace{0.2cm}
\noindent
$\bullet$~  \underline{$e^-e^+\to\gamma\gamma$}\\
The  leading-log contributions to the  triangle amplitudes are:
\bqa
&&
N^{Tri}_1=N^{Tri}_2~\to~ \alpha^2
\Bigg \{ \Big ({1\over t}+{1\over u}\Big )\Big (2
\Big[\ln{s\over M^2_{\gamma}} \Big ]
(P_L+P_R) \nonumber \\
&& + {(2s^2_W-1)^2\over 2s^2_Wc^2_W}\Big [\ln{s\over \mzd}\Big ]P_L
 +~2{s^2_W\over c^2_W}\Big [\ln{s\over \mzd}\Big ]P_R \Big )
\nonumber \\
&& +\Big ({1\over s^2_W}\Big [{1\over2t}\ln^2{|t|\over \mwd}
+{1\over2u}\ln^2{|u|\over \mwd}\Big ]
-{2\over s^2_W s}\Big [\ln{s\over \mwd}\Big ]\Big )P_L
\nonumber\\
&&+\Big( { (1+2c^2_W) \over 2s^2_Wc^2_W}
\Big [\ln{s\over M^2_{MSSM}}\Big ]
P_L+{2\over c^2_W}\Big [\ln{s\over M^2_{MSSM}}\Big ]P_R \Big )
\Big ({1\over t}+{1\over u} \Big ) \Bigg \} ~~,
\label{N-tri-leading-gg1} \\
&&
N^{Tri}_4~\to~ \alpha^2
\Bigg \{ \Big ({1\over t}-{1\over u}\Big )
\Big (-~2\Big [\ln{s\over M^2_{\gamma}}\Big ](P_L+P_R)
\nonumber \\
&& -~{(2s^2_W-1)^2\over 2s^2_Wc^2_W}\Big [\ln{s\over \mzd}\Big ]P_L
-~2{s^2_W\over c^2_W}\Big [\ln{s\over \mzd}\Big ]P_R \Big )
\nonumber\\
&&+ {1\over s^2_W}\Big [{1\over t}\ln^2{|t|\over \mwd}
-{1\over u}\ln^2{|u|\over \mwd}\Big ] P_L
\nonumber\\
&&+\Big ( {1+2c^2_W\over 2s^2_Wc^2_W}\Big [\ln{s\over M^2_{MSSM}}\Big ]
P_L+{2\over c^2_W}\Big [\ln{s\over M^2_{MSSM}}\Big ]P_R \Big)
\Big ({1\over t}-{1\over u}\Big )\Bigg \} ~~,
\label{N-tri-leading-gg4}
\eqa
where the four lines in each equation  are induced by
 the photon (with ultraviolet mass
$M_{\gamma}$),  $Z$, $W$ and the $MSSM$ sectors, respectively.

\vspace{0.5cm}
\noindent
$\bullet$ ~ \underline{$e^-e^+\to Z \gamma $}\\
The triangle amplitudes receiving leading-log contributions are:
\bqa
&&
N^{Tri}_1~\to~ \alpha^2
\Bigg \{\Big ({1\over t}+{1\over u}\Big )
\Big ( -~{(2s^2_W-1)\over s_Wc_W}
\Big [\ln{s\over M^2_{\gamma}}\Big ]P_L
-~2{s_W\over c_W}\Big [\ln{s\over M^2_{\gamma}}\Big ]P_R
\nonumber\\
&&-~ {(2s^2_W-1)^3\over 4s^3_Wc^3_W}\Big [\ln{s\over \mzd}\Big ]P_L
-~2{s^3_W\over c^3_W}\Big [\ln{s\over \mzd}\Big ]P_R
-~{3\over 4s^3_Wc_W}\Big [\ln{s\over \mwd}\Big ]P_L \Big )\nonumber\\
&&-\Big [~{2s^2_W-1\over 4s^3_Wc_W}~ \Big ({1\over t}
\ln^2{|t|\over \mwd}+~{1\over u} \ln^2{|u|\over \mwd} \Big )
+{2c_W\over s^3_W s} \ln{s\over \mwd} \Big ]P_L
\nonumber\\
&&+\Big ({(1-2s^2_W)(1+2c^2_W)\over 4s^3_Wc^3_W}~
\Big [\ln{s\over M^2_{MSSM}}\Big ]
P_L-{2s_W\over c^3_W}\Big [\ln{s\over M^2_{MSSM}}\Big ]P_R \Big  )
\Big ({1\over t}+{1\over u}\Big ) \Bigg \} ~~,
\label{N-tri-leading-Zg1}
\\[0.4cm]
&&
N^{Tri}_2~\to~ \alpha^2
\Bigg \{ \Big ({1\over t}+{1\over u}\Big )
\Big ( -~{(2s^2_W-1)\over s_Wc_W}
\Big [\ln{s\over M^2_{\gamma}}\Big ]P_L
-~2{s_W\over c_W}\Big [\ln{s\over M^2_{\gamma}}\Big ]P_R
\nonumber\\
&&-~{(2s^2_W-1)^3\over 4s^3_Wc^3_W}\Big [\ln{s\over \mzd}\Big ]P_L
-~2{s^3_W\over c^3_W} \Big [\ln{s\over \mzd}\Big ]P_R
+{1\over4s^3_Wc_W}\Big [\ln{s\over \mwd}\Big ]P_L \Big )\nonumber\\
&&+\Big [\Big ({c_W\over 2s^3_W})~\Big({1\over t}\ln^2{|t|\over \mwd}
+~{1\over u}\ln^2{|u|\over \mwd}\Big )
-{2c_W\over s^3_W s } \ln{s\over \mwd} \Big ]P_L
\nonumber\\
&&+\Big ({(1-2s^2_W)(1+2c^2_W)\over4s^3_Wc^3_W}
\Big [\ln{s\over M^2_{MSSM}}\Big ]
P_L-{2s_W\over c^3_W}\Big [\ln{s\over M^2_{MSSM}}\Big ]P_R \Big )
\Big ({1\over t}+{1\over u}\Big ) \Bigg\} ~~,
\label{N-tri-leading-Zg2}
\\[0.4cm]
&&
N^{Tri}_4~\to~ \alpha^2
\Bigg \{\Big ({1\over t}-{1\over u} \Big )
\Big ({(2s^2_W-1)\over s_Wc_W}
\Big [\ln{s\over M^2_{\gamma}}\Big ]P_L
+~2{s_W\over c_W}\Big [\ln{s\over M^2_{\gamma}}\Big ]P_R
\nonumber\\
&&+~ {(2s^2_W-1)^3\over 4s^3_Wc^3_W}\Big [\ln{s\over \mzd}\Big ]P_L
+~2{s^3_W\over c^3_W}\Big [\ln{s\over \mzd}\Big ]P_R
+{1\over4s^3_Wc_W}\Big [\ln{s\over \mwd}\Big ]P_L \Big )\nonumber\\
&&+~{(1-4c^2_W)\over 4s^3_Wc_W}\Big (-{1\over t}\ln^2{|t|\over \mwd}
+{1\over u} ln^2{|u|\over \mwd}\Big )P_L \nonumber\\
&&
+\Big ({(1-2s^2_W)(1+2c^2_W)\over4s^3_Wc^3_W}
\Big [\ln{s\over M^2_{MSSM}}\Big ]P_L
-{2s_W\over c^3_W}\Big [\ln{s\over M^2_{MSSM}}\Big ]P_R \Big)
\Big ({1\over t}-{1\over u} \Big )\Bigg \} ~~,
\label{N-tri-leading-Zg4}
\\[0.4cm]
&&
N^{Tri}_5~\to~ \alpha^2 \Bigg \{ {1\over u}
\Big \{{(2s^2_W-1)\over s_Wc_W}
\Big [\ln{s\over M^2_{\gamma}}\Big ]P_L
+ ~2{s_W\over c_W}\Big [\ln{s\over M^2_{\gamma}}\Big ]P_R
\nonumber\\
&&+~ {(2s^2_W-1)^3\over 4s^3_Wc^3_W}\Big [\ln{s\over \mzd}\Big ]P_L
+~2{s^3_W\over c^3_W}\Big [\ln{s\over \mzd}\Big ]P_R\nonumber\\
&&+ \Big ({c_W\over2s^3_W}~\ln^2{|u|\over \mwd}
+~{1\over s^3_Wc_W}~\ln{s\over \mwd}\Big )P_L \Big \}
+{c_W\over s^3_W s}\Big [\ln{s\over \mwd}\Big ] P_L
\nonumber\\
&&+{(u-s)\over us}\Big ({(4c^2_W-1) \over 4s^3_Wc_W}
\ln^2{|u|\over \mwd}
+{1\over4s^3_Wc_W} \ln{s\over \mwd} \Big )P_L\nonumber\\
&&-~{1\over s}\Big ({(4c^2_W-1)\over 4s^3_Wc_W}ln^2{|t|\over \mwd}
+{1\over4s^3_Wc_W} \ln{s\over \mwd} \Big )P_L\nonumber\\
&&-~{1\over u}\Big ({(1-2s^2_W)(1+2c^2_W)\over4s^3_Wc^3_W}
\Big [ \ln{s\over M^2_{MSSM}}\Big ]
P_L-{2s_W\over c^3_W}\Big [\ln{s\over M^2_{MSSM}}\Big ]P_R \Big)
\Bigg \} ~~,
\label{N-tri-leading-Zg5}
\\[0.4cm]
&&
N^{Tri}_6~\to ~ -   {2 \alpha^2 \over s}
\Bigg \{\Big ({1\over t}+{1\over u}\Big )
\Big ( {(2s^2_W-1)\over s_Wc_W}
\Big [\ln{s\over M^2_{\gamma}}\Big ]P_L
+~2{s_W\over c_W}\Big [\ln{s\over M^2_{\gamma}}\Big ]P_R
\nonumber\\
&&+~{(2s^2_W-1)^3\over 4 s^3_Wc^3_W}\Big [\ln{s\over \mzd}\Big ]P_L
+~2{s^3_W\over c^3_W}\Big [\ln{s\over \mzd}\Big ]P_R
+{1\over4s^3_Wc_W}\Big [\ln{s\over \mwd}\Big ]P_L \Big)
\nonumber\\
&&-~ {(1-4c^2_W)\over 4s^3_Wc_W}\Big [{1\over t} ln^2{|t|\over \mwd}
+{1\over u} ln^2{|u|\over \mwd}\Big ]P_L
\nonumber\\
&&+\Big ({(1-2s^2_W)(1+2c^2_W)\over4s^3_Wc^3_W}
\Big [\ln{s\over M^2_{MSSM}}\Big ]P_L
-{2s_W\over c^3_W}\Big [\ln{s\over M^2_{MSSM}}\Big ]P_R\Big )
\Big ({1\over t}+{1\over u}\Big ) \Bigg \} ~~ .
\label{N-tri-leading-Zg6}
\eqa

\vspace{0.5cm}
\noindent
$\bullet$ ~ \underline{$e^-e^+\to ZZ$}\\
The triangle amplitudes   receiving leading-log contributions are:
\bqa
&&
N^{Tri}_1=N^{Tri}_2~\to~\alpha^2
\Bigg \{\Big ({1\over t}+{1\over u}\Big )
\Big ({(2s^2_W-1)^2\over 2s^2_Wc^2_W}
\Big [\ln{s\over M^2_{\gamma}}\Big ]P_L
+~2{s^2_W\over c^2_W}\Big [\ln{s\over M^2_{\gamma}}\Big ]P_R
\nonumber\\
&&+~{(2s^2_W-1)^4\over 8s^4_Wc^4_W}\Big [\ln{s\over \mzd}\Big ]P_L
+~2{s^4_W\over c^4_W}\Big [\ln{s\over \mzd}\Big ]P_R
+{2s^2_W-1\over 4s^4_W c^2_W}
\Big [\ln{s\over \mwd}\Big ]P_L \Big )
\nonumber\\
&&-~{(2s^2_W-1)\over 2s^4_W}\Big[{1\over2t}ln^2{|t|\over \mwd}
+{1\over2u}ln^2{|u|\over \mwd}\Big ] P_L
-{2c^2_W\over s^4_W s} \Big [ \ln{s\over \mwd} \Big ] P_L
\nonumber\\
&&+\Big ({(2s^2_W-1)^2(1+2c^2_W)\over8s^4_Wc^4_W}
\Big [\ln{s\over M^2_{MSSM}}\Big ]P_L
+{2s^2_W\over c^4_W}\Big [\ln{s\over M^2_{MSSM}}\Big ]P_R \Big )
\Big ({1\over t}+{1\over u}\Big ) \Bigg \} ~~,
\label{N-tri-leading-ZZ1}
\\[0.4cm]
&&
N^{Tri}_4~\to~ \alpha^2
\Bigg \{ \Big ({1\over t}-{1\over u}\Big )
\Big (-~ {(2s^2_W-1)^2\over 2s^2_Wc^2_W}
\Big [\ln{s\over M^2_{\gamma}}\Big ]P_L
-~2{s^2_W\over c^2_W}\Big [\ln{s\over M^2_{\gamma}}\Big ]P_R
\nonumber\\
&&-~{(2s^2_W-1)^4\over 8 s^4_Wc^4_W}\Big [\ln{s\over \mzd}\Big ]P_L
-~2{s^4_W\over c^4_W}\Big [\ln{s\over \mzd}\Big ]P_R
-{(2s^2_W-1)\over 4s^4_W c^2_W}
\Big [ \ln{s\over \mwd}\Big ]P_L \Big )
\nonumber\\
&&-~{(2s^2_W-1)\over 2s^4_W}\Big [{1\over t}ln^2{|t|\over \mwd}
-{1\over u}\ln^2{u\over \mwd} \Big ]P_L
\nonumber\\
&&+\Big ({(2s^2_W-1)^2(1+2c^2_W)\over8s^4_Wc^4_W}
\Big [\ln{s\over M^2_{MSSM}}\Big ]P_L
+{2s^2_W\over c^4_W}\Big [\ln{s\over M^2_{MSSM}}\Big ]P_R \Big )
\Big ({1\over t}-{1\over u}\Big ) \Bigg \} ~~,
\label{N-tri-leading-ZZ4}
\\[0.4cm]
&&
N^{Tri}_5~\to~ \alpha^2 \Bigg \{-~{1\over u}
~\Big ( {(2s^2_W-1)^2\over 2s^2_Wc^2_W}
\Big [\ln{s\over M^2_{\gamma}}\Big ]P_L
+~{ 2 s^2_W\over c^2_W}\Big [\ln{s\over M^2_{\gamma}}\Big ]P_R
\nonumber\\
&&+~{(2s^2_W-1)^4\over 8 s^4_Wc^4_W}\Big [\ln{s\over \mzd}\Big ]P_L
+~{2 s^4_W\over c^4_W}\Big [\ln{s\over \mzd}\Big ]P_R \Big )
\nonumber\\
&&+\Big ({(u-s)\over us}\Big [-~{(2s^2_W-1)\over 2s^4_W}
\ln^2{|u| \over \mwd}\Big ] -{(2s^2_W-1) \over 4s^4_W c^2_W}
\ln{s\over \mwd}\Big )P_L
\nonumber\\
&&+~{(2s^2_W-1)\over 2s^4_W s }\Big [\ln^2{|t|\over \mwd}\Big ]P_L
-~{(2s^2_W-1)\over 4s^4_W u} \ln^2{|u|\over \mwd}P_L
+{c^2_W\over s^4_W s}\Big [\ln{s\over \mwd}\Big ]P_L
\nonumber\\
&&-~{1\over u} \Big ({(2s^2_W-1)^2(1+2c^2_W)\over8s^4_Wc^4_W}
\Big [\ln{s\over M^2_{MSSM}}\Big ]P_L
+{2s^2_W\over c^4_W}\Big [\ln{s\over M^2_{MSSM}}\Big ]P_R \Big)
\Bigg \} ~~,
\label{N-tri-leading-ZZ5}
\\[0.4cm]
&&
N^{Tri}_6=-N^{Tri}_8~\to (-~{2\alpha^2 \over s})
\Bigg \{\Big ({1\over t}+{1\over u}\Big )
\Big (-{(2s^2_W-1)^2\over 2s^2_Wc^2_W}
\Big [\ln{s\over M^2_{\gamma}}\Big ]P_L
-~{2s^2_W\over c^2_W}\Big [\ln{s\over M^2_{\gamma}}\Big ]P_R
\nonumber\\
&&-~{(2s^2_W-1)^4\over 8 s^4_Wc^4_W}\Big [\ln{s\over \mzd}\Big ]P_L
-~{2 s^4_W\over c^4_W}\Big [\ln{s\over \mzd}\Big ]P_R
-{(2s^2_W-1) \over 4s^4_W c^2_W}
\Big [\ln{s\over \mwd}\Big ]P_L \Big )
\nonumber\\
&&-~{(2s^2_W-1)\over 2s^4_W}\Big [{1\over t}\ln^2{|t|\over \mwd}
+{1\over u}\ln^2{|u|\over \mwd}\Big ]P_L
\nonumber\\
&&+\Big ({(2s^2_W-1)^2(1+2c^2_W)\over8s^4_Wc^4_W}
\Big [\ln{s\over M^2_{MSSM}}\Big ]P_L
+{2s^2_W\over c^4_W}\Big [\ln{s\over M^2_{MSSM}}\Big ]P_R \Big)
\Big ({1\over t}+{1\over u}\Big ) \Bigg \} ~~,
\label{N-tri-leading-ZZ6}
\\[0.4cm]
&&
N^{Tri}_7~\to~ \alpha^2 \Bigg \{-~{1\over t}
\Big ({(2s^2_W-1)^2\over 2s^2_Wc^2_W}
\Big [\ln{s\over M^2_{\gamma}}\Big ]P_L
+~{2 s^2_W\over c^2_W}\Big [\ln{s\over M^2_{\gamma}}\Big ]P_R
\nonumber\\
&&+~{(2s^2_W-1)^4\over 8s^4_Wc^4_W}\Big [\ln{s\over \mzd}\Big ]P_L
+~{2 s^4_W\over c^4_W}\Big [\ln{s\over \mzd}\Big ]P_R \Big)\nonumber\\
&&
+\Big ({(t-s)\over ts}\Big [-~{(2s^2_W-1)\over 2s^4_W}
\ln^2{|t|\over \mwd}
\Big ]-{(2s^2_W-1)\over 4s^4_W c^2_W t}\ln{s\over \mwd}\Big ) P_L
\nonumber\\
&&
+~{(2s^2_W-1)\over 2s^4_W s} \Big [ln^2{|u|\over \mwd}\Big ]P_L
-~{(2s^2_W-1)\over 4s^4_W t}
\ln^2{|t|\over \mwd}P_L
+{c^2_W\over s^4_W  s}\Big [\ln{s\over \mwd}\Big ]P_L
\nonumber\\
&&-~{1\over t}\Big ({(2s^2_W-1)^2(1+2c^2_W)\over8s^4_Wc^4_W}
\Big [\ln{s\over M^2_{MSSM}}\Big ]P_L
+{2s^2_W\over c^4_W}\Big [\ln{s\over M^2_{MSSM}}\Big ]P_R \Big)
~\Bigg \} ~~ . \label{N-tri-leading-ZZ7}
\eqa\\

Concerning the NAGC couplings discussed in Section 3.3, and
 calculated in  \cite{NAGCt1}
and (\ref{f5Z-neutralino}), we note that in the asymptotic regime
 they are always found to vanish; \ie
$f^{\gamma,~Z}_5\to 0 $,  ~ $h^{\gamma,~Z}_3 \to 0$.


\vspace{2.cm}
\renewcommand{\theequation}{F.\arabic{equation}}
\renewcommand{\thesection}{F.\arabic{section}}
\setcounter{equation}{0}
\setcounter{section}{0}

{\Large \bf Appendix F: Asymptotic contributions from box diagrams}\\

Leading $\ln s$ and $\ln^2s$ order contributions only arise from
 SM boxes of the types $k=1,2,3$ defined in Sec.3.4.  The
 purely MSSM boxes, which  are of types $(k=4,5,6)$,  provide only
 subleading logarithmic contributions like $\ln (s/ t)$. The SM or
MSSM contributions from the type $k=7$ box, vanish asymptotically
like $M^2/s$.

The box amplitudes receiving leading-log contributions are:

\vspace{0.5cm}
\noindent
$\bullet$ ~ \underline{$e^-e^+\to\gamma\gamma$}
\bqa
N^{Box}_j&=& \alpha^2
\Big \{~\bar{N}^{1,Box}_j(M_{\gamma})[P_L+P_R]+\bar{N}^{1,Box}_j(\mz)
\Big [{(2s^2_W-1)^2\over4s^2_Wc^2_W}P_L+{s^2_W\over c^2_W}P_R \Big ]
\nonumber\\
&&
+\bar{N}^{2,Box}_j(\mw){1\over2s^2_W}P_L +{\rm "sym"} \Big \} ~~,
\eqa

\noindent
$\bullet$ ~ \underline{$e^-e^+\to Z \gamma $}
\bqa
&& N^{Box}_j= \alpha^2 \Big \{~\Big (\bar{N}^{1,Box}_j(M_{\gamma})
\Big [{(1-2s^2_W)\over2s_Wc_W}P_L-{s_W\over c_W}P_R \Big]
+\bar{N}^{1,Box}_j(\mz)
\Big [{(1-2s^2_W)^3\over8s^3_Wc^3_W}P_L-{s^3_W\over c^3_W}P_R\Big ]
\nonumber\\
&& +\bar{N}^{2,Box}_j(\mw)
{c_W\over2s^3_W}P_L~+~{\rm "sym"~}\Big )
+\bar{N}^{3,Box}_j(\mw) ~{1\over4s^3_Wc_W}P_L~~{\rm (no~"sym") }
\Big \} ~~ ,
\eqa

\noindent
$\bullet$ ~ \underline{$e^-e^+\to ZZ$}
\bqa
&& N^{Box}_j= \alpha^2 \Big \{~\bar{N}^{1,Box}_j(M_{\gamma})
\Big [{(2s^2_W-1)^2\over4s^2_Wc^2_W}P_L+{s^2_W\over c^2_W}P_R \Big ]
+\bar{N}^{1,Box}_j(\mz)
\Big [{(2s^2_W-1)^4\over16s^4_Wc^4_W}P_L+{s^4_W\over c^4_W}P_R \Big ]
\nonumber\\
&&
+\bar{N}^{1,Box}_j(\mw)~{1\over8s^4_Wc^2_W}P_L
+\bar{N}^{2,Box}_j(\mw)~{c^2_W\over2s^4_W}P_L
+\bar{N}^{3,Box}_j(\mw)~{1\over4s^4_W}P_L
+{\rm "sym"} \Big \} ~~.
\eqa
The meaning of $+{\rm "sym"}$ in the above
equations has been given in
Section.3.4, while discussing (\ref{+sym}).

\vspace{0.5cm}
\noindent
{\bf \underline{Tables of asymptotic values for $\bar N^{k,~Box}_j$}}\\
The complete Box contributions to  $\bar N^{Box}_j$ have
been analytically calculated and used in the numerical
calculations presented in this paper. Since, these expressions are
enormous though, we refrain from giving them in the text and only
list their asymptotic leading-log contribution.
Below we write for each such form,
in a first step ($\to$) the full logarithmic expressions obtained
from the asymptotic expansion of the
Passarino-Veltman $D_{ij}$ functions;
and in a second step ($\Longrightarrow$) the leading
($\ln{s\over M^2}$, $\ln^2{|s|\over M^2}$,
$\ln^2{|t|\over M^2}$, $\ln^2{|u|\over M^2}$) terms
\cite{Denner-asym}.

\vspace{0.5cm}
\noindent
$\bullet$~ {\bf Box type $k=1$; $(abcd)=(Vfff)$}
\bqa
 \bar{N}^{1,~Box}_1 &= &\bar{N}^{1,~Box}_2 \to
{1\over tu^2}\Big [-(s^2+tu)\ln^2{|t|\over s}+u^2\ln^2{s\over M^2}
+2tu \ln{s\over M^2}
\nonumber \\
&& -(2t^2+4s^2+6st)\ln{|t|\over M^2}\Big ]
~\Longrightarrow~
~{1\over t} \Big [\ln^2{s\over M^2}-4\ln{s\over M^2} \Big ] ~~ ,
\nonumber \\
 \bar{N}^{1,~Box}_{3} &\to & -~{4\over tu^3}\Big [t(s-u)\ln^2{|t|\over s}
+(3s^2+4st+t^2)\ln{s\over |t|}\Big ]
~\Longrightarrow~0 ~~, \nonumber \\
 \bar{N}^{1,~Box}_4 & \to &
{1\over tu^2}\Big [(tu-s^2)\ln^2{|t|\over s}+u^2\ln^2{s\over \mzd}
+2tu \ln{s\over |t|}\Big ]
~\Longrightarrow~~{1\over t}
\ln^2{s\over M^2}~~, \nonumber \\
\bar{N}^{1,~Box}_5 &\to&
-~{1\over su^2}\Big [(s^2+tu)\ln^2{|t|\over s}+u^2\ln^2{s\over M^2}
-(4t^2+6s^2+10st)\ln{s\over |t|} \Big ]
\nonumber\\
&&~\Longrightarrow~~~
-~{1\over s} \ln^2{s\over M^2} ~~, \nonumber \\
\bar{N}^{1,~Box}_6&\to&{2\over stu^3}
\Big [-(s^3+t^3+2st^2)\ln^2{|t|\over s}
-u^3\ln^2{s\over M^2}\nonumber\\
&& -(12st^2+4t^3+8s^2t)\ln{s\over |t|}
   \Big ]
~\Longrightarrow~~~-~{2\over st} \ln^2{s\over M^2} ~~,
\nonumber \\
\bar{N}^{1,~Box}_7&\to&
{1\over stu}\Big [(st-s^2+t^2)\ln^2{|t|\over s}
+(s^2-t^2)\ln^2{s\over M^2}
-4tuln{s\over M^2}-4u^2\ln{|t|\over M^2}\Big ]
\nonumber\\
&&~\Longrightarrow~~~
{(t-s)\over st}
\ln^2{s\over M^2}+{4\over t}\ln{s\over M^2}
~~, \nonumber \\
\bar{N}^{1,~Box}_8&\to&
{2\over stu^2}\Big [-(t^2+s^2+3st)\ln^2{|t|\over s}
+u^2\ln^2{s\over M^2}
-(4t^2+6s^2+10st)\ln{s\over |t|}\Big ]
\nonumber \\
&&~\Longrightarrow~~~
~{2\over st}
\ln^2{s\over M^2}
~~, \nonumber \\
\bar{N}^{1,~Box}_9&\to&
~~~0 ~~. \label{Box-asym-1Vfff}
\eqa

\vspace{0.5cm}
\noindent
$\bullet$ ~ {\bf Box type $k=2$; $(abcd)=(fVVV)$}
\bqa
\bar{N}^{2,~Box}_1 &= &\bar{N}^{2,~Box}_2 \to
{1\over stu^2}\Big [-s^2(s+2t)\ln^2{|t|\over s}+su^2(\ln^2{s\over M^2}
+\ln^2{|t|\over M^2})
\nonumber\\
&&-2s(s^2+3st+2t^2)\ln{|t|\over M^2}
+2t(t^2+3st+2s^2)\ln{s\over M^2}\Big ]
\nonumber\\
&&~\Longrightarrow~~~{1\over t}
\Big (\ln^2{s\over M^2}+\ln^2{|t|\over M^2}\Big )
+2{(t-s)\over st}\ ln{s\over M^2}~~, \nonumber \\
\bar{N}^{2,~Box}_3&=&\to~~-~{4\over tu^3}
\Big [t(s+2t)\ln^2{|t|\over s}
+(3s^2+8st+5t^2)\ln{|t|\over s}\Big ]
\nonumber\\
&&~\Longrightarrow~~~0
~~, \nonumber \\
\bar{N}^{2,~Box}_4&\to&
{1\over2stu^2}\Big [(s^3+4s^2t+6st^2+4t^3)\ln^2{|t|\over s}\nonumber\\
&&
+u^2(2s+t)\ln^2{s\over M^2}+8u^2t\ln^2{|t|\over M^2}
+4ut^2\ln{s\over M^2}
+4us(s+2t)\ln{|t|\over M^2}\Big ]
\nonumber\\
&&\Longrightarrow~ {(2s+t)\over 2st}
\ln^2{s\over M^2}+{4\over s}\ln^2{|t|\over M^2}
+{2u\over st}\ln{s\over M^2}]
~~, \nonumber \\
\bar{N}^{2,~Box}_5&\to&
{1\over su^2}\Big [-s(s+2t)\ln^2{|t|\over s}
+4u^2\ln^2{|t|\over M^2}
+u(s-t)\ln{s\over M^2}+u(4s+6t)\ln{|t|\over M^2}\Big ]
\nonumber\\
&&\Longrightarrow~{4\over s}\ln^2{|t|\over M^2}
-~{5\over s}\ln{s\over M^2}
~~, \nonumber \\
\bar{N}^{2,~Box}_6&\to&
{2\over stu^3}\Big [(-3ts^2-s^3-4st^2-4t^3)\ln^2{|t|\over s}
\nonumber\\
&&
-u^3\ln^2{s\over M^2}+(6t^3+6st^2-2s^3-2ts^2)\ln{|t|\over M^2}
\nonumber\\
&&
-(4ts^2+12st^2+8t^3)\ln{s\over M^2}\Big ]
\nonumber\\
&&
\Longrightarrow~~ {2\over st}
(-ln^2{s\over M^2}+2ln{s\over M^2})
~~, \nonumber \\
\bar{N}^{2,~Box}_7&\to&
{1\over stu}\Big [-s(s+2t)\ln^2{|t|\over s}-us\ln^2{s\over M^2}
\nonumber\\
&&
+(s^2+5st+4t^2)\ln^2{|t|\over M^2}-3tu \ln{s\over M^2}
-(2s^2+t(8s+6t)\ln{|t|\over M^2}\Big ]
\nonumber\\
&&
\Longrightarrow~~ -~{1\over t}
\ln^2{s\over M^2}+{4u+3s\over st}\ln^2{|t|\over M^2}
-{3u+s\over st}\ln{s\over M^2}
~~, \nonumber \\
\bar{N}^{2,~Box}_8&\to&
{2\over stu^2}\Big [-(s+2t)^2\ln^2{|t|\over s}+u^2\ln^2{s\over M^2}
+2u(3s+4t)\ln{s\over M^2}-2u(2s+3t)\ln{|t|\over M^2} \Big ]
\nonumber\\
&&
\Longrightarrow~~
-~{2\over st} (-ln^2{s\over M^2}+2ln{s\over M^2})
~~, \nonumber \\
\bar{N}^{2,~Box}_9&\to&~~0 ~~. \label{Box-asym-2fVVV}
\eqa

\vspace{0.5cm}
\noindent
{\bf Box type $k=3$; $(abcd)=(VffV)$}
\bqa
\bar{N}^{3,~Box}_1&\to&
{1\over tu}\Big [-s \ln^2{|t|\over u}
-t \ln^2{|t|\over M^2}-u \ln^2{|u|\over M^2}
+5t \ln{|u|\over M^2}+5u \ln{|t|\over M^2}\Big ]
\nonumber\\
&&\Longrightarrow~~  -~{1\over t}
\ln^2{|u|\over M^2}-~{1\over u}
\ln^2{|t|\over M^2}-{5s\over ut}\ln{s\over M^2}
~~, \nonumber \\
\bar{N}^{3,~Box}_2&\to&
{1\over tu} \Big [-s\ln^2{|t|\over u}+s\Big (\ln^2{|u|\over M^2}
+\ln^2{|t|\over M^2}\Big )
+t \ln{|u| \over M^2}+u \ln{|t|\over M^2}\Big ]
\nonumber\\
&&\Longrightarrow~~{s\over tu}\Big [ \ln^2{u\over M^2}+
ln^2{t\over M^2}-ln{s\over M^2}\Big ]
~~, \nonumber \\
\bar{N}^{3,~Box}_3&\to&
{12\over tu}\ln{u\over t}~~~~~~~~~\Longrightarrow~~0
~~, \nonumber \\
\bar{N}^{3,~Box}_4&\to&
{1\over tu}\Big [(u-t)\ln^2{\Big |{t\over u}\Big |}
+t \ln^2{|t|\over M^2}
-u \ln^2{|u|\over M^2}+u \ln{|t|\over M^2}-t \ln{|u|\over M^2}\Big ]
\nonumber\\
&&\Longrightarrow~~ {1\over u}
\ln^2{|t|\over M^2}-~{1\over t}\ln^2{|u|\over M^2}
+{u-t\over ut} \ln{s\over M^2}
~~, \nonumber \\
\bar{N}^{3,~Box}_5&\to&
{1\over tus^3}\Big [(t^4+u^3t+3t^3u+3t^2u^2)
\ln^2\Big |{t\over u}\Big| \nonumber\\
&&
+s^2tu \ln^2{|u|\over M^2}+s^2t(s-u)\ln^2{|t|\over M^2}
+5s^2tu \ln{|t|\over M^2}+5s^2t^2 \ln{|u|\over M^2}\Big ]
\nonumber \\
&&\Longrightarrow~~{1\over s}\ln^2{|u|\over M^2}
+{s-u\over us}\ln^2{|t|\over M^2}
-~{5\over u}\ln{s\over M^2}
~~, \nonumber \\
\bar{N}^{3,~Box}_6&\to&
{2\over s^2tu}\Big [s^2\ln^2\Big |{t\over u}\Big |
+us \ln^2{|u|\over M^2}
+st \ln^2{|t|\over M^2}+(6t^2+7tu+u^2)\ln{|t|\over M^2}
\nonumber\\
&& +5ts \ln{|u|\over M^2}\Big ]
~~ \Longrightarrow~~{2\over st}
\ln^2{|u|\over M^2}+{2\over su}
\ln^2{|t|\over M^2}+{2\over tu}\ln{s\over M^2})
~~, \nonumber \\
\bar{N}^{3,~Box}_7&\to&
-~{1\over uts^2}\Big [(u^4+t^3u+3u^3t+3u^2t^2)
\ln^2\Big |{t\over u}\Big |
+su^2\ln^2{|t|\over M^2}
\nonumber\\
&& -su(u+2s)\ln^2{|u|\over \mwd}
-5stu \ln{|u|\over M^2}+u(u^2-3tu-4t^2)\ln{|t|\over M^2}\Big ]
\nonumber\\
&&\Longrightarrow~~ -~{u\over st}
\ln^2{|t|\over M^2}+{u+2s\over st}\ln^2{|u|\over M^2}
-{1\over t}\ln{s\over M^2}]
~~, \nonumber \\
\bar{N}^{3,~Box}_8&\to&
{2\over s^2tu}\Big [-s^2\ln^2\Big |{t\over u}\Big |
-us \ln^2{|u|\over M^2}-ts \ln^2{|t|\over M^2}
\nonumber\\
&& -6us \ln \Big |{t\over u}\Big |+ts \ln{|u|\over M^2}
+us \ln{|t|\over M^2}\Big]
\nonumber\\
&&\Longrightarrow~~-~{2\over st}
\ln^2{|u|\over M^2}-{2\over su}
\ln^2{|t|\over M^2}-{2\over tu}\ln{s\over M^2})
~~, \nonumber \\
\bar{N}^{3,~Box}_9&\to&
~~0 ~~.
\label{Box-asym-3VffV}
\eqa

\vspace{0.5cm}
\noindent
{\bf Box type $k=4$; $(abcd)=(Sfff)$}.\\
In sect.3.4 we have separated this box contribution into 4 parts
coming from different combinations of the kinetic and massive
parts of the three fermion propagators.
The $k=4B,~4C,~4D$ parts asymptotically vanish like $M^2/s$.
\bqa
\bar{N}^{4A,~Box}_1& = &\bar{N}^{4A,~Box}_2=\bar{N}^{4A,~Box}_4
\to
-~{1\over 2u^2}\Big [t \ln^2{|t|\over s}-2u ln{s\over |t|}\Big ]~~
~~~\Longrightarrow~0 ~~,
\nonumber \\
\bar{N}^{4A,~Box}_{3}&\to&~{2\over tu^3}\Big [t^2\ln^2{|t|\over s}
-(3t^2+4st+s^2)\ln{|t|\over s}\Big ]
~~~~~~~ \Longrightarrow~0  ~~ , \nonumber \\
\bar{N}^{4A,~Box}_5&\to&
-~{1\over 2su^2}
\Big [t^2\ln^2{|t|\over s}-2(2t^2+3st+s^2)\ln{|t|\over s}\Big ]
~~~~~~~~\Longrightarrow~~0 ~~, \nonumber \\
\bar{N}^{4A,~Box}_6&\to&
-~{1\over stu^3}\Big [t^2(s-t)\ln^2{|t|\over s}
-4ut^2\ln{|t|\over s} \Big ]
~~~~~~~~\Longrightarrow~~~0 ~~ , \nonumber \\
\bar{N}^{4A,~Box}_7&\to&-~{1\over 2su}
\Big [t\ln^2{|t|\over s}+4u\ln{|t|\over s} \Big ]
~~~~~~~\Longrightarrow~~~0 ~~ , \nonumber \\
\bar{N}^{4A,~Box}_8&\to&
~{1\over stu^2}\Big [t^2\ln^2{|t|\over s}
-2(2t^2+s^2+3st)\ln{|t|\over s}~
\Big ] ~~~~~~~~\Longrightarrow~~~0
~~ , \nonumber \\
\bar{N}^{4A,~Box}_9&\to&~~~0 ~~ . \label{Box-asym-4Sfff}
\eqa

\vspace{0.5cm}
\noindent
{\bf Box type $k=5$; $(abcd)=(fSSS)$}
\bqa
\bar{N}^{5,~Box}_1& =& \bar{N}^{5,~Box}_2=\bar{N}^{5,~Box}_4
\to
-~{1\over 2u^2}\Big [-s \ln^2{|t|\over s}+2u \ln{|t| \over s}\Big ]
~~~~~~\Longrightarrow~~~0 ~~, \nonumber \\
\bar{N}^{5,~Box}_{3}&\to&-~{2\over tu^3}\Big [ts \ln^2{|t|\over s}
+(t^2-s^2)\ln{|t|\over s} \Big ]
~~~~~~~~~\Longrightarrow~~~0
~~ , \nonumber \\
\bar{N}^{5,~Box}_5&\to&
~{1\over 2u^2}\Big [t \ln^2{|t|\over s}+2u\ln{|t|\over s} \Big ]
~~~~~~\Longrightarrow~~~0
~~ , \nonumber  \\
\bar{N}^{5,~Box}_6&\to&
-~{1\over tu^3}\Big [t(t-s)\ln^2{|t|\over s}
+4ut \ln{|t|\over s} \Big ]
~~~~~~~\Longrightarrow~~~0
~~ , \nonumber  \\
\bar{N}^{5,~Box}_7&\to&~{1\over 2u} \ln^2{|t|\over s}
~~~~~~~\Longrightarrow~~~0
~~ , \nonumber  \\
\bar{N}^{5,~Box}_8&\to&
-~{1\over tu^2}\Big [t \ln^2{|t|\over s}+2u \ln{|t|\over s}\Big ]
~~~~~~~~\Longrightarrow~~~0
~~ , \nonumber  \\
\bar{N}^{5,~Box}_9&\to&
~~~0 ~~.
\label{Box-asym-5fSSS}
\eqa

\vspace{0.5cm}
\noindent
{\bf Box type $k=6$; $(abcd)=(SffS)$}\\
In Section 3.4 we have separated the contribution of this box into
two parts: $6A$ coming from the kinetic part of the fermion propagators
and  $6B$ coming from their massive part.
The $k=6B$ part asymptotically vanish like $M^2/s$.
\bqa
\bar{N}^{6A,~Box}_1& = &\bar{N}^{6A,~Box}_2=\bar{N}^{6A,~Box}_4
\to  ~~ 0 ~~, \nonumber \\
\bar{N}^{6A,~Box}_{3}&\to&-~{2\over tu}
\ln\Big |{t\over u}\Big | ~~~~~~~\Longrightarrow~ 0 ~~ , \\
\bar{N}^{6A,~Box}_5&\to&
~{1\over s} \ln\Big |{t\over u}\Big |
~~~~~~~\Longrightarrow~0 ~~ , \nonumber \\
\bar{N}^{6A,~Box}_6&\to&
-~{2\over stu} \Big [t \ln\Big |{t\over u}\Big | \Big ]
~~~~~~~\Longrightarrow~0 ~~ , \nonumber \\
\bar{N}^{6A,~Box}_7&\to&-~{1\over s} \ln\Big |{t\over u}\Big |
~~~~~~~\Longrightarrow~0 ~~ , \nonumber \\
\bar{N}^{6A,~Box}_8&\to&
-~{2\over stu}\Big [u \ln\Big |{t\over u}\Big | \Big ]
~~~~~~~\Longrightarrow~0 ~~ , \nonumber \\
\bar{N}^{6A,~Box}_9&\to&
~~0 ~~ .
\label{Box-asym-6BSffS}
\eqa

\clearpage
\newpage

\begin{figure}[th]
\[
\epsfig{file=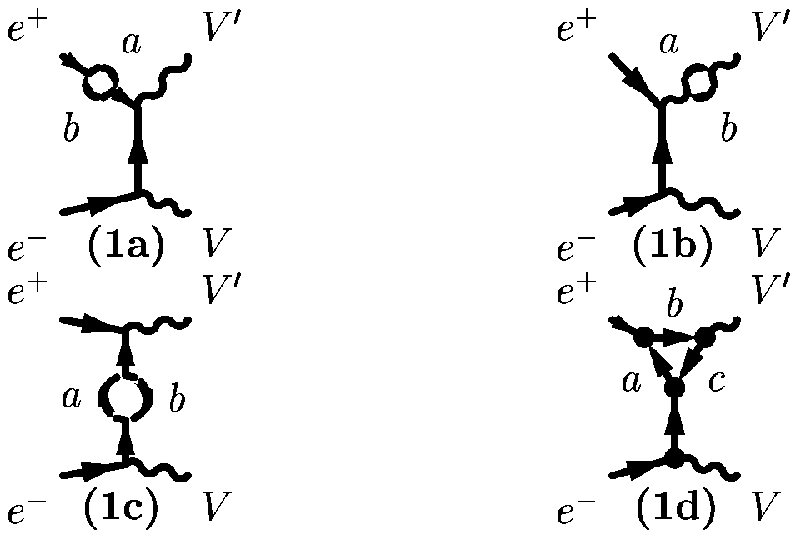,height=8.5cm,width=13cm}
\]
\[
\epsfig{file=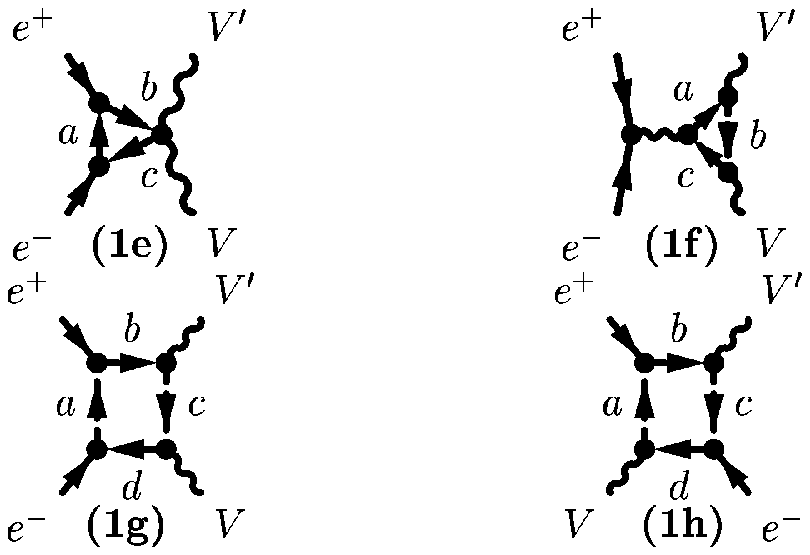,height=8.5cm,width=13cm}
\]
\caption[1]{Diagrams at one loop}
\end{figure}

\newpage

\begin{figure}[p]
\vspace*{-3cm}
\[
\hspace{-0.5cm}\epsfig{file=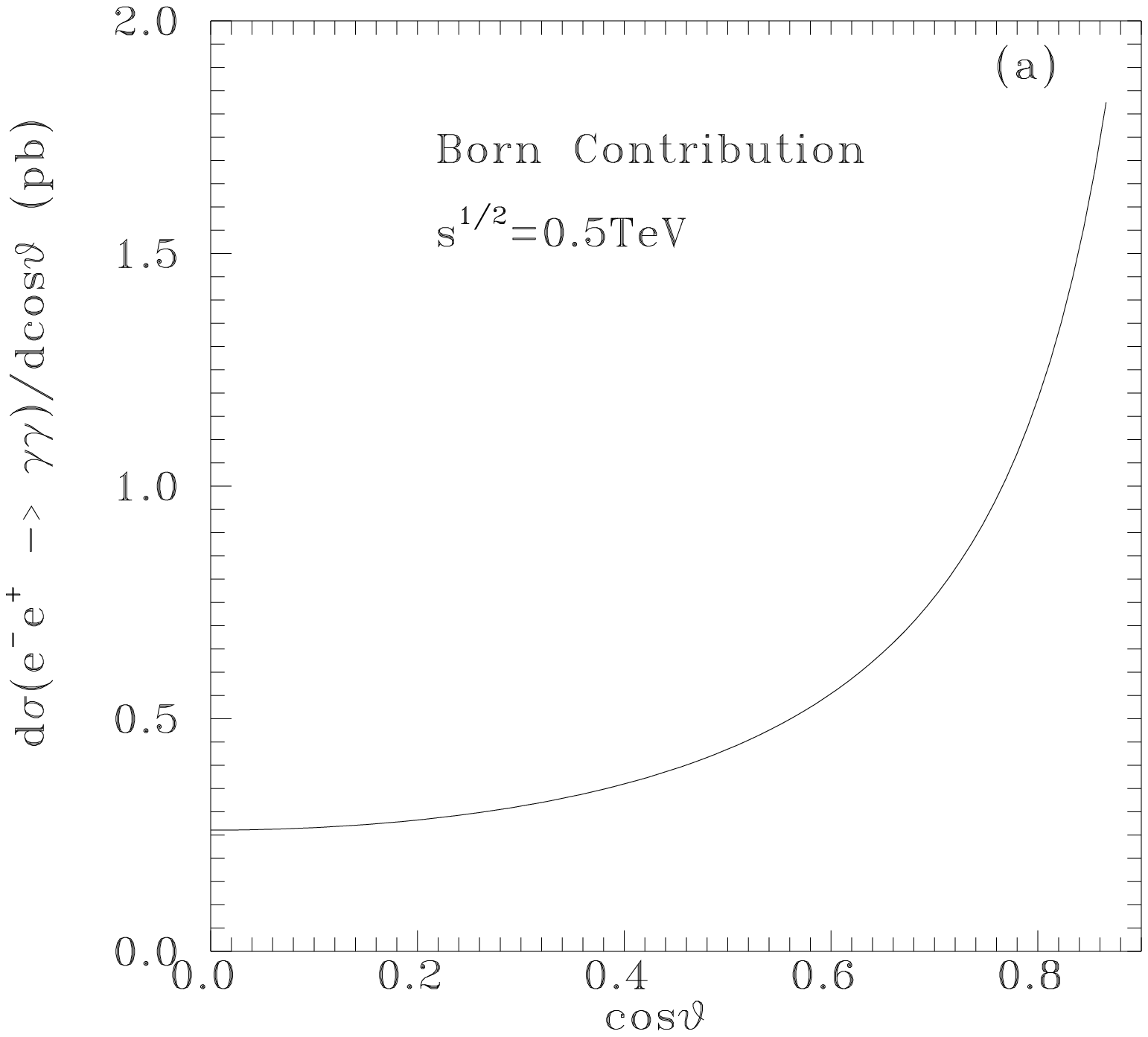,height=7cm}\hspace{0.5cm}
\epsfig{file=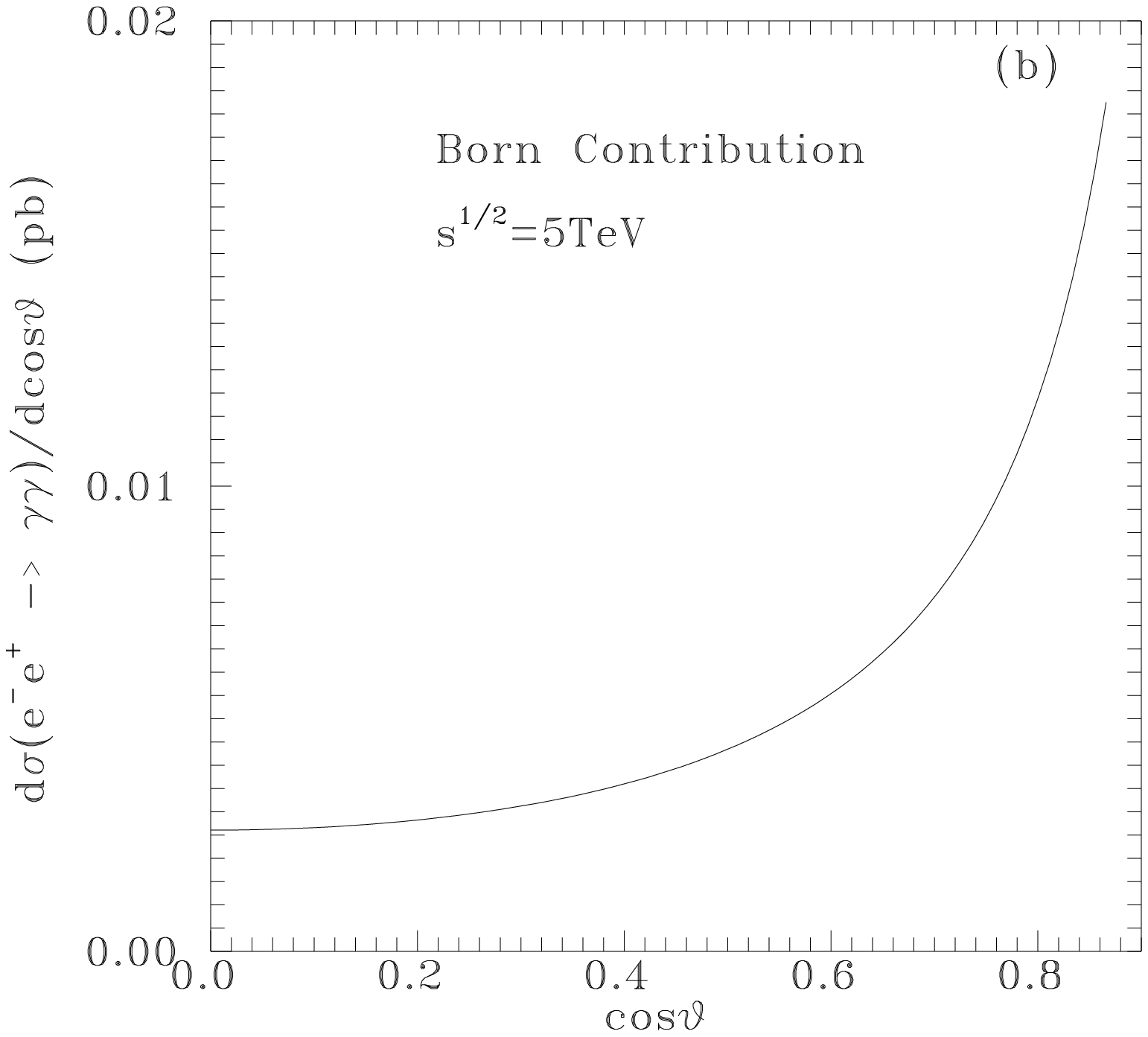,height=7cm}
\]
\vspace*{0.5cm}
\[
\hspace{-0.5cm}\epsfig{file=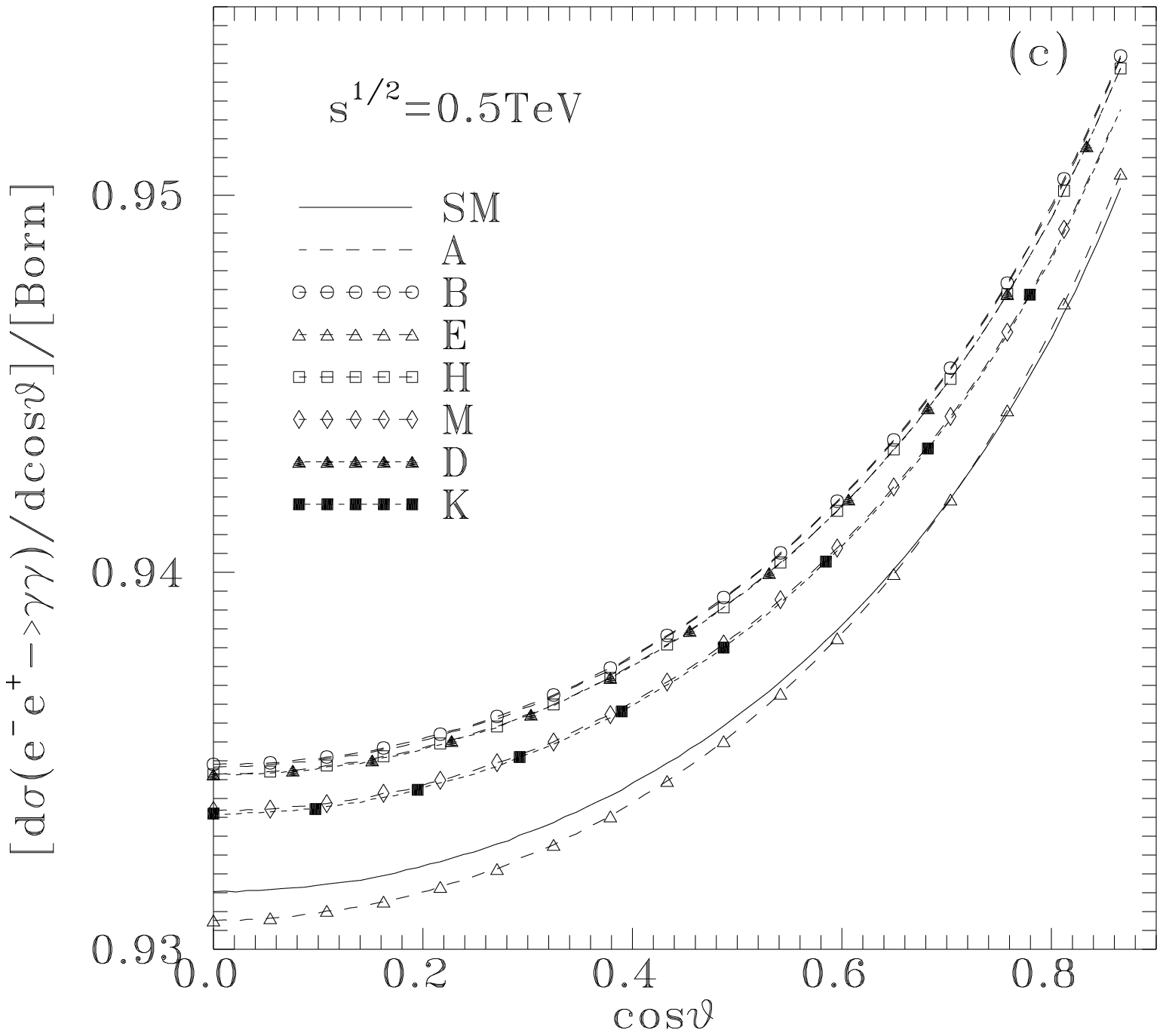,height=7cm}\hspace{0.5cm}
\epsfig{file=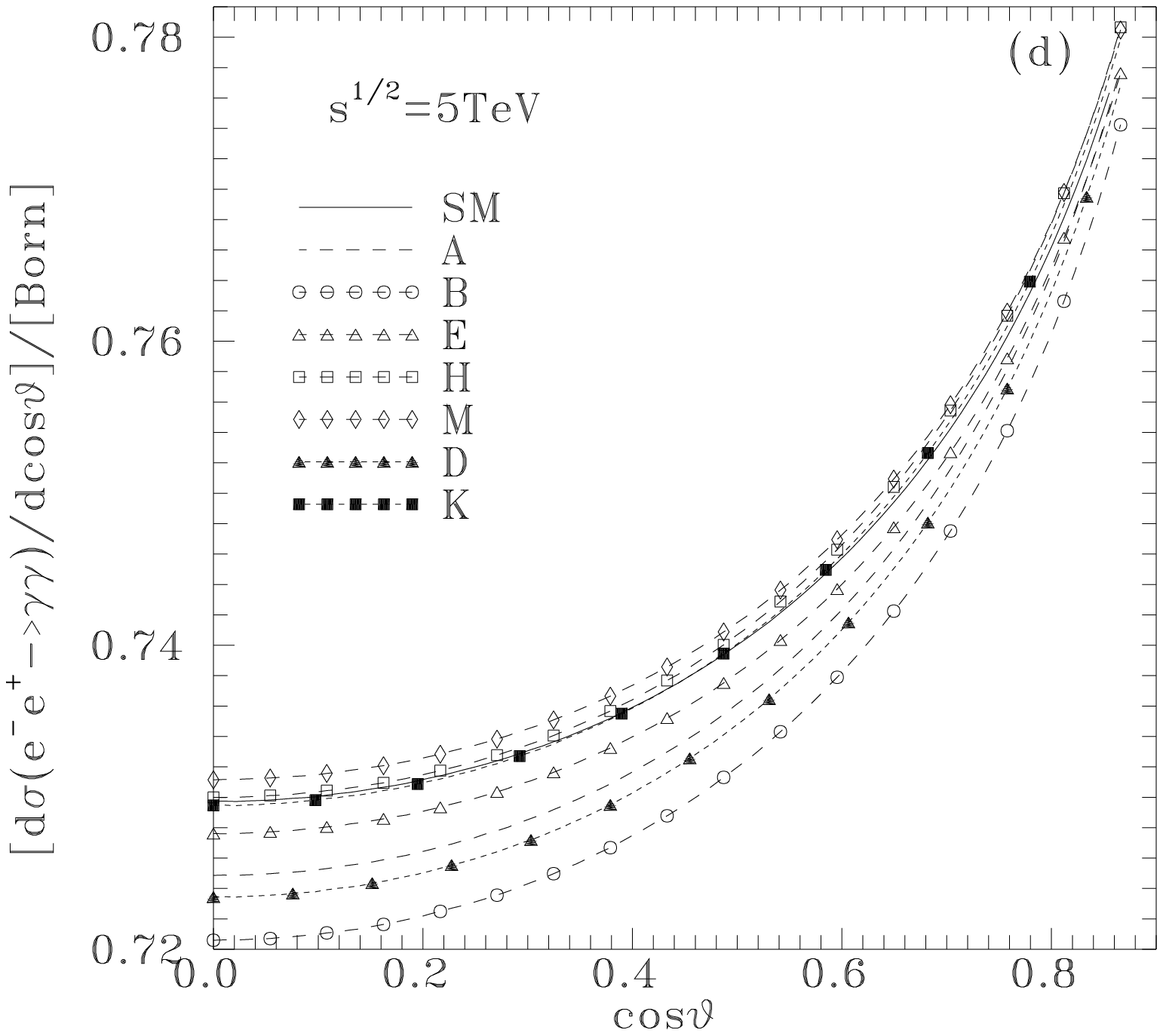,height=7cm}
\]
\vspace*{-0.5cm}
\caption[1]{The unpolarized differential
cross section for $e^-e^+ \to \gamma \gamma $. In (a) and (b) the
Born contributions are given at 0.5TeV and 5TeV respectively;
while in (c) and (d) the radiative corrections to them  are
respectively indicated for  SM and
a representative subset
of the benchmark MSSM models of \cite{Ellis-bench}.}
\label{gg-differential-fig}
\end{figure}

\clearpage
\newpage

\begin{figure}[t]
\vspace*{-3cm}
\[
\hspace{-0.5cm}\epsfig{file=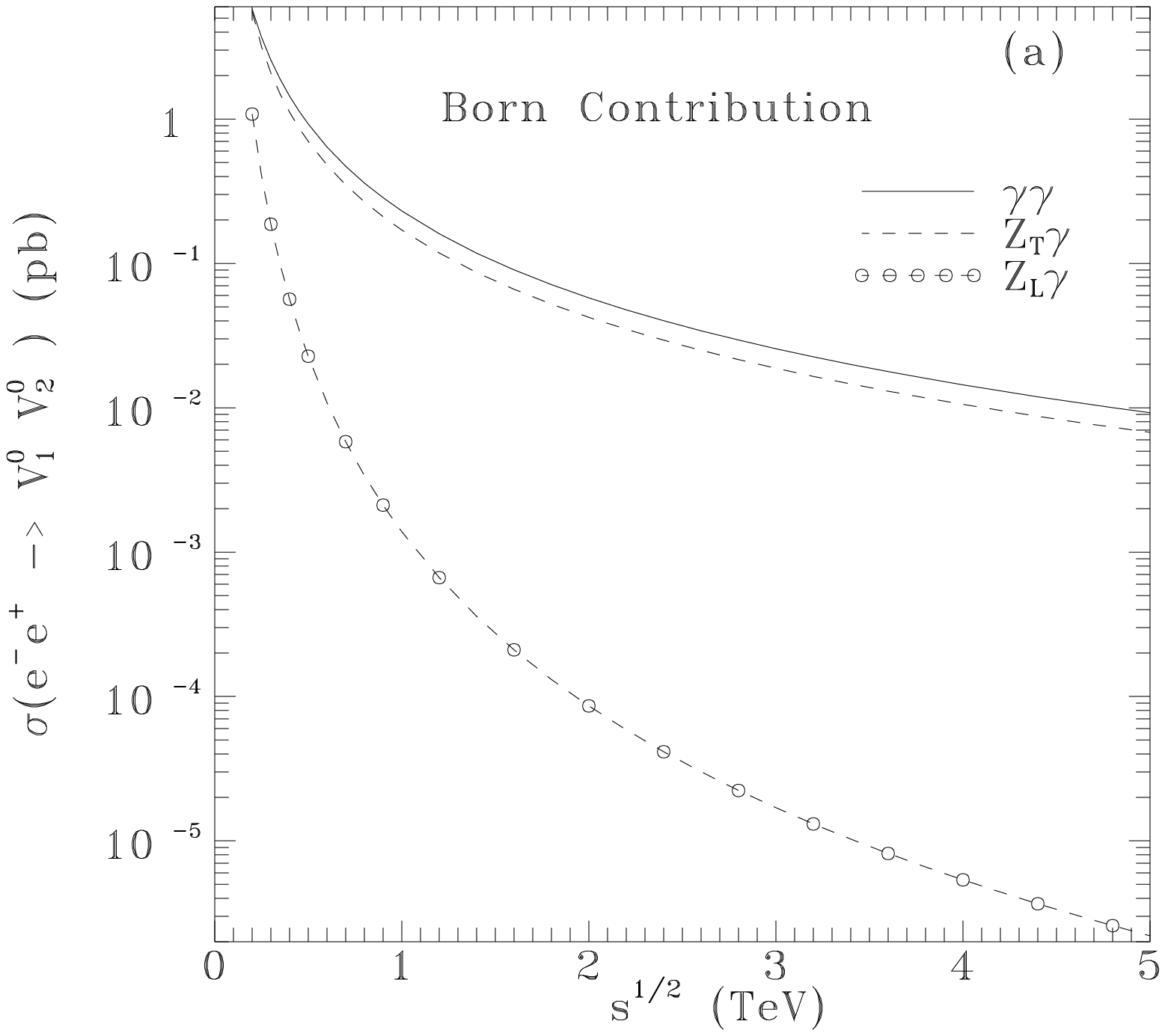,height=7cm}\hspace{0.5cm}
\epsfig{file=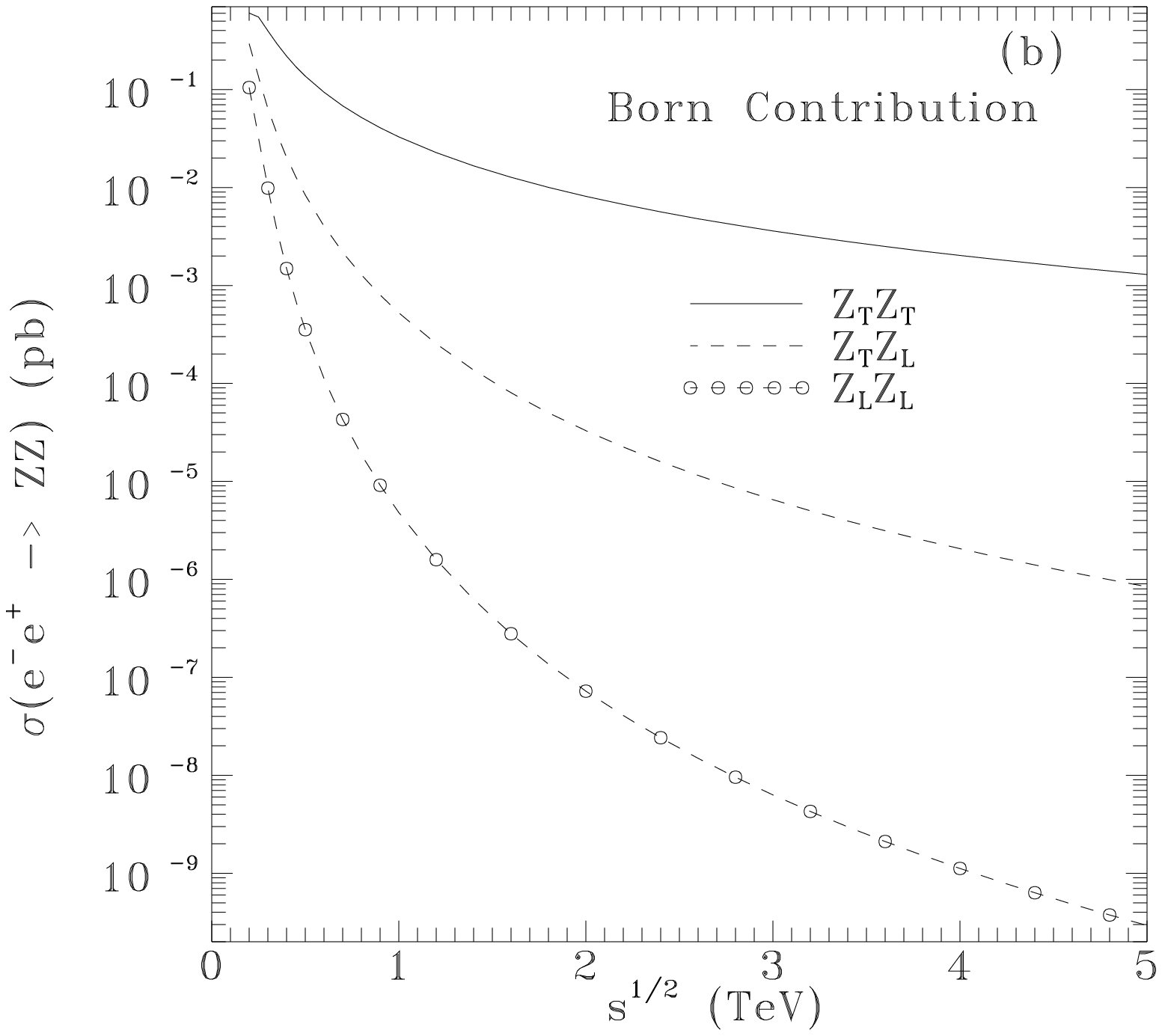,height=7cm}
\]
\vspace*{-0.5cm}
\caption[1]{The integrated Born cross section in the region
$30^o<\theta< 150^o$, for
$e^-e^+ \to \gamma \gamma, ~Z\gamma $ (a) and
$e^-e^+ \to ZZ $ (b), with transverse or
longitudinal $Z$-states, as a function of the energy.}
 \label{Born-integrated-fig}
\end{figure}

\begin{figure}[b]
\vspace*{-4cm}
\[
\epsfig{file=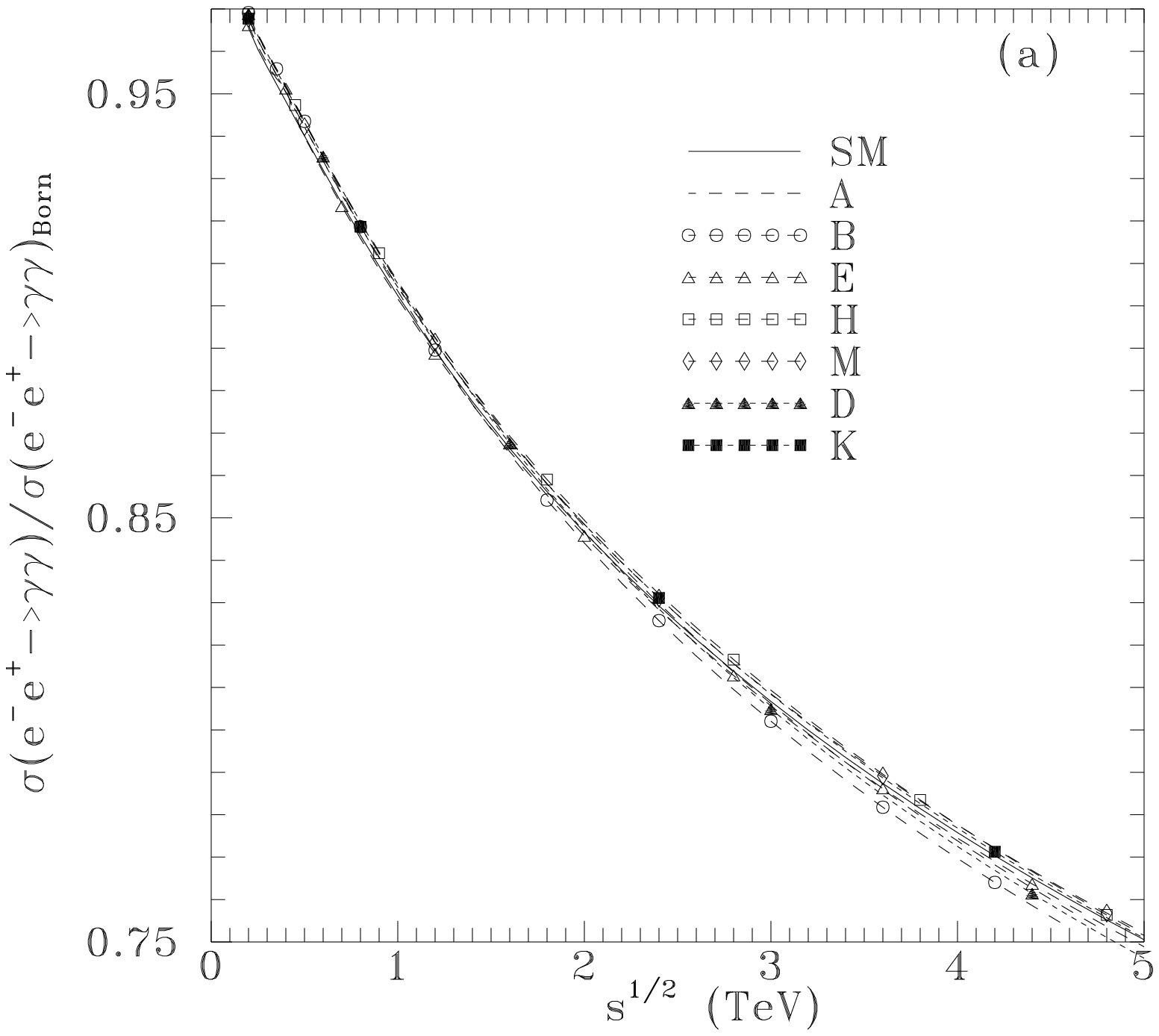,height=7cm}\hspace{0.5cm}
\epsfig{file=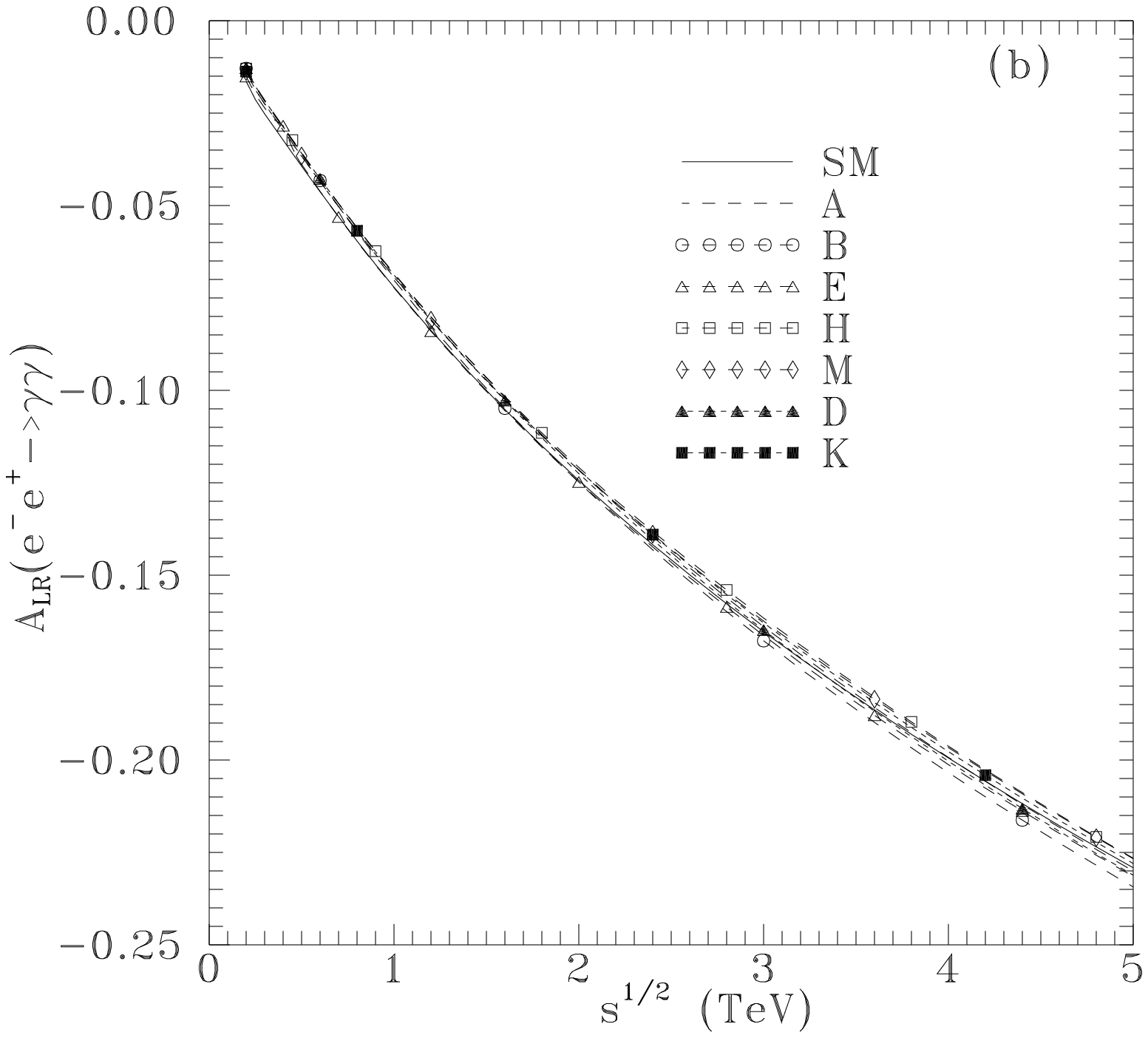,height=7cm}
\]
\vspace*{-0.5cm}
\caption[1]{The ratio of the unpolarized integrated
$\sigma(e^-e^+\to \gamma \gamma)$ cross section
to the Born cross section (a),
and the $A_{LR}$ asymmetry (b), as a
function of the energy, for  SM and a representative subset
of the benchmark MSSM models of \cite{Ellis-bench}.}
 \label{gg-sig-ratio-fig}
\end{figure}

\clearpage
\newpage

\begin{figure}[p]
\vspace*{-3cm}
\[
\hspace{-0.5cm}\epsfig{file=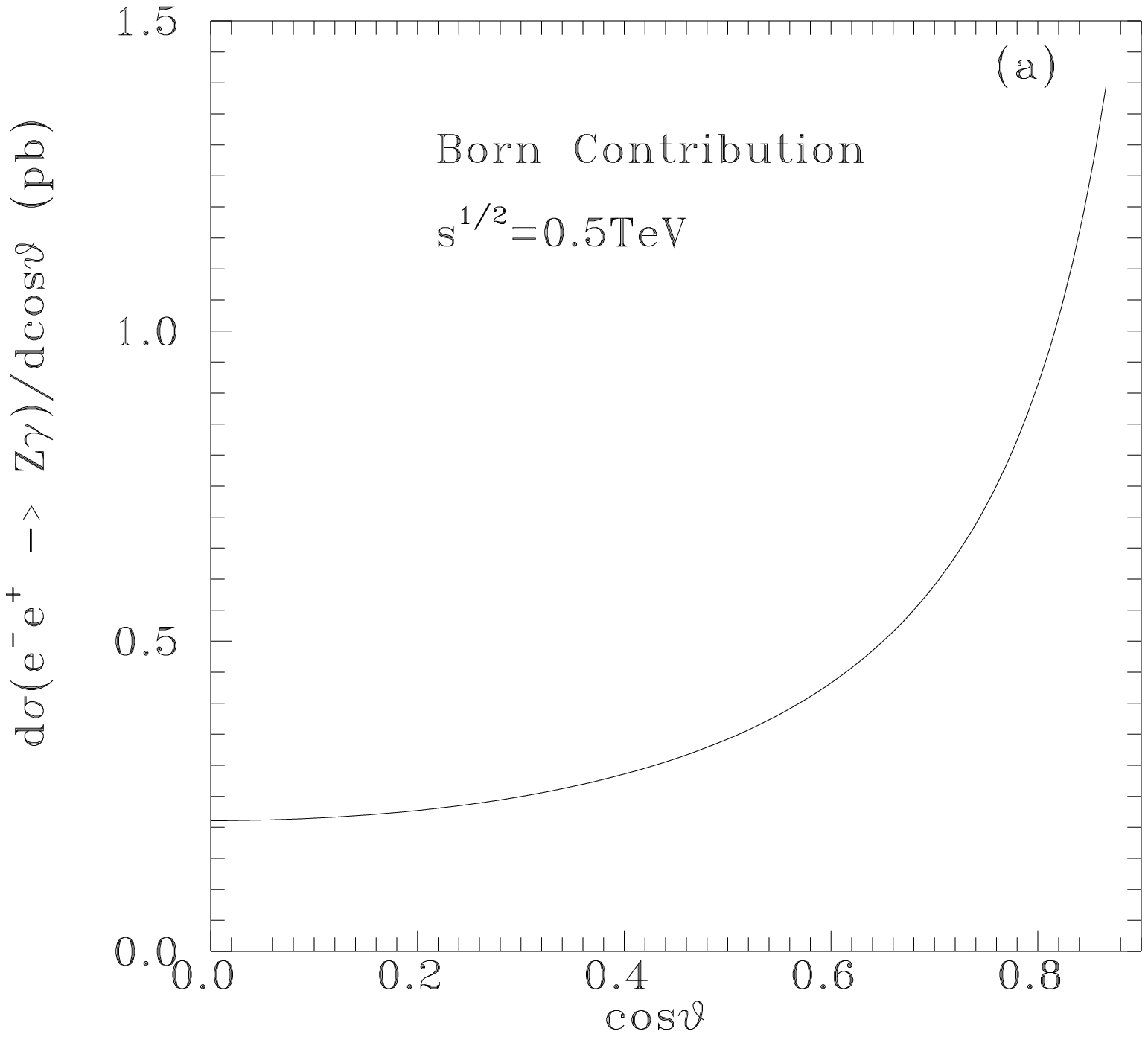,height=7cm}\hspace{0.5cm}
\epsfig{file=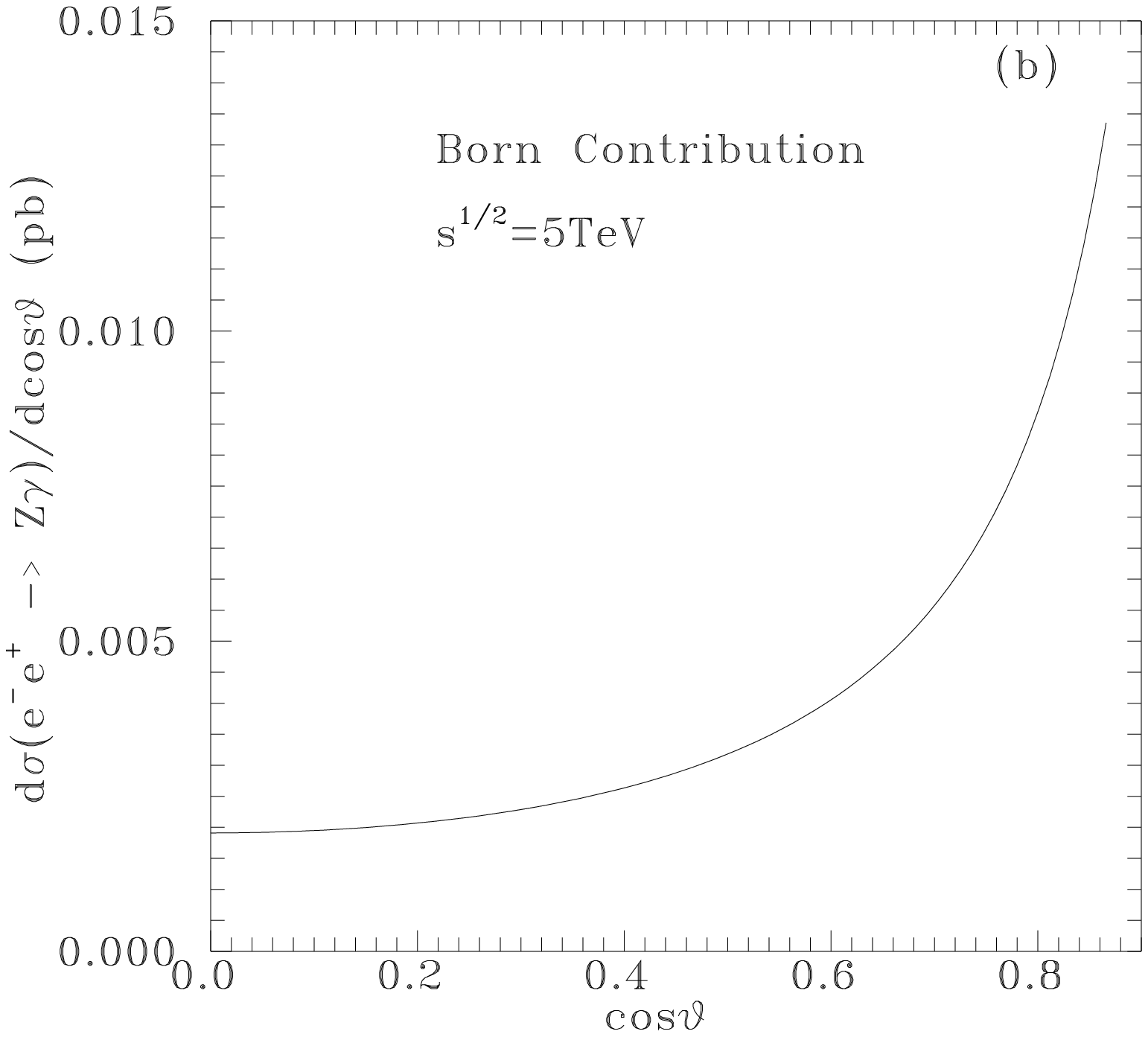,height=7cm}
\]
\vspace*{0.5cm}
\[
\hspace{-0.5cm}\epsfig{file=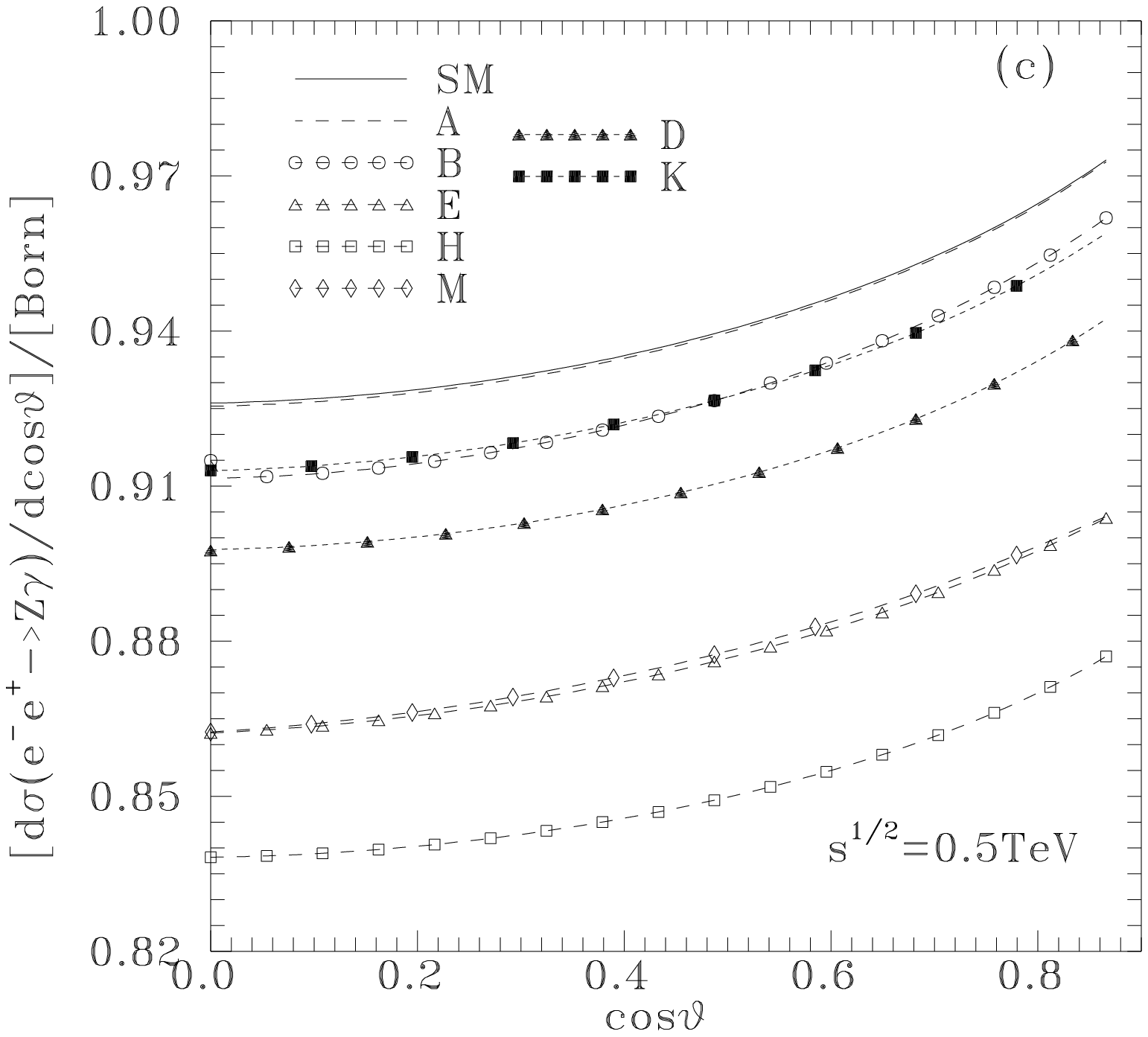,height=7cm}\hspace{0.5cm}
\epsfig{file=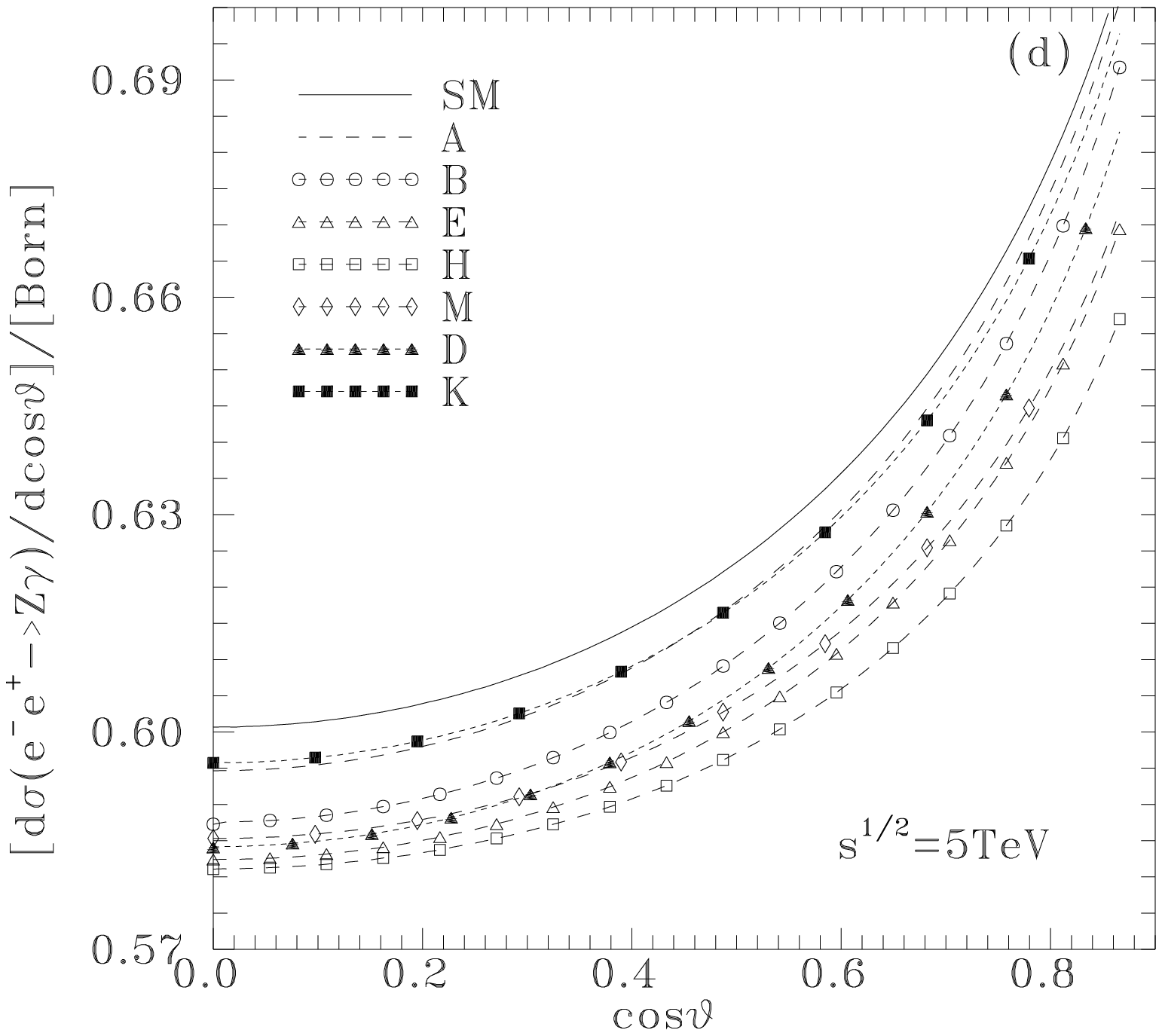,height=7cm}
\]
\vspace*{-0.5cm}
\caption[1]{The unpolarized differential
cross section for $e^-e^+ \to Z \gamma $. In (a) and (b) the
Born contribution at 0.5TeV and 5TeV respectively are given;
while in (c) and (d) the radiative corrections to it are
respectively indicated for  SM and
a representative subset
of the benchmark MSSM models of \cite{Ellis-bench}.}
\label{Zg-differential-fig}
\end{figure}

\clearpage
\newpage

\begin{figure}[p]
\vspace*{-3cm}
\[
\hspace{-0.5cm}\epsfig{file=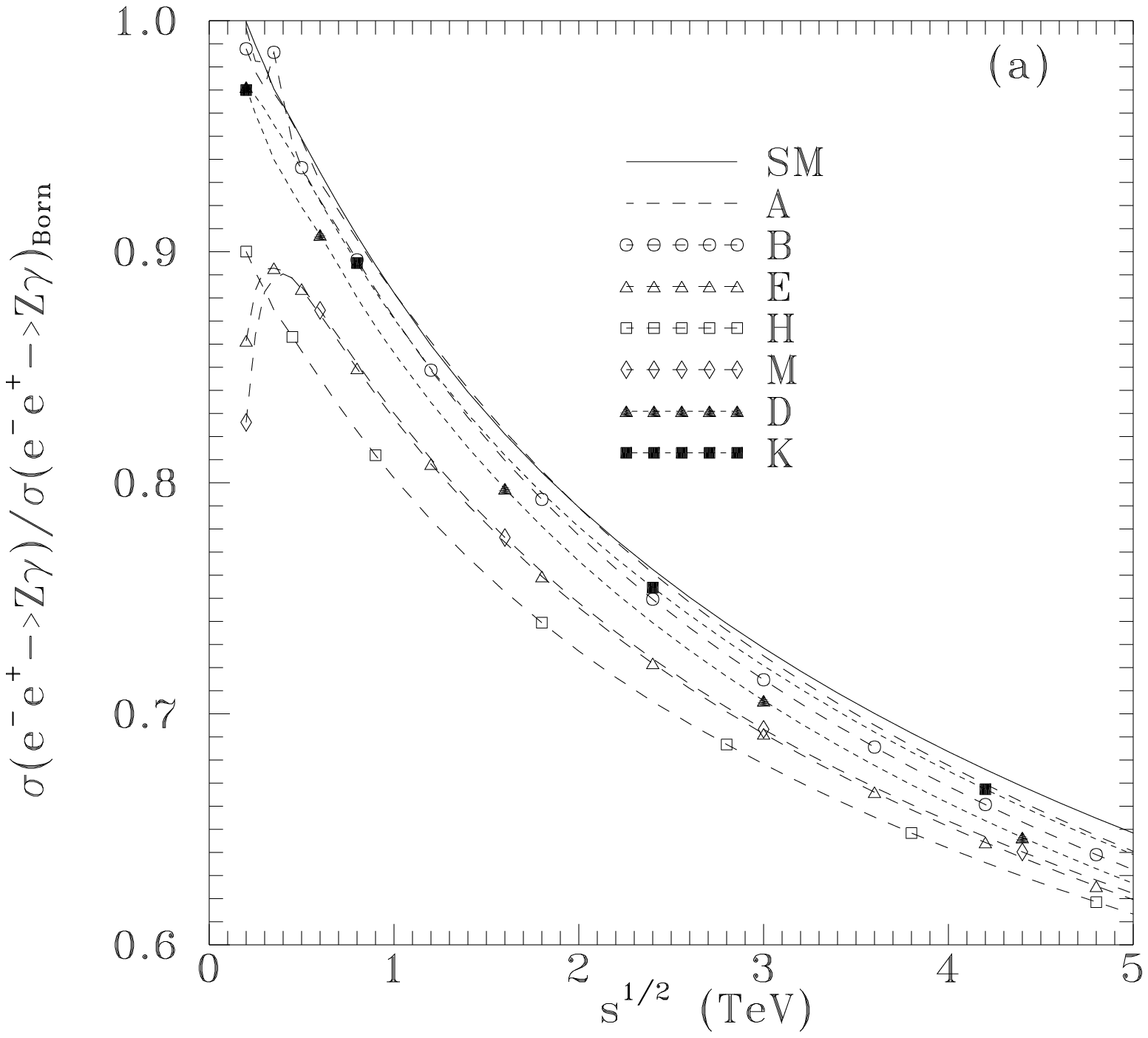,height=7cm}\hspace{0.5cm}
\epsfig{file=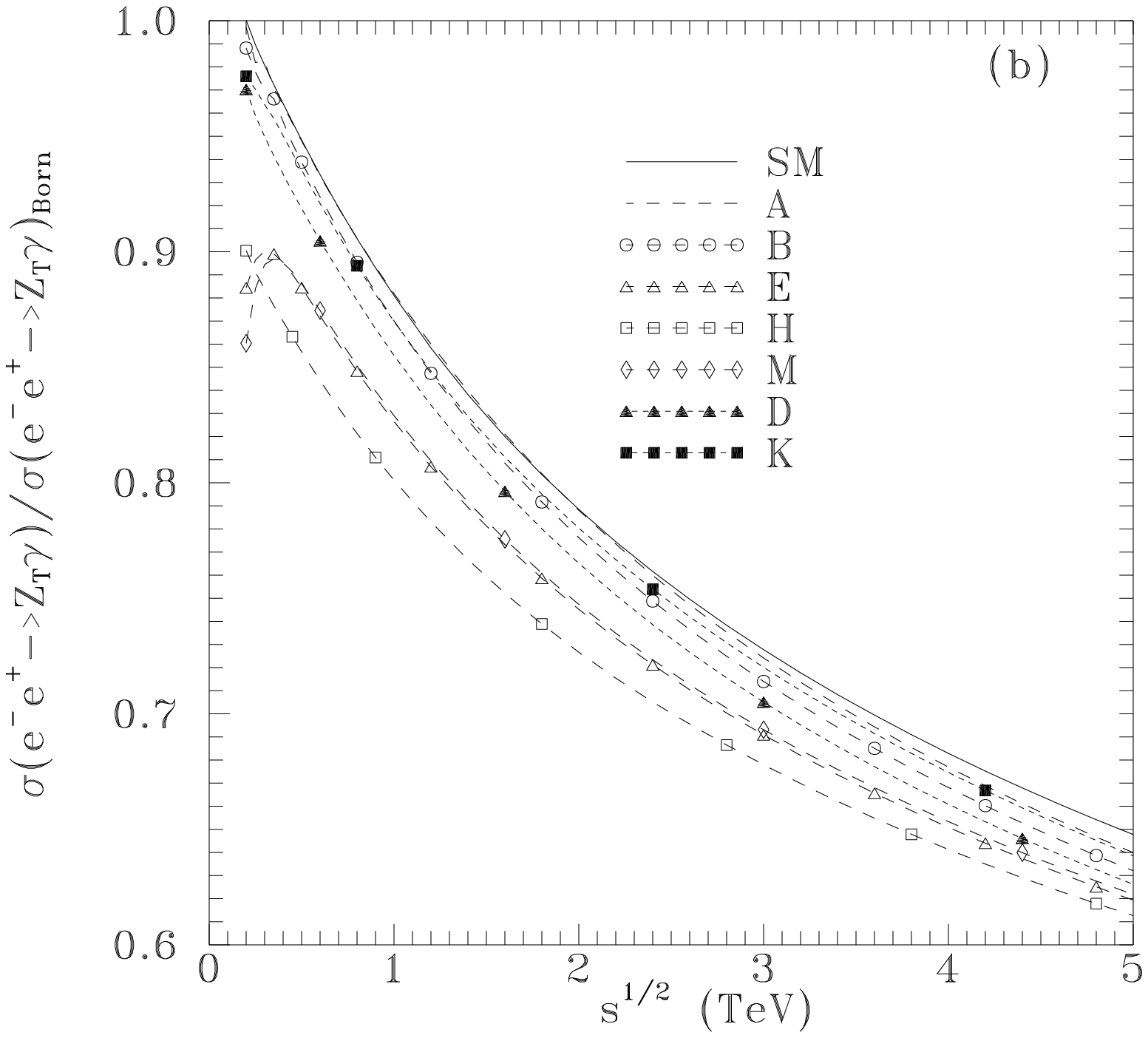,height=7cm}
\]
\vspace*{0.5cm}
\[
\epsfig{file=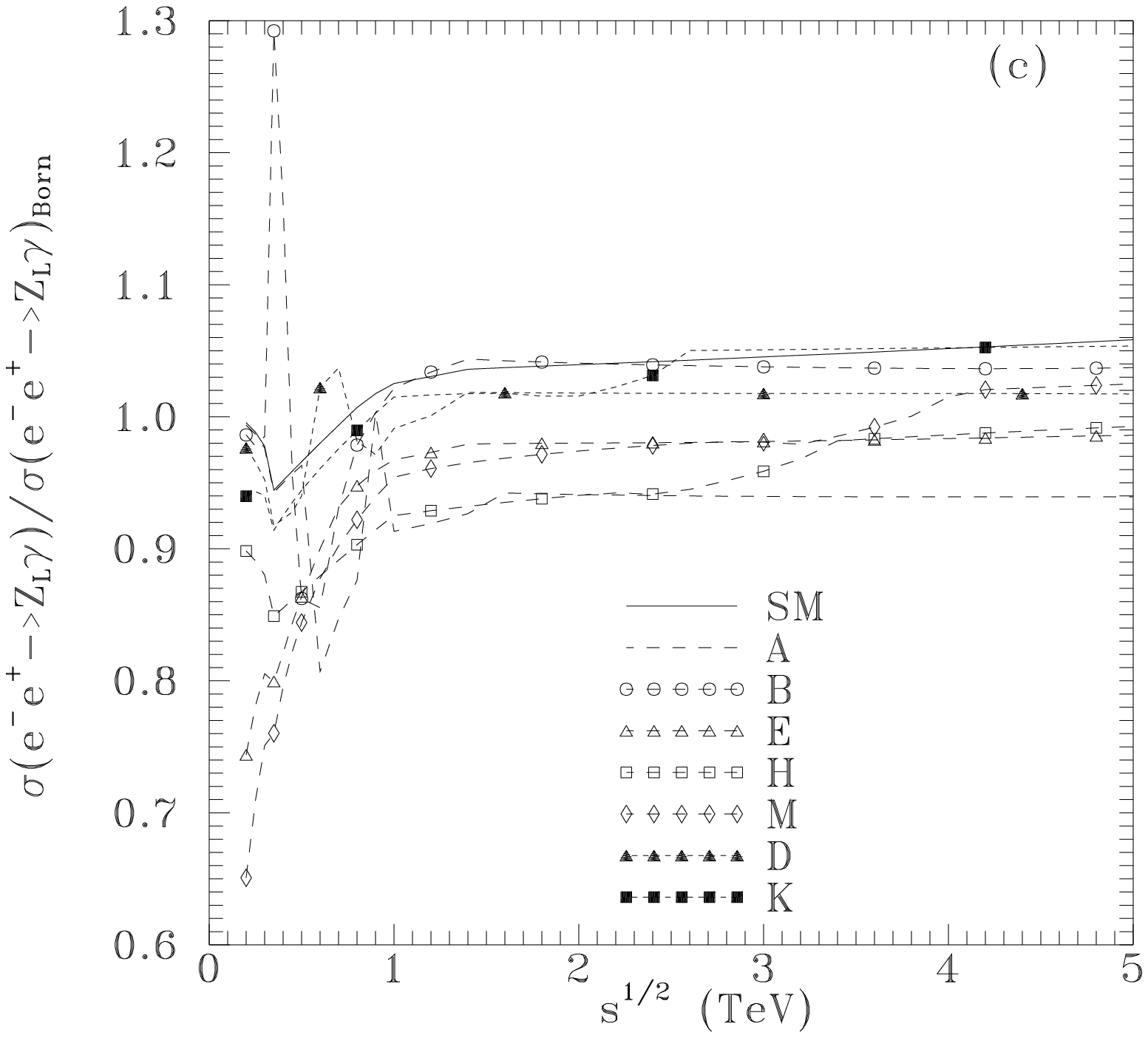,height=7cm}\hspace{0.5cm}
\epsfig{file=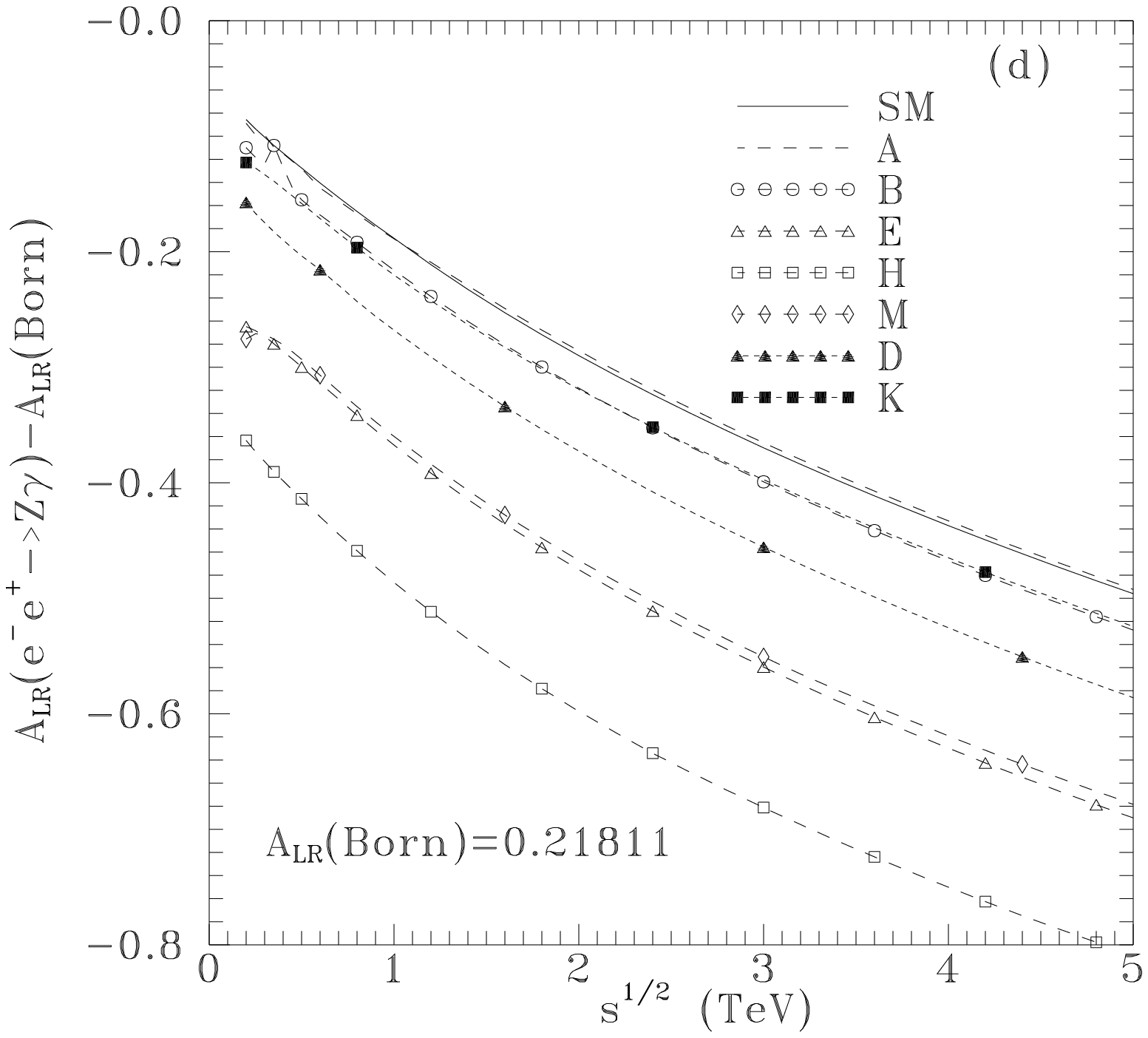,height=7cm}\]
\vspace*{-0.5cm}
\caption[1]{The radiative corrections to  the integrated
$\sigma(e^-e^+\to Z \gamma )$ cross section,
for unpolarized $Z$ (a), transverse $Z$ (b), or longitudinal  $Z$
(c) states, as a function of the energy  for SM
and a set  MSSM models of \cite{Ellis-bench}.
In (d)  the radiative correction to the $A_{LR}$ asymmetry,
(where all final gauge polarizations are summed over),  is
also given.}
\label{Zg-sig-ratio-fig}
\end{figure}

\clearpage
\newpage

\begin{figure}[p]
\vspace*{-3cm}
\[
\hspace{-0.5cm}\epsfig{file=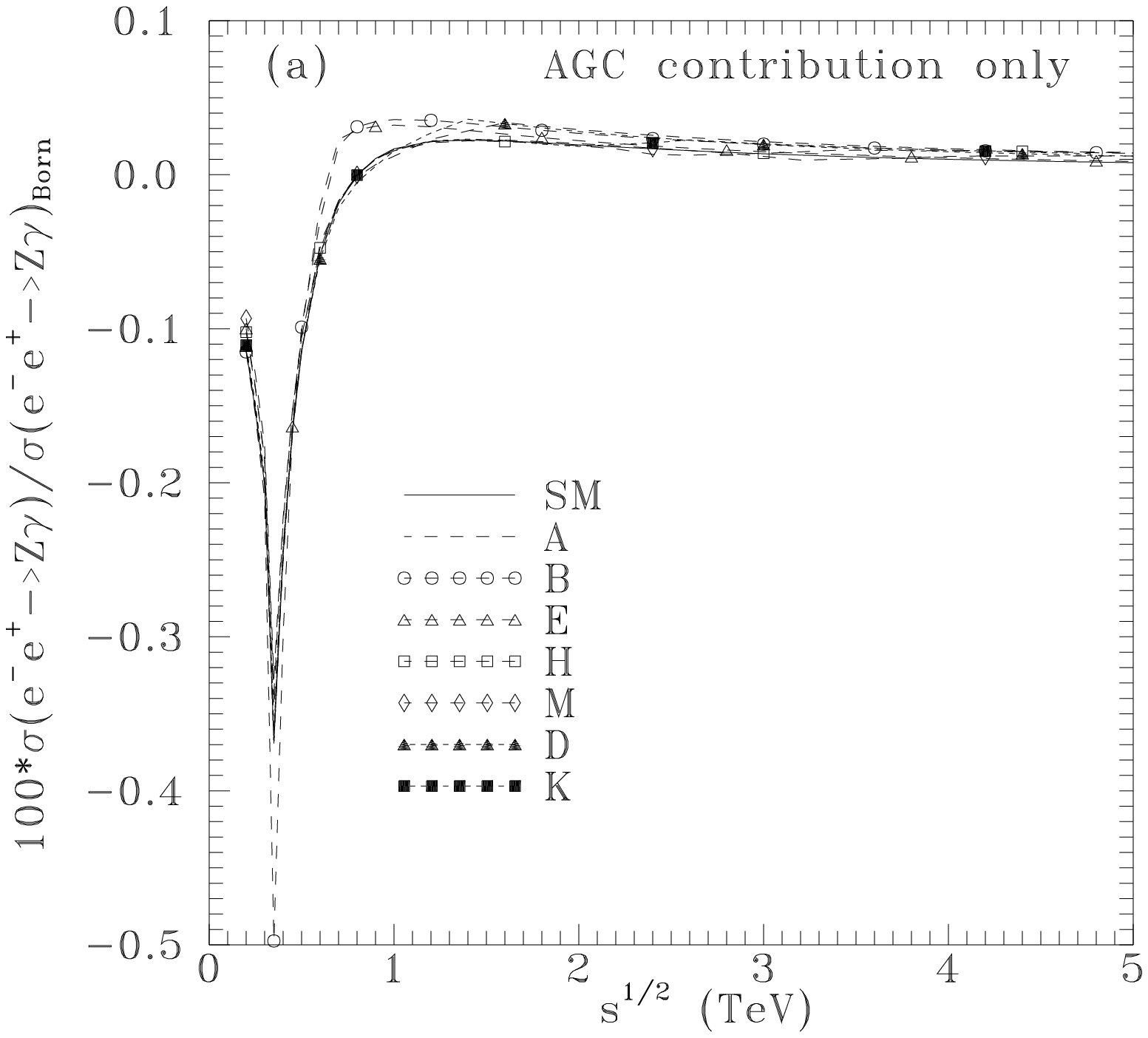,height=7cm}\hspace{0.5cm}
\epsfig{file=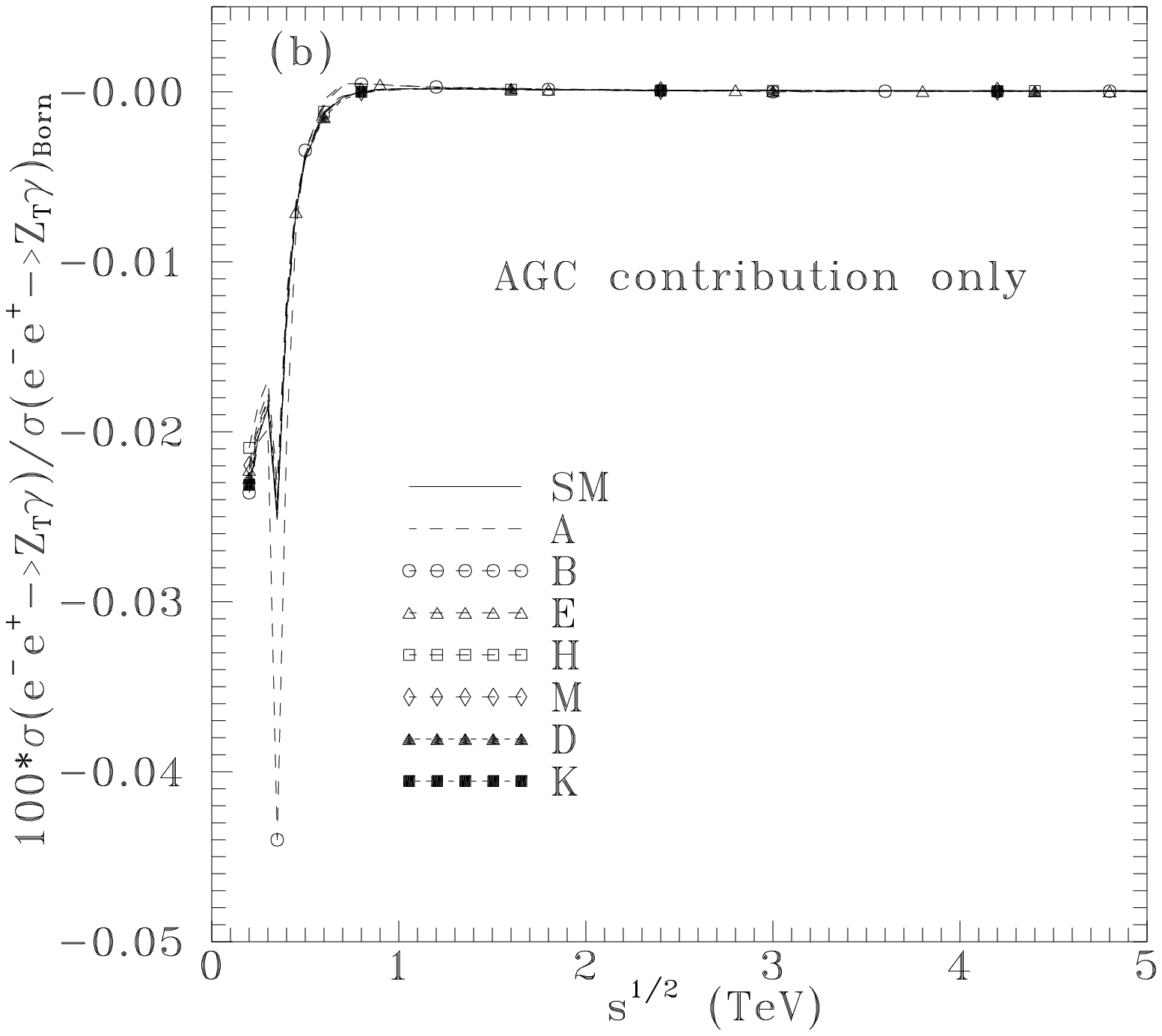,height=7cm}\]
\vspace*{0.5cm}
\[
\hspace{-0.5cm}\epsfig{file=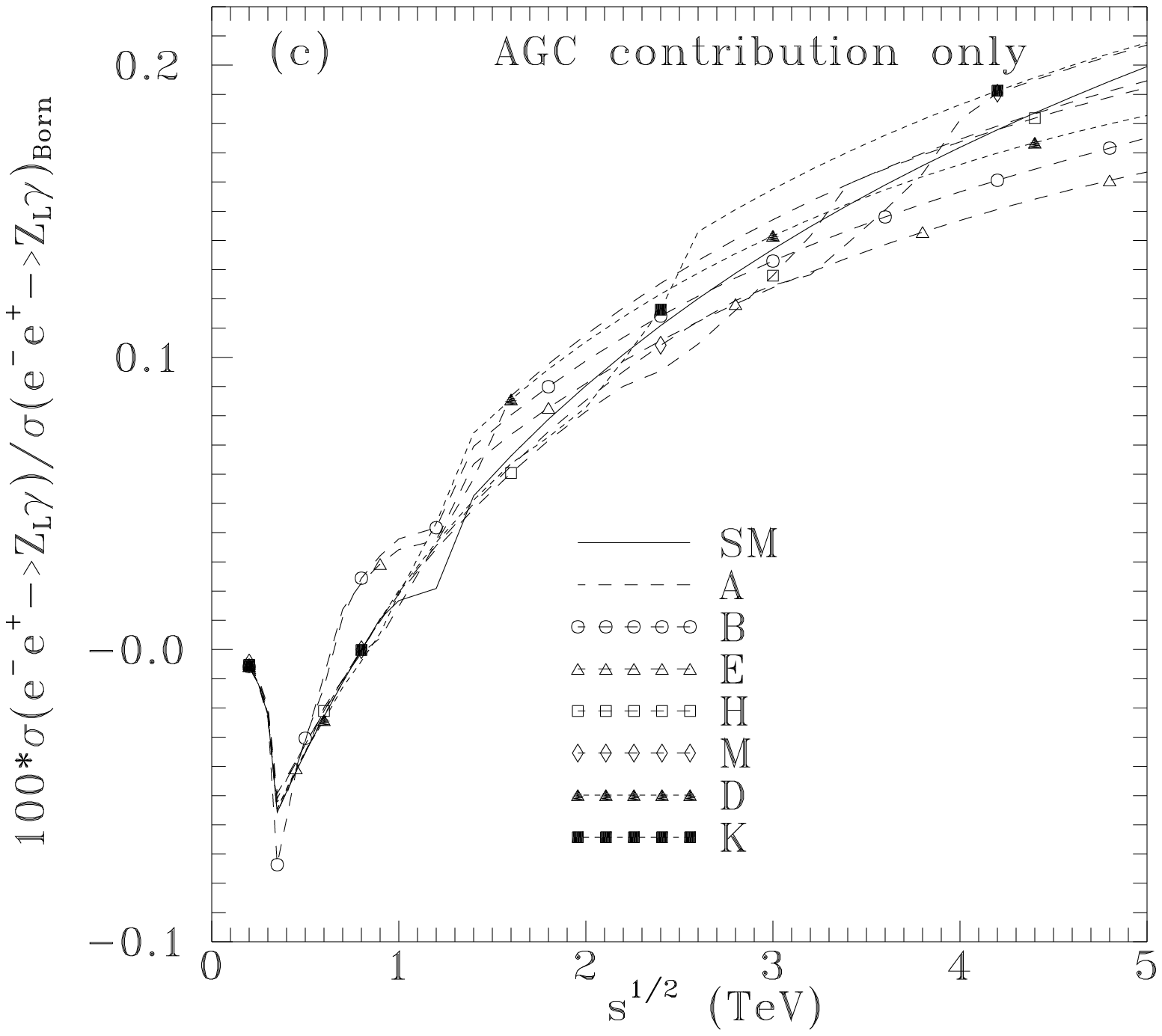,height=7cm}\hspace{0.5cm}
\epsfig{file=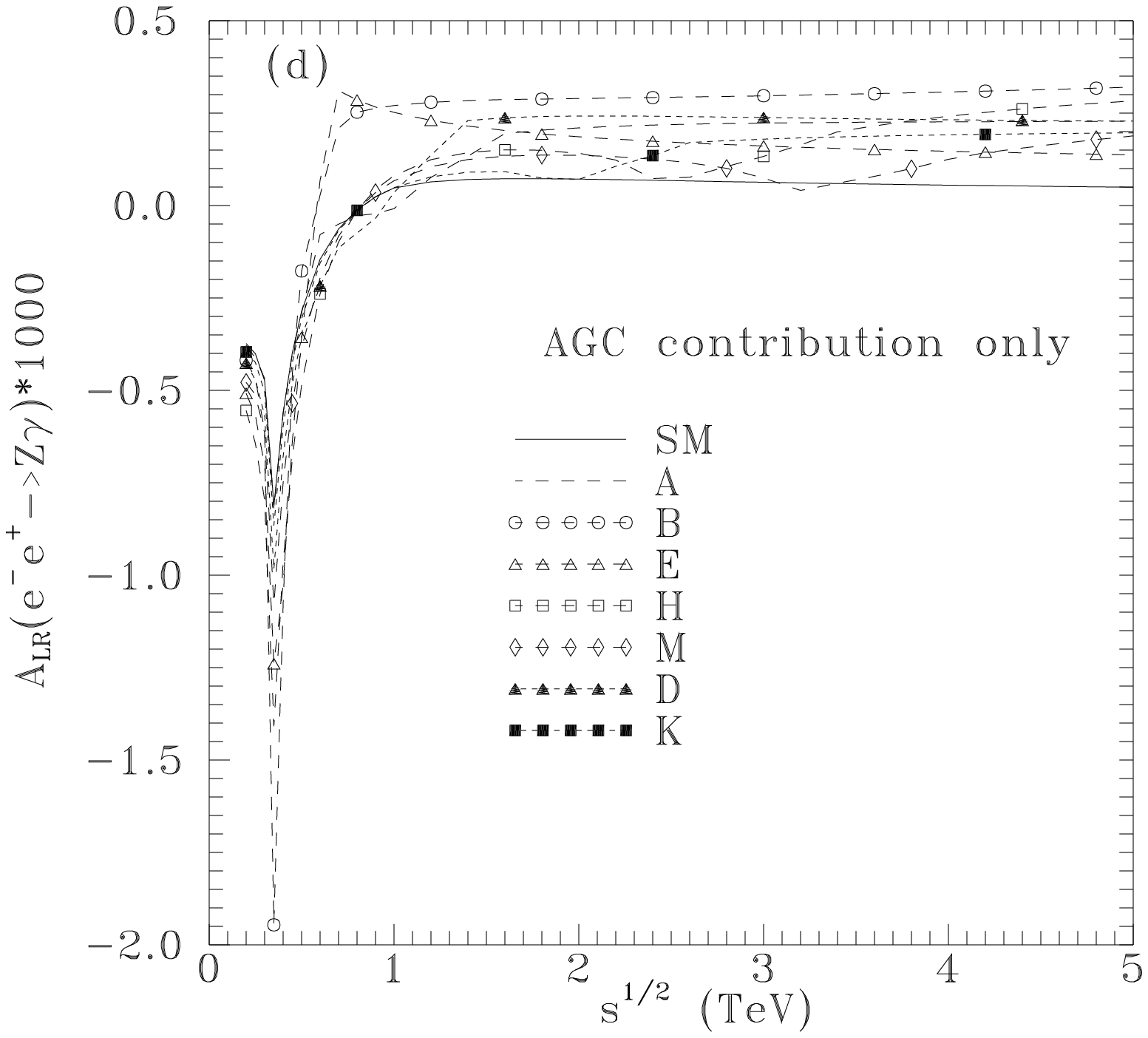,height=7cm}\]
\vspace*{0.5cm}
\caption[1]{NAGC contributions to the
unpolarized (a), TT (b), LT (c) $e^+e^-\to Z\gamma$ cross sections,
and to the  $A_{LR}$ asymmetry (d), as a function of the energy.}
\label{Zg-NAGC-fig}
\end{figure}

\clearpage
\newpage

\begin{figure}[p]
\vspace*{-3cm}
\[
\hspace{-0.5cm}\epsfig{file=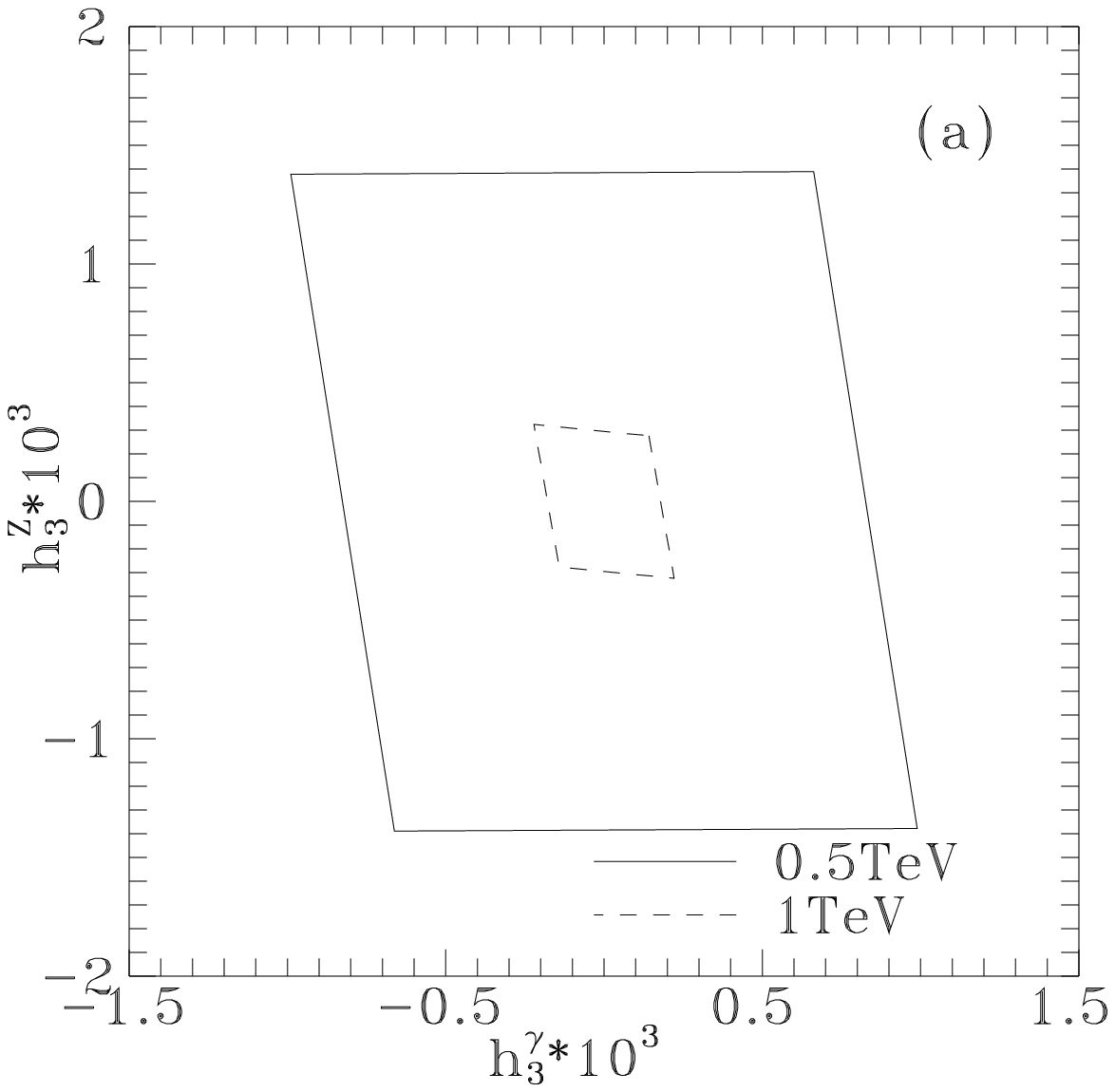,height=7cm}\hspace{0.5cm}
\epsfig{file=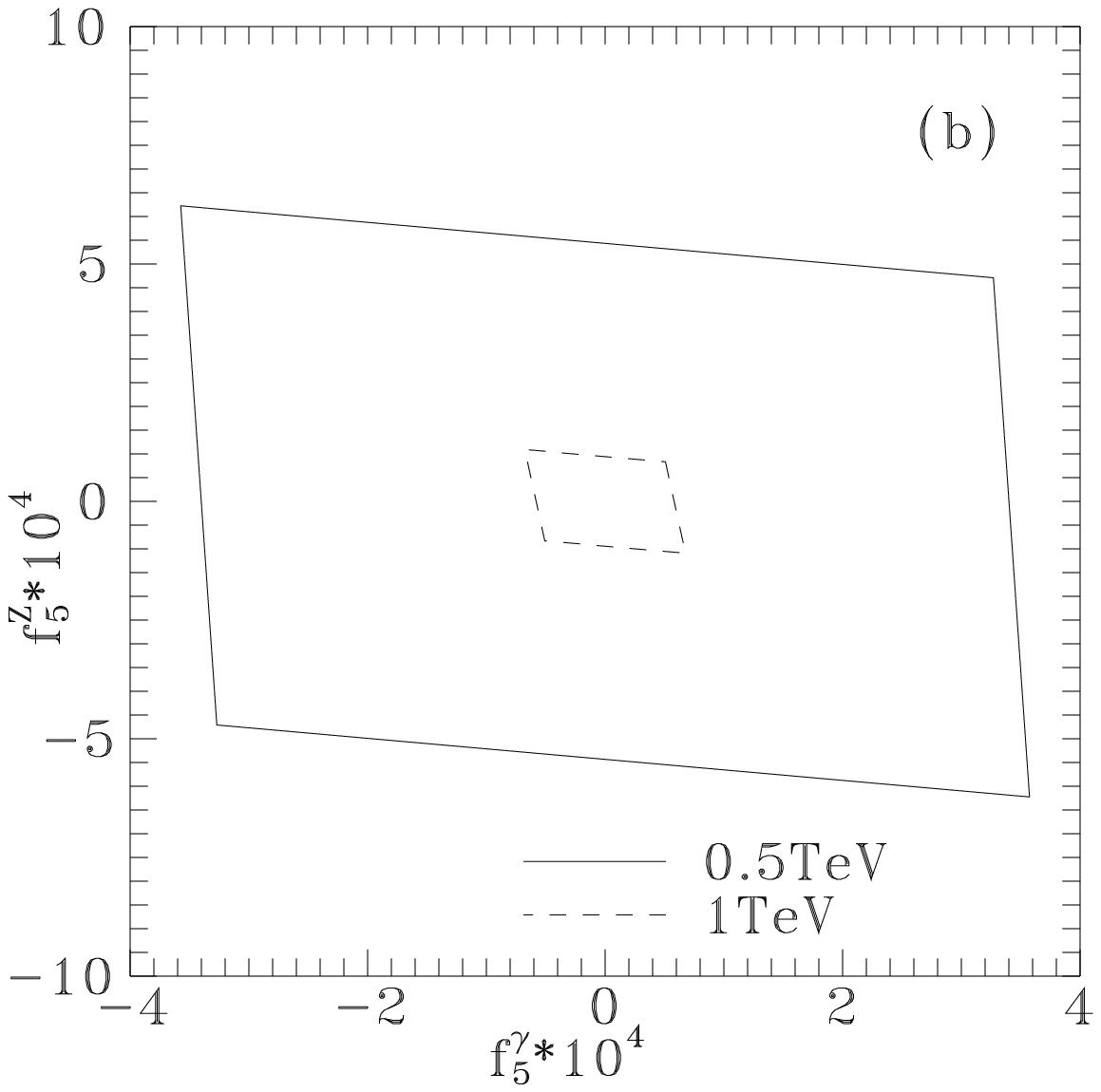,height=7cm}
\]
\vspace*{0.5cm}
\caption[1]{The NAGC limits obtained from $\sigma_{unp}$ and $A_{LR}$
in the $e^+e^-\to Z\gamma$ process (a), and in the
$e^+e^-\to ZZ$ process (b), assuming an accuracy of $1\%$ on
these observables.}
\label{NAGClim-fig}
\end{figure}

\clearpage
\newpage

\begin{figure}[p]
\vspace*{-3cm}
\[
\hspace{-0.5cm}\epsfig{file=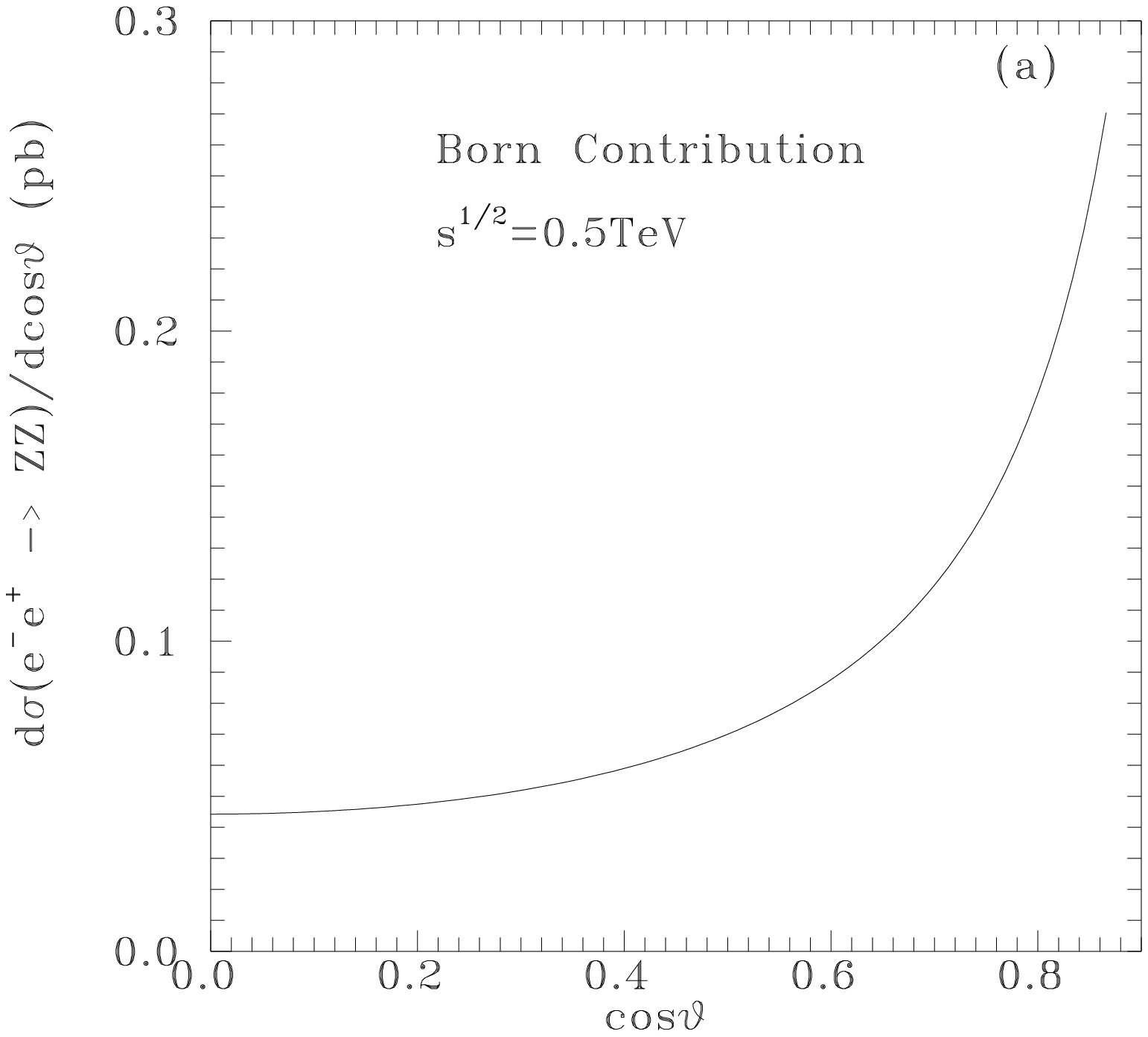,height=7cm}\hspace{0.5cm}
\epsfig{file=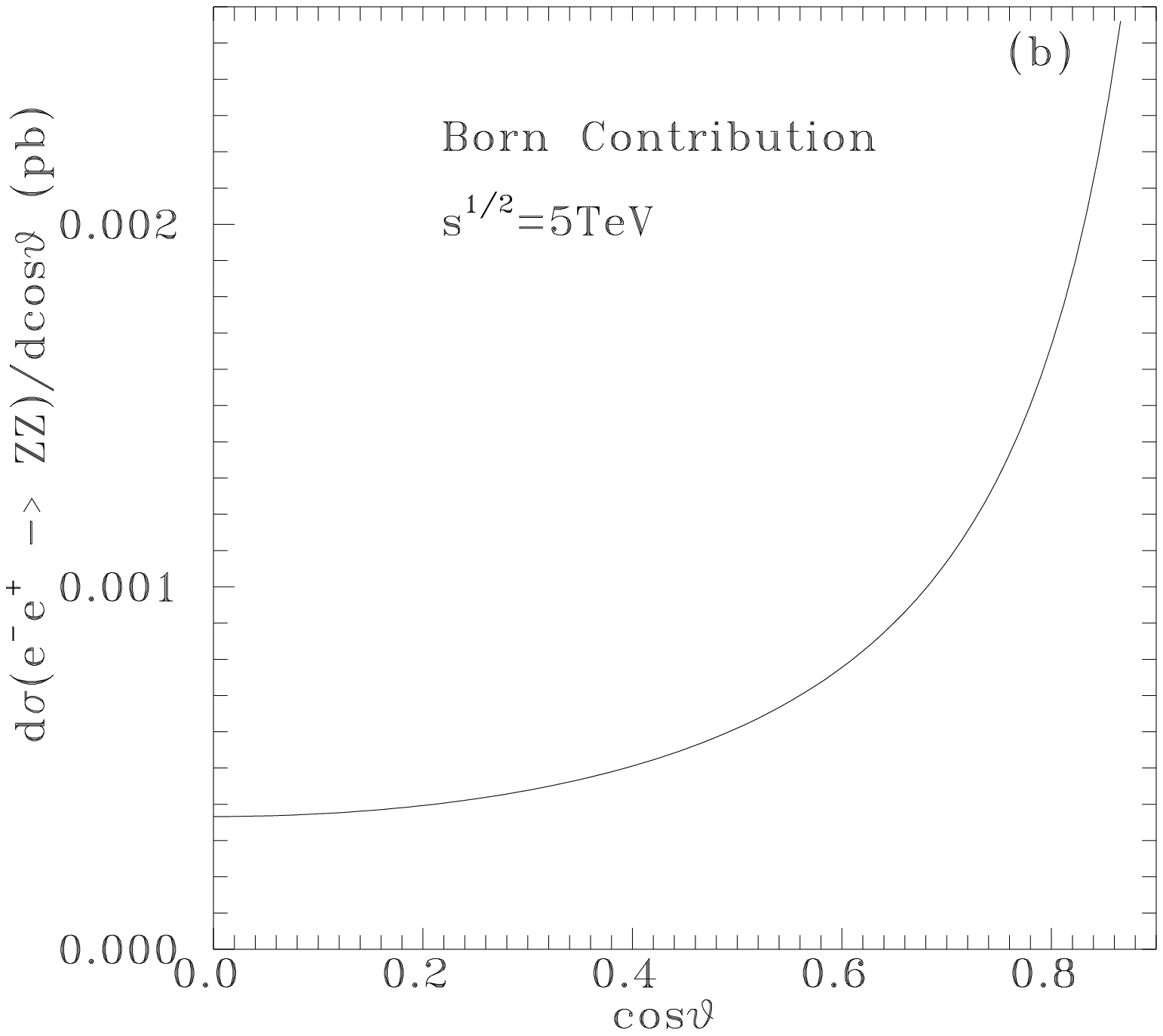,height=7cm}
\]
\vspace*{0.5cm}
\[
\hspace{-0.5cm}\epsfig{file=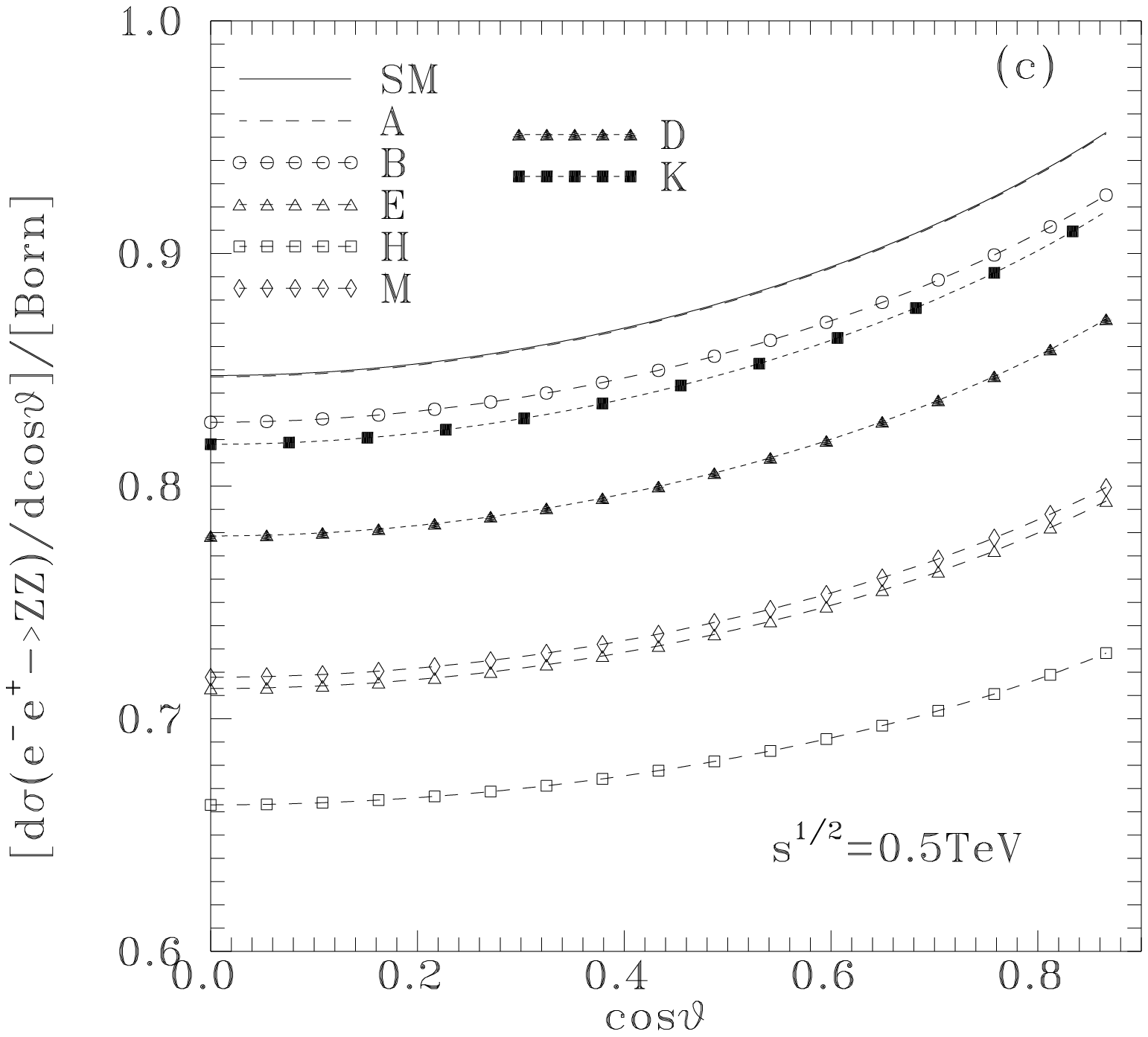,height=7cm}\hspace{0.5cm}
\epsfig{file=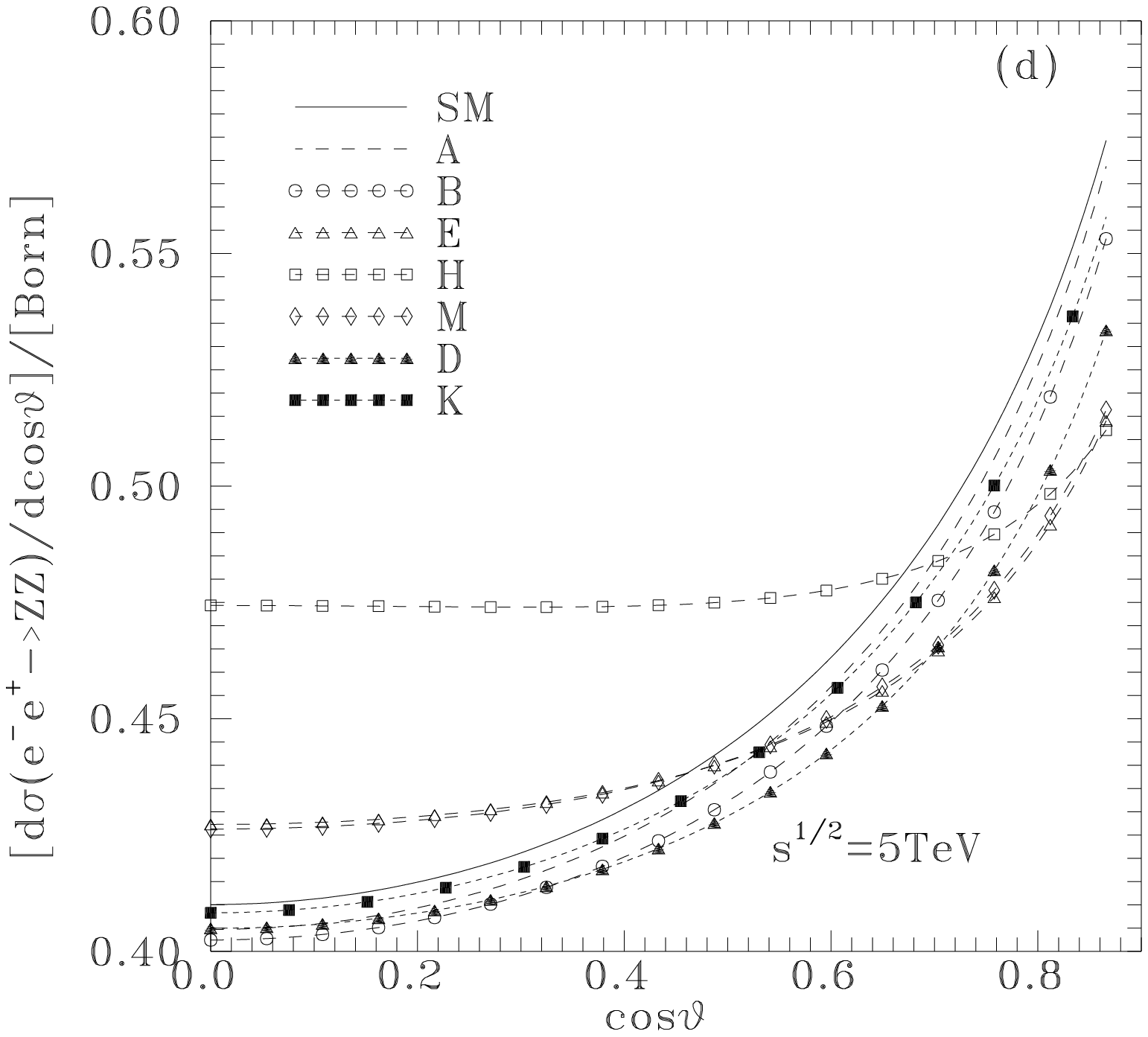,height=7cm}
\]
\vspace*{-0.5cm}
\caption[1]{The unpolarized differential
cross section for $e^-e^+ \to ZZ $. In (a) and (b) the
Born contribution at 0.5TeV and 5TeV respectively are given;
while in (c) and (d) the radiative corrections to it are
respectively indicated for  SM and
a representative subset
of the benchmark MSSM models of \cite{Ellis-bench}.}
\label{ZZ-differential-fig}
\end{figure}

\clearpage
\newpage

\begin{figure}[p]
\vspace*{-3cm}
\[
\hspace{-0.5cm}\epsfig{file=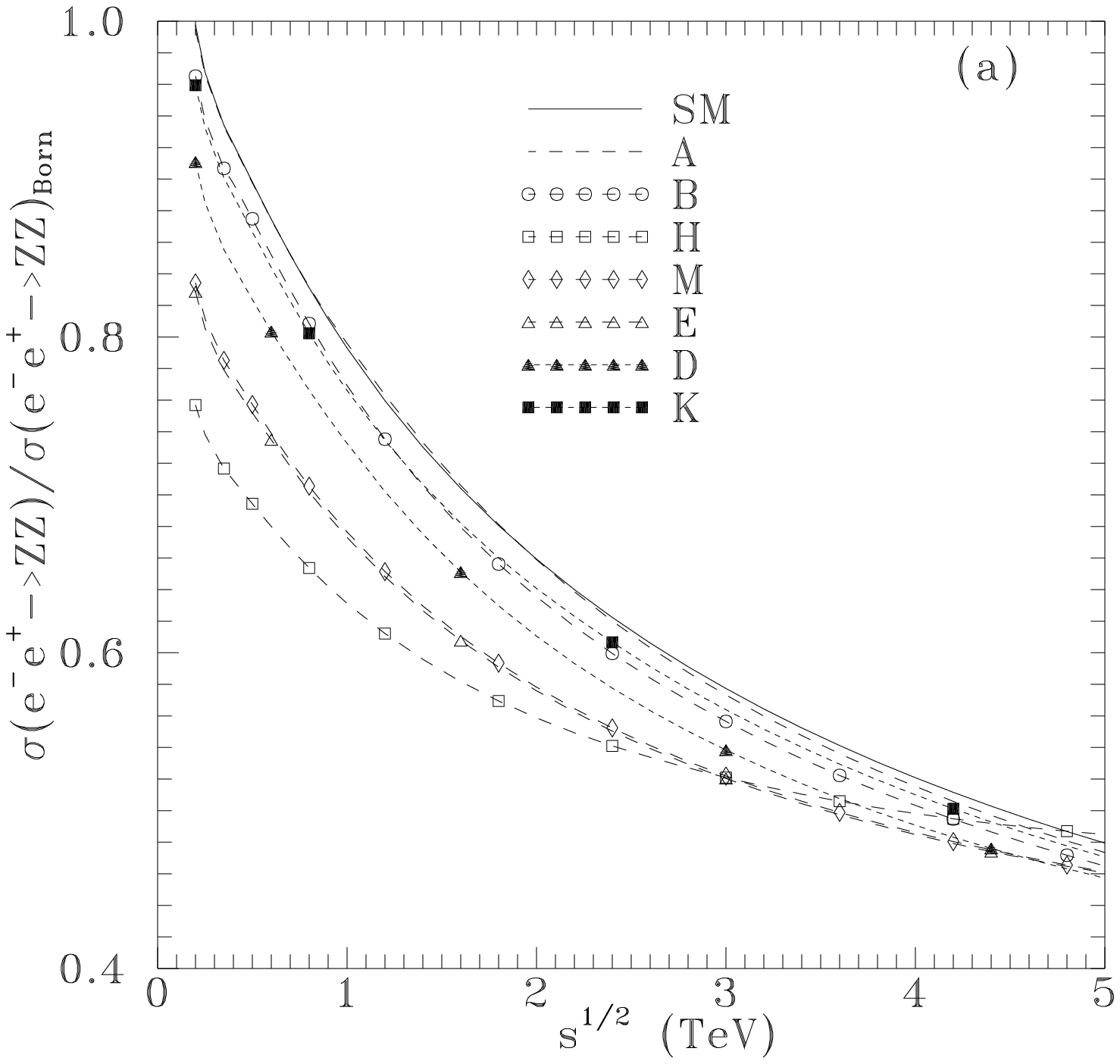,height=7cm, width=7cm}\hspace{0.5cm}
\epsfig{file=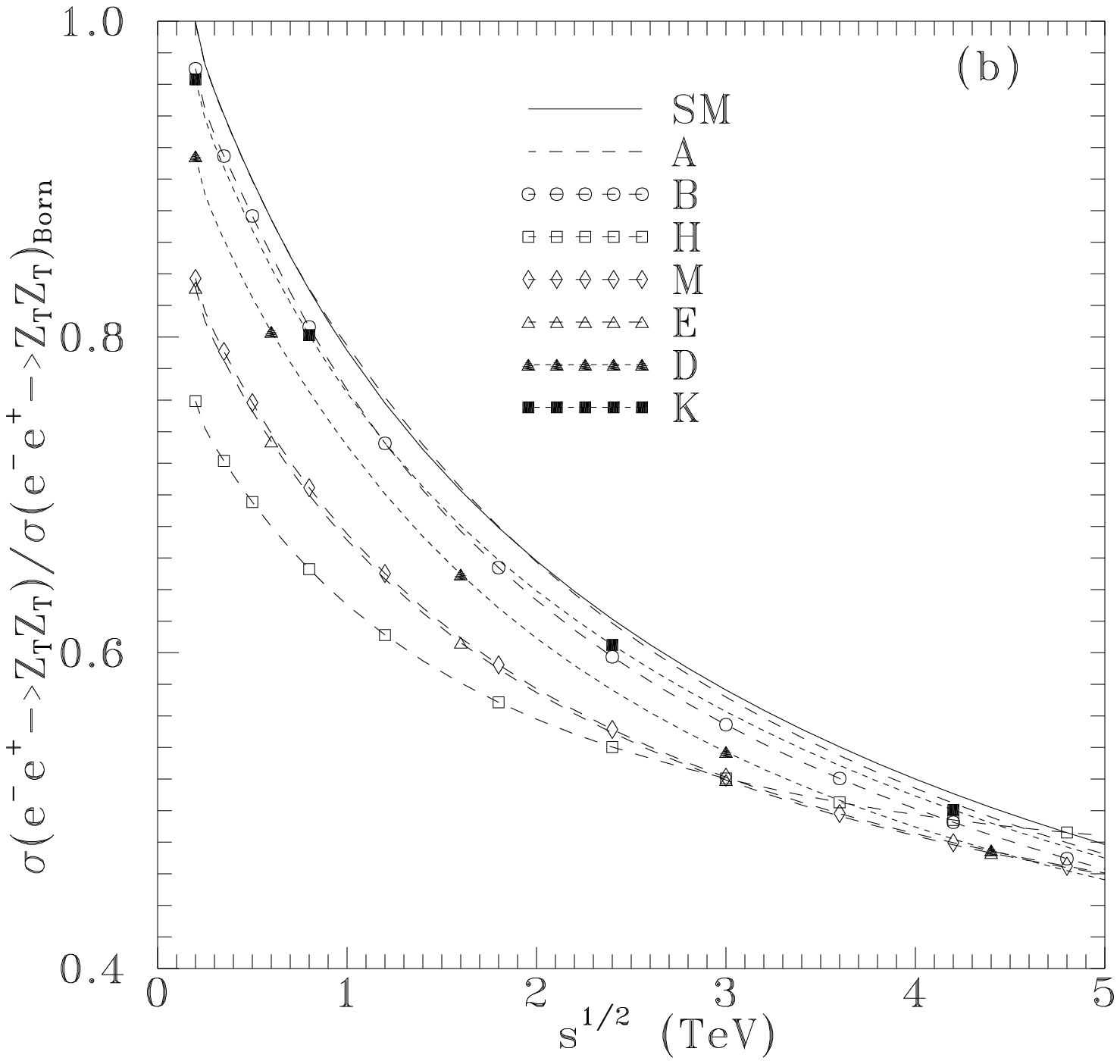,height=7cm, width=7cm}
\]
\vspace*{0.5cm}
\[
\hspace{-0.5cm}\epsfig{file=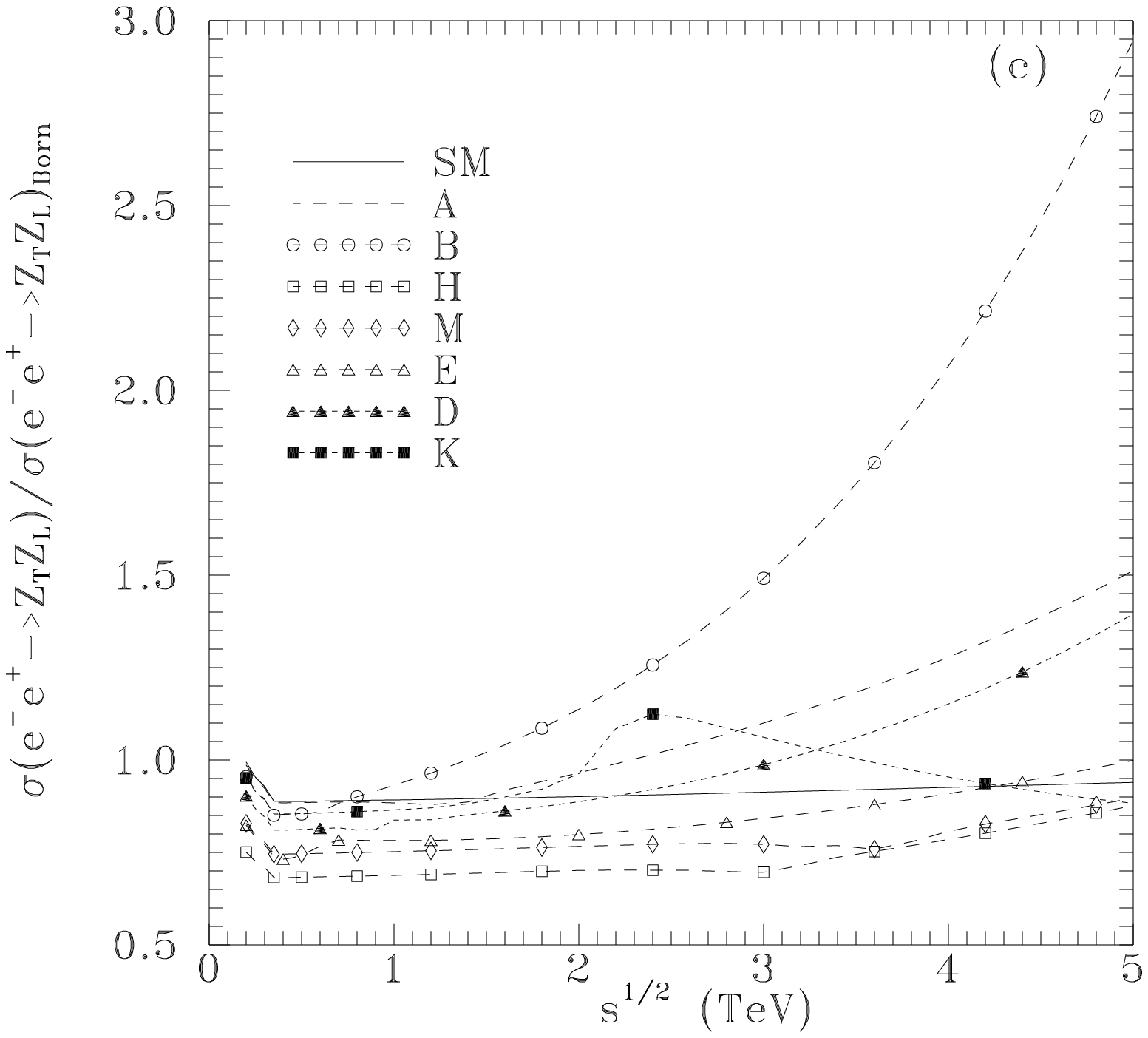,height=7cm, width=7cm}\hspace{0.5cm}
\epsfig{file=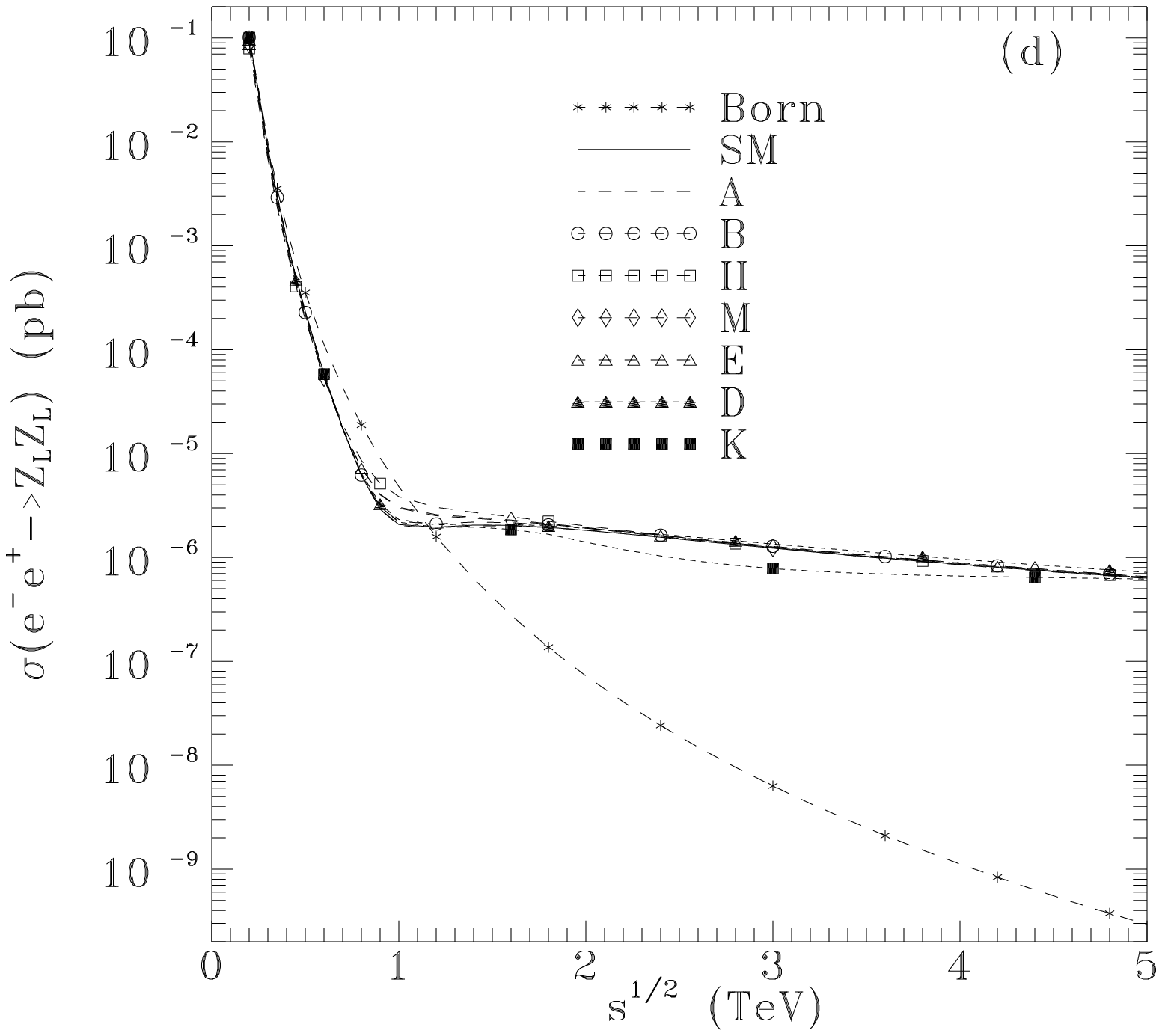,height=7cm, width=7cm}
\]
\vspace*{-0.5cm}
\caption[1]{The ratio integrated
$\sigma(e^-e^+\to Z Z )$ to the Born cross section,
for unpolarized $ZZ$ (a), transverse $Z_TZ_T$ (b),
  $Z_TZ_L$ (c) final states,and the
$\sigma(e^-e^+\to Z _LZ_L )$ cross section (d)
as a function of the energy. The results correspond to
  SM, and a representative subset
of the benchmark MSSM models of \cite{Ellis-bench}.}
\label{ZZ-sig-ratio-fig}
\end{figure}

\clearpage
\newpage

\begin{figure}[p]
\vspace*{-3cm}
\[
\hspace{-0.5cm}\epsfig{file=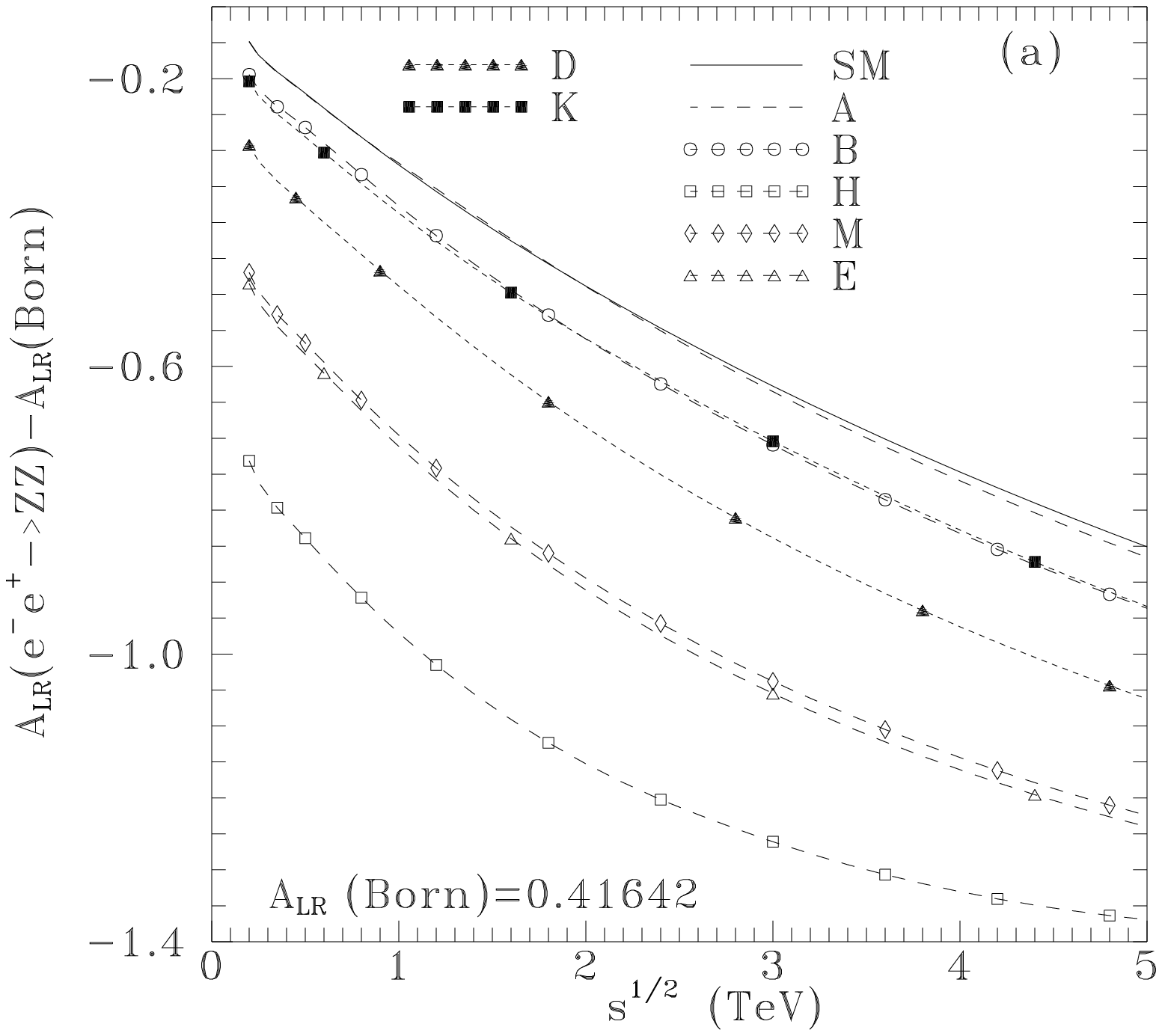,height=7cm}
\hspace{0.5cm}\epsfig{file=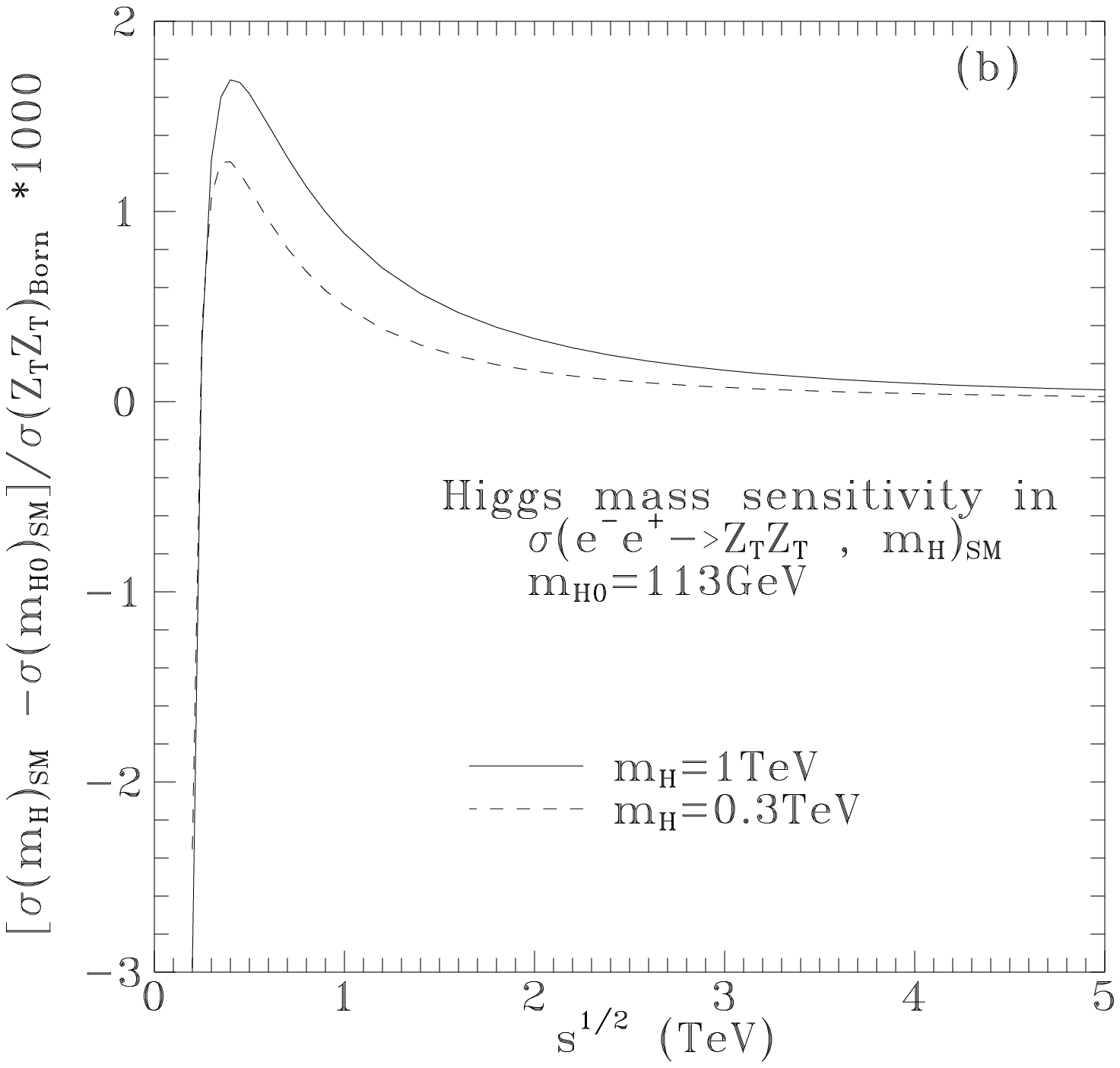,height=7cm}
\]
\vspace*{0.5cm}
\[\hspace{-0.5cm}\epsfig{file=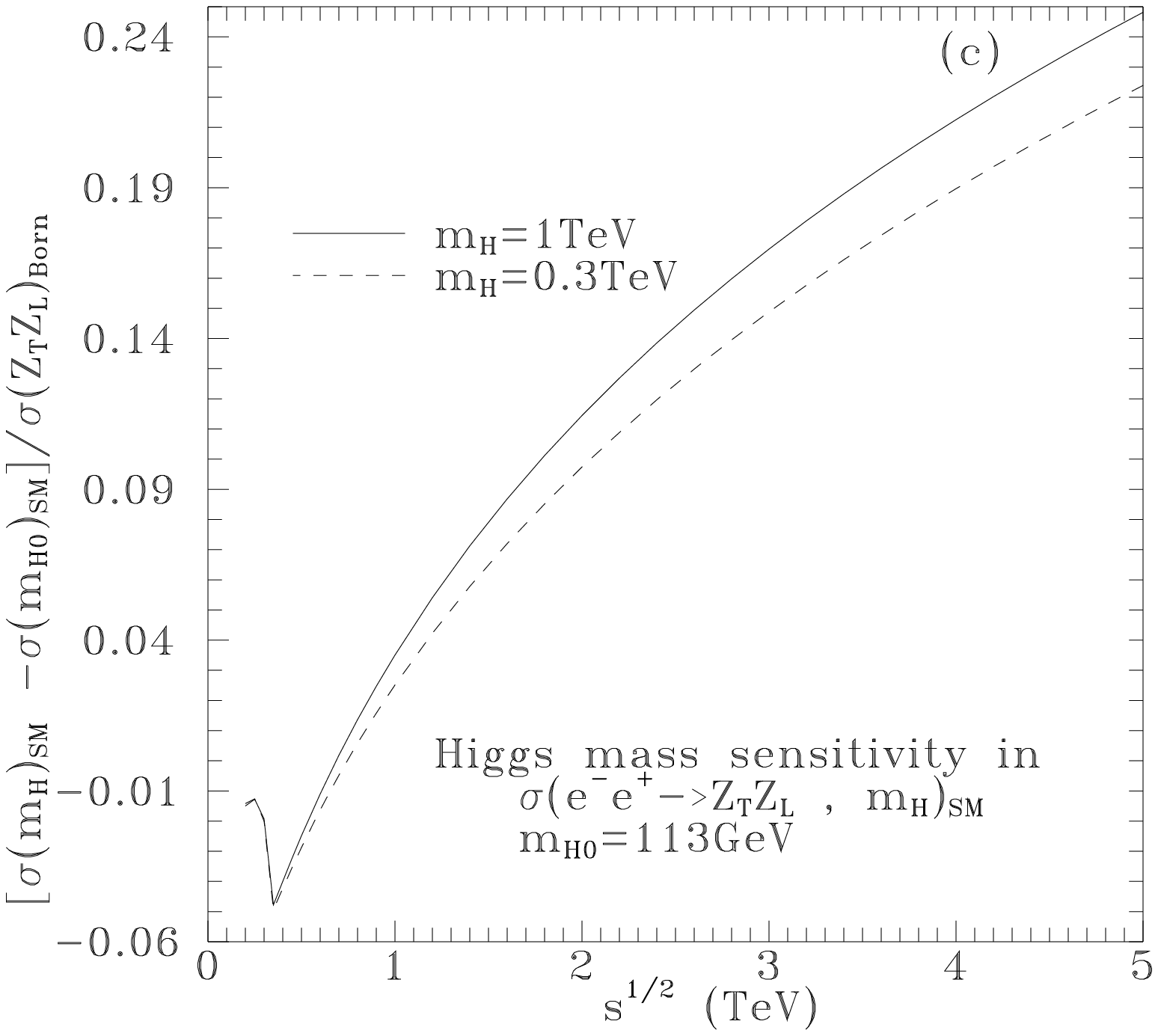,height=7cm}
\hspace{0.5cm}\epsfig{file=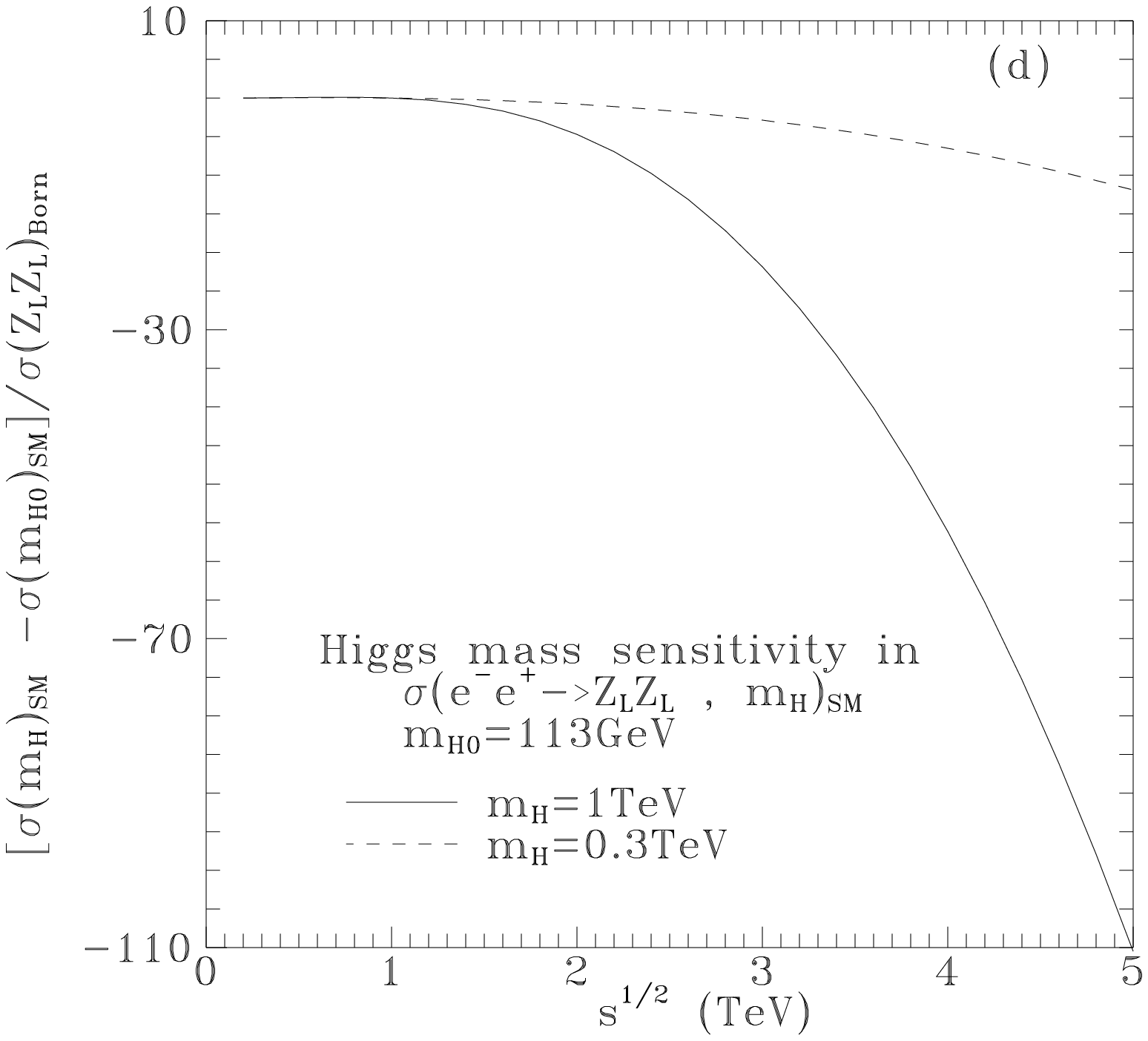,height=7cm}
\]
\vspace*{-0.5cm}
\caption[1]{the $A_{LR}$ asymmetry in $e^+e^-\to ZZ$ (a) and
the Higgs box contribution to the TT (b),
TL (c) and LL (d) $e^+e^-\to ZZ$ cross sections
as a function of the energy.}
\label{ZZ-higgs-fig}
\end{figure}

\clearpage
\newpage

\begin{figure}[p]
\vspace*{-3cm}
\[
\hspace{-0.5cm}\epsfig{file=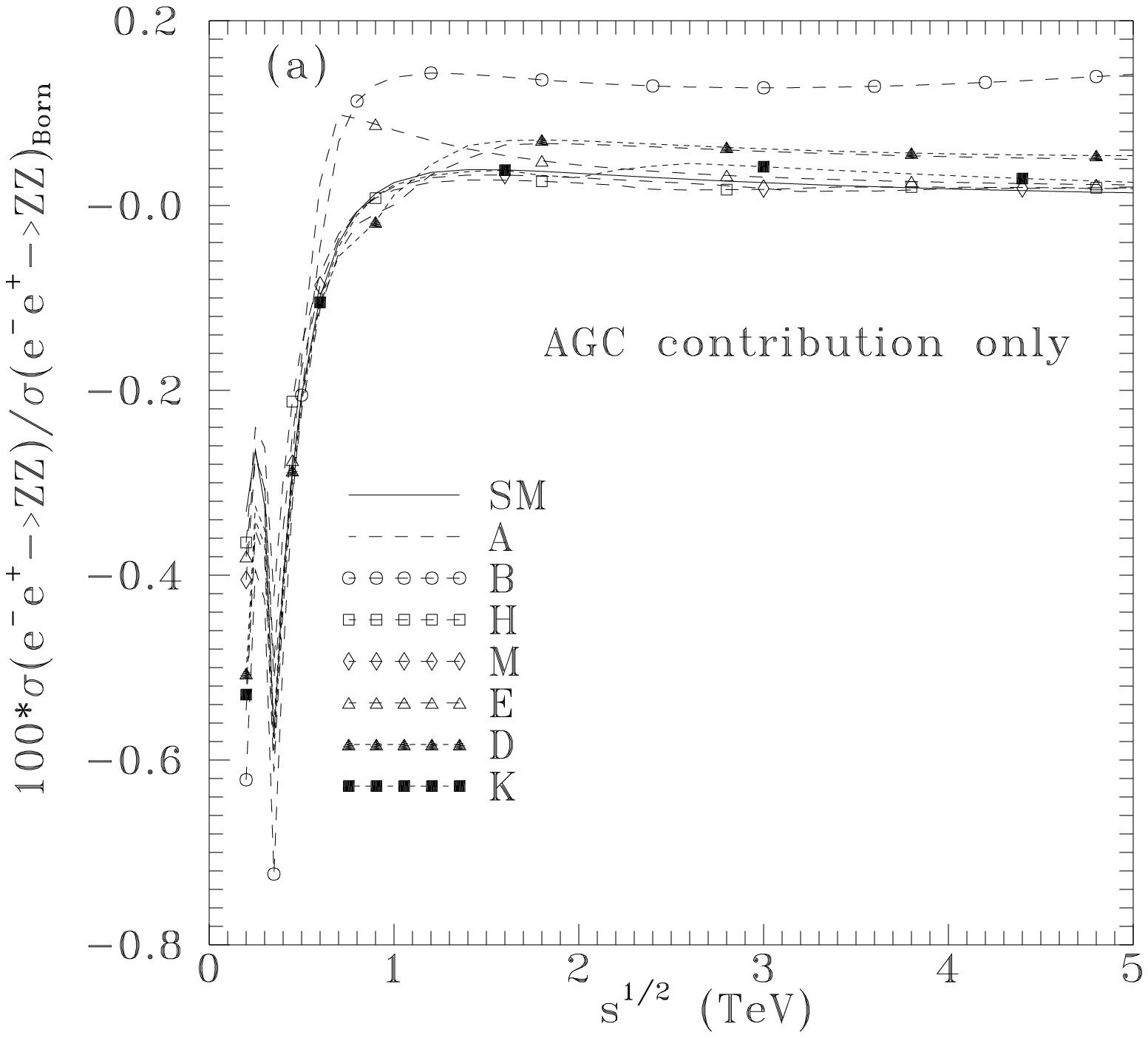,height=7cm}
\hspace{0.5cm}\epsfig{file=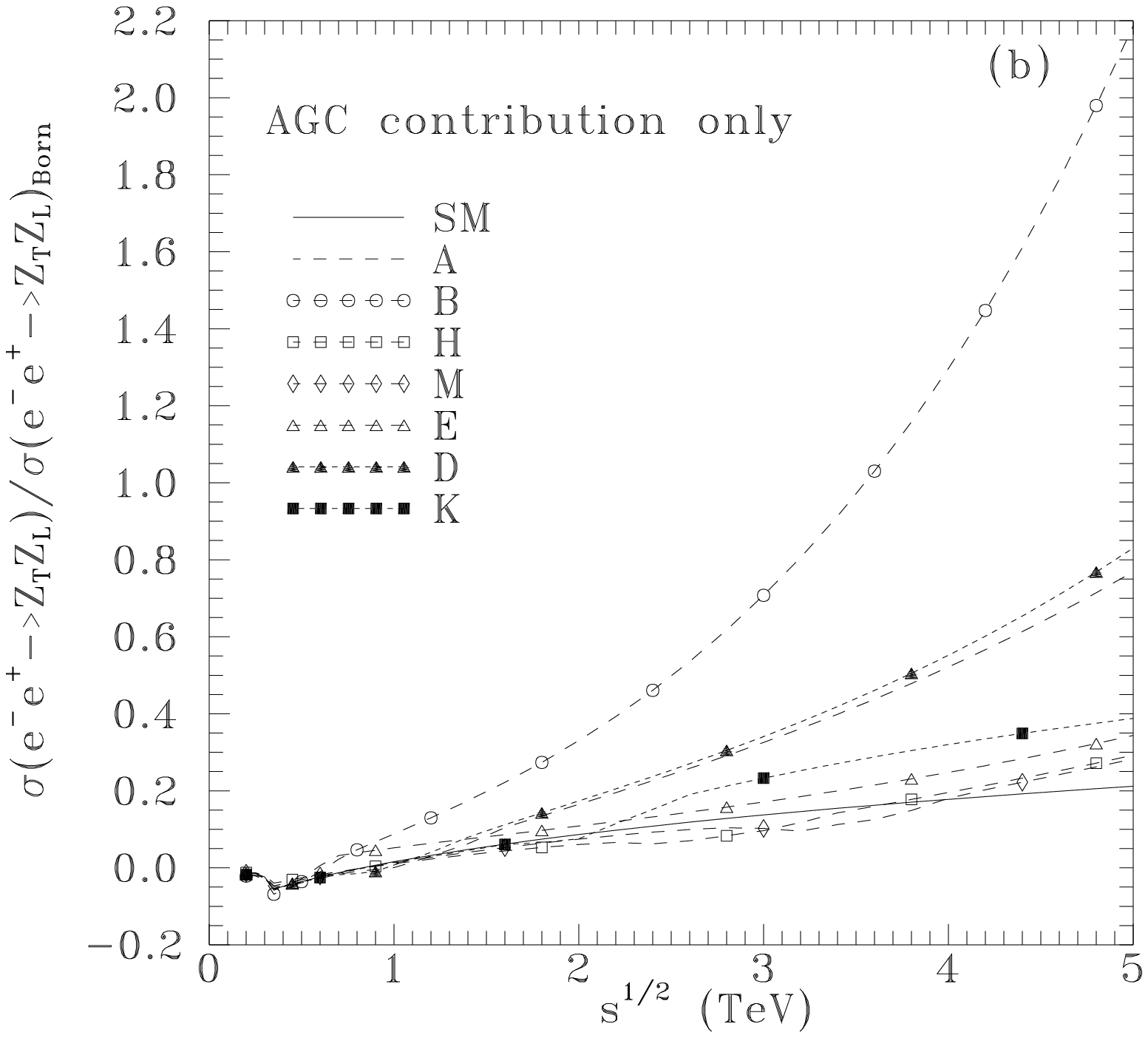,height=7cm}\]
\vspace*{0.5cm}
\[
\hspace{-0.5cm}\epsfig{file=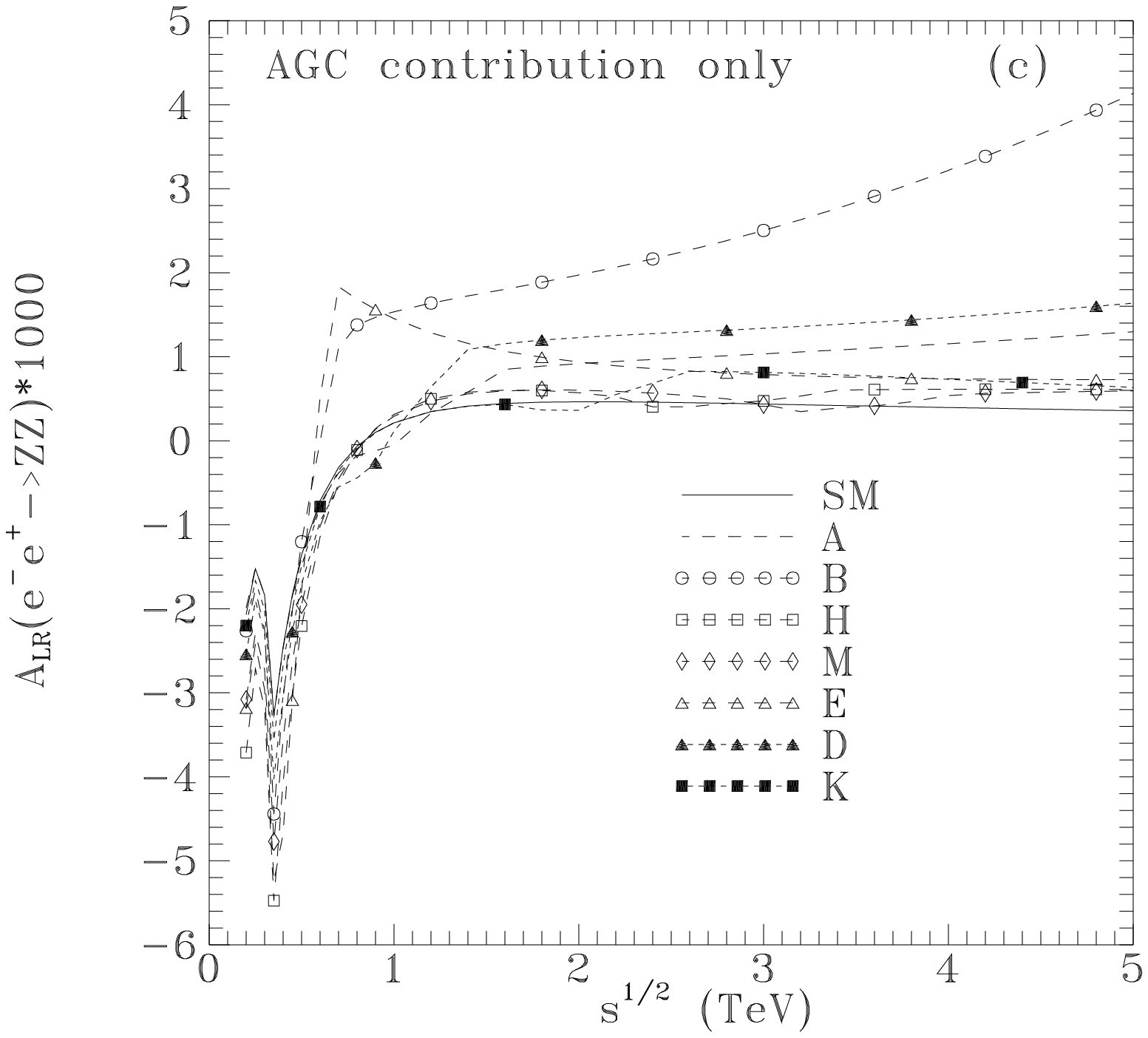,height=7cm}
\]
\vspace*{-0.5cm}
\caption[1]{NAGC contributions to the
unpolarized (a) and  TL  ~$e^+e^-\to ZZ$ cross sections (b),
and to the $A_{LR}$ asymmetry (c), as a function of the energy.}
\label{ZZ-NAGC-fig}
\end{figure}

\end{document}